# Local Symmetry Breaking in Skyrmion-Hosting Centrosymmetric Hexagonal Compounds

*Anupam K. Singh[1], Krishna K. Dubey[1], Parul Devi[2], Pritam Das[3], Martin Etter[4], Ola. G. Grendal[5], Catherine Dejoie[5], Andrew Fitch[5], Anatoliy Senyshyn[6], Seung-Cheol Lee[3], Satadeep Bhattacharjee[3*], Sanjay Singh[1*], and Dhananjai Pandey[1]*

[1]School of Materials Science and Technology, Indian Institute of Technology (Banaras Hindu University), Varanasi 221005, India
Email: ssingh.mst@itbhu.ac.in

[2]Dresden High Magnetic Field Laboratory, Helmholtz-Zentrum Dresden – Rossendorf, Bautzner Landstr. 400, 01328 Dresden, Germany

[3]Indo-Korea Science and Technology Center (IKST), Bangalore 560065, India
Email: s.bhattacharjee@ikst.res.in

[4]Deutsches Elektronen-Synchrotron (DESY), 22607 Hamburg, Germany

[5]The European Synchrotron Radiation Facility (ESRF), 71 Avenue des Martyrs, CS40220, Grenoble Cedex 9 38043, France

[6]Forschungsneutronenquelle Heinz Maier-Leibnitz (FRM-II), Technische Universität München, Lichtenbergstrasse 1, D-85747 Garching b. München, Germany

**Abstract**

Dzyaloshinskii-Moriya interaction (DMI) in the noncentrosymmetric crystals plays a crucial role to stabilize the unique topologically stable chiral spin-textures, such as skyrmions which has been proposed to host tremendous potential for next generation electronics. The recent discovery of such spin-textures in the globally centrosymmetric crystals with vanishing DMI has been debatable, whereas theoretical studies suggest non-vanishing DMI even if there is local inversion symmetry breaking in an otherwise globally centrosymmetric crystal. We present here the first experimental evidence of local symmetry breaking by a systematic crystal structure study of two skyrmion-hosting centrosymmetric hexagonal compounds, MnNiGa and MnPtGa, using the atomic pair distribution function method. Our results reveal that the local structure of both compounds corresponds to the noncentrosymmetric trigonal ($P3m1$), while average global structure remains centrosymmetric hexagonal ($P6_3/mmc$). These findings are also supported by theoretical calculations, elucidating the local DMI plays an important role behind stabilization of topological spin-textures in such centrosymmetric compounds. Our results in conjunction with the recent theoretical predictions, provide a rationale for the genesis of skyrmions in centrosymmetric materials in terms of non-vanishing DMI due to local inversion symmetry breaking, and open a new prospective for a systematic search of skyrmions and other topological phenomena in a vast family of centrosymmetric materials.

## 1. Introduction

The phenomenological theories of macroscopic symmetry breaking leading to the emergence of an order parameter below a critical transition temperature ($T_C$) has played a key role in explaining a rich variety of phenomena ranging from superconductivity[1-2] and superfluidity[1-3] to displacive ferroelectricity[4], magnetism[5] and quantum phase transitions[6-7]. In recent years, there is a growing realization that local symmetry breaking can also provide a microscopic insight towards understanding several exotic emergent phenomena in condensed matter and materials physics[8-11]. For example, the issue of energy scales in the dual/single-orbital ordered state associated with the low temperature Peierls transition in some of the exciting materilas for e.g., $NaTiSi_2O_6$, $CuIr_2S_4$ and $MgTi_2O_4$ has been resolved in terms of the local symmetry breaking-induced orbital degeneracy lifting present well above the transition temperature[12-13]. Similarly, local symmetry breaking and consequent Rashba spin-orbit coupling in compounds like $CeRh_2As_2$ has been shown to lead to field-induced transition from spin-singlet to spin-triplet superconducting states with unusually high critical fields (~14 T), opening new vistas for spin-orbitronics and topological phenomena[14-15]. In the fascinating relaxor ferroelectrics[16-17] and strain glasses[18] also, local symmetry breaking has been shown to hold the key to the understanding of critical glassy freezing and ergodic symmetry breaking. Even in the displacive ferroelectric systems, like $BaTiO_3$ and the related compound $KNbO_3$, where signatures of macroscopic symmetry breaking have been established by conventional diffraction techniques[19], it is now known that local structure in these compounds remains rhombohedral regardless of the macroscopic symmetry-broken states like tetragonal or orthorhombic[20-21].

Skyrmions are an exotic class of particle-like topologically stable spin-textures in which the spins swirl around a unit sphere radially leading to spin-vortex on nanometric length scales[22-24]. Ever since their discovery, the magnetic skyrmions have received vast attention due to their potential as high-density information carriers useful for data processing and storage in the spintronic devices operating at very low currents density[22-25]. The emergence of chiral skyrmionic textures is generally believed to be stabilized through a competition between the Heisenberg exchange interactions and the asymmetric Dzyaloshinskii-Moriya interactions (DMI)[26-27], and as such the search for skyrmions was initially limited to the noncentrosymmetric materials (e.g., B20 magnets, etc.)[28-30]. However, skyrmionic textures were subsequently discovered in a few centrosymmetric materials as well for which the DMI is supposed to be effectively zero[31-34]. In the absence of DMI, the emergence of the skyrmions in such centrosymmetric structures were attributed to the competition between the magnetic anisotropy energy and dipole-dipole interaction energy[35-37], whereas some recent studies reveal even short-range geometrical frustration and itinerant electron-mediated interactions can stabilize such topological spin-textures[38-39]. Thus, an apparent physical mechanisms behind stabilization of such spin-textures in centrosymmetric compounds has remain elusive and controversial[38].

More recently, the $Ni_2In$-type centrosymmetric hexagonal compounds with $P6_3/mmc$ space group (SG)[40-42], especially MnNiGa and MnPtGa, have generated considerable interest as the former exhibits super-stable biskyrmions in a wide temperature range (16–340 K)[43-44] with a high ferromagnetic (FM) transition temperature $T_C$ (~ 350 K)[43-46], while the latter hosts robust Neel-type skyrmions in a fairly wide temperature range 5-220 K[47]. In view of the absence of the DMI in such centrosymmetric compounds, the crystal structure and the origin of biskyrmionic and Neel-type skyrmionic textures, respectively, has been under intense debate[36-37, 43-51]. Recent theoretical predictions suggest that the local inversion symmetry breaking in an otherwise centrosymmetric crystal can also induce DMI[52-57] and hence provide a possible explanation for the observed biskyrmions and skyrmions in terms of competition between exchange interactions and DMI[52-57]. However, the conventional structure determination/refinement techniques using single-crystals[47] or polycrystalline samples[48], based on the analysis of the intensities of the Bragg peaks only, can not capture the local symmetry breaking as they provide information about the average long-range ordered (LRO) structure only[58]. The signatures of the deviations from the average LRO structure in the local or short-range (SR) regime are not present in the Bragg peaks, observed at the reciprocal lattice nodes, but are hidden in the diffuse scattering extending beyond the reciprocal lattice nodes[59]. The possible signatures of deviations from the average structure in the SR regime, and hence the local symmetry breaking in centrosymmetric compounds, can be obtained by the atomic pair distribution function (PDF)[60-63], as it utilizes the total scattering, which includes both the Bragg and the diffuse scatterings. Indeed, experimentally, the atomic PDFs are obtained after Fourier inversion of the high-$Q$ powder diffraction data[60-63]. This PDF approach provides information about the real-space structure in the SR (1 to 3 unit cells), medium-range (MR) and long-range (LR) regimes simultaneously[16, 60, 64-65]. The power of atomic PDF has been demonstrated for explaining a rich variety of phenomena in terms of local disorder/symmetry breaking in recent times[12-13, 65-69].

Despite the recent theoretical predictions about the role of local symmetry breaking induced-DMI[52-57] and the debatable mechanism behind the stabilization of topological spin-textures in inversion symmetric crystals[38-39], a direct experimental evidence of local symmetry breaking-induced DMI and understanding its impact on the stabilization of topological spin-textures in average centrosymmetric materials is still required. We present here the results of the first experimenl proof of local symmetry breaking by atomic PDF study on the $Ni_2In$-type centrosymmetric hexagonal compounds (MnNiGa and MnPtGa) using the high-$Q$, high energy and high flux synchrotron x-ray powder diffraction (SXRPD) data. Our reciprocal space Rietveld refinement results, using the high-$Q$ SXRPD data, confirm that the average structure remains hexagonal in the centrosymmetric $P6_3/mmc$ SG in the 400 to 100 K range in agreement with the previous reports on both compounds[43, 45, 48, 51]. In contrast, our real-space Rietveld refinements, using the experimental atomic PDFs, reveal for the first time that the local structure in the SR regime for both compounds corresponds to the noncentrosymmetric trigonal structure in the $P3m1$ SG, while the average structure in the MR+LR regimes remains centrosymmetric in the $P6_3/mmc$ SG. Our findings acquire special significance in understanding the physics of skyrmions in the so-called centrosymmetric crystals in the light of the recent theoretical predictions about the DMI induced by local inversion symmetry breaking[52-57] and suggest that local DMI plays a key role behind the stabilization of topological spin-textures in average centrosymmetric MnNiGa and MnPtGa compounds. Further, our results also resolve the recent controversy about the average LRO structure of MnPtGa, obtained by reciprocal space structure refinement techniques using single-crystal[47] and powder diffraction data[48]. We believe that our findings offer a new benchmark to understand the debatable mechanism behind the stabilization of topological spin-textures in such centrosymmetric compounds and open a perspetctive for a systematic study about skyrmionic textures and other related toplogical phenomena in a large family of centrosymmetric materials, which were hitherto ignored based on the general belief about the absence of DMI and hence the skyrmions in such materials.

## 2. Results and Discussion

The details of sample preparation, experimental measurements and data analysis are given in the Supporting Information (SI). Both the compounds (MnNiGa and MnPtGa) were synthesized using standard arc-melting technique. Chemical composition was confirmed by energy dispersive x-rays analysis. The magnetization measurements were performed using a Physical Properties Measurement System to detect the phase transition. High-$Q$ as well as high-resolution SXRPD and powder neutron diffraction data were used for crystallographic studies.

### 2.1. Temperature-dependent magnetization

Temperature dependence of the magnetization M(T) of MnNiGa and MnPtGa measured on a zero-field cooled sample during the warming cycle under 100 Oe field, are shown in Figure 1 for the temperature range 2-400 K. Coming from the higher temperature paramagnetic (PM) side for MnNiGa, one observes a sudden rise in M(T) around 355 K which corresponds to the PM to the FM phase transition with $T_{\mathrm{C}} \sim 347$ K. On further lowering the temperature the M(T) starts decreasing around 210 K due to a spin reorientation transition (SRT)[45] with $T_{\mathrm{SRT}} \sim 200$ K leading to a helical magnetic structure. The values of $T_{\mathrm{C}}$ and $T_{\mathrm{SRT}}$ were obtained by taking minima and maxima of temperature derivative of M(T) data. The observed behavior of M(T) and the two magnetic transition temperatures are in good agreement with the previous report on MnNiGa[43, 45]. Similar to MnNiGa, the compound MnPtGa also exhibits a PM to the FM phase transition but with a lower transition temperature $T_C \sim 235$ K, followed by a SRT at a temperature $T_{\mathrm{SRT}} \sim 188$ K. The value of $T_{\mathrm{SRT}}$ for MnPtGa is also in good agreement with the existing report[48] whereas the $T_{\mathrm{C}}$ is slightly higher in our case.

### 2.2. Global crystal structure analysis

Rietveld refinements using high-resolution SXRPD data confirm that the average LRO structure of both the compounds corresponds to the hexagonal $P6_3/mmc$ SG (see Figure S1 and the related discussion in the SI). This was additionally confirmed by refinement using powder neutron diffraction data for MnPtGa (see Figure S2 and related discussion in the SI). The reciprocal space Rietveld refinements carried out using the high-$Q$ SXRPD data also supports the average LRO hexagonal structure for MnNiGa in the temperature range of 400-100 K. This can be seen from the excellent fits between the observed and calculated profiles, shown for three representative temperatures 400, 300 and 100 K in Figure S3a, S3b and S3c, respectively, of the SI. Table S1 of the SI lists the refined structural parameters at the three characteristic temperatures obtained after Rietveld refinement using the high-$Q$ SXRPD data. We note that considering the anisotropy thermal parameters in the Rietveld refinement (Figure S3d of the SI) reveals slightly higher thermal parameter values for Ni and Ga, indicating some extent of hidden disorder, which is comprehensively discussed using real space atomic PDF analysis in the next section.

### 2.3. Local crystal structure analysis

As discussed earlier, the reciprocal space Rietveld refinement ignores the diffuse scattering resulting from the deviation from the average LRO structure and as such cannot capture the local disorder in the SR regime[58]. In order to capture the deviations from the average LRO structure, we obtained the experimental atomic PDFs G(r) of MnNiGa by taking the Fourier transform of the reduced structure functions F($Q$) given in Figure S4 of the SI for the three representative temperatures 400, 300 and 100 K, following the procedure given in the SI. The PDFs so obtained are shown in Figure S5 of the SI. We note that for the PDF data quality assessment, the structure function S($Q$) is shown in Figure S6 with related discussion in the SI. The real-space structure refinements (PDF refinements) were first carried out using the hexagonal SG $P6_3/mmc$ corresponding to the average structure at the three selected temperatures. The initial values of lattice parameters, atomic displacement parameters (ADPs) and atomic positions were taken from the reciprocal space Rietveld refined average LRO hexagonal structure at the respective temperatures given in Table S1 of the SI. The lattice parameters and ADPs were varied during the PDF refinements, as all the atoms in the asymmetric unit for this SG occupy the special Wyckoff positions only. The results of such a real-space refinement using the experimental PDF pattern in the 2.1-30 Å range at 400 K, which is representative of the crystal structure in the PM phase region, is shown in Figure 2a for isotropic ADPs ($U_{iso}$). The magnified views for selected SR, MR and LR regimes are shown in Figure 2b, 2c and 2d, respectively. It is evident from these figures that the fits between the observed and calculated PDF patterns are unsatisfactory in all the regimes. In contrast, consideration of anisotropic ADPs ($U_{aniso}$) for the Mn, Ni and Ga atoms leads to excellent fits in the MR and LR regimes, as can be seen from the magnified view shown in Figure 2g and Figure 2h, respectively. The $R_w$ factor is also lowered drastically from 13.3% to 6.4%. Such a large decrease in the value of $R_w$, upon consideration of the anisotropic ADPs in the PDF refinement is a signature of the presence of significant anisotropy in the crystal structure, as discussed in the literature for other systems also[70-71]. A similar behaviour was also observed below the $T_C$ in the FM region as well as below the SRT temperature $T_{SRT}$, as can be seen from Figure S7 of the SI for the other two representative temperatures 300 K (< $T_C$) and 100 K (< $T_{SRT}$). The parameters obtained after the PDF refinements in the $r = 2.1$ to $r = 30$ Å range at 400, 300, and 100 K, are given in Tables S2a and S3a of the SI for isotropic and anisotropic ADPs, respectively, while such parameters obtained after the PDF refinements in the SR ($r = 2.1$ to $r = 5.25$ Å) are given in Tables S2b and S3b of the SI.

Although the fit for the SR regime has also improved using anisotropic ADPs, a close visual inspection reveals slight misfit just below the shoulder of the 5 Å peak in Figure 2f (see the encircled region of the difference PDF). Provided that we did not find any compelling evidence of anti-site defects from the real-space PDF and reciprocal space Rietveld refinements including the neutron diffraction data, the possibility of anti-site defects behind the small misfit arises in Figure 2e, f is discarded. In order to get further insight into the possible role of $U_{\text{aniso}}$ in the SR, MR and LR regimes, real-space refinements were carried out considering PDFs up to different length scales ($r$) varying from $r_{\text{max}} = 4.41$ to 30 Å (i.e., '$r$'-dependent PDF refinement) and the results for the Mn atom are shown in Figure 2i. We note that the lattice parameters and correlation parameter were kept fixed to the value obtained from the full range refinement ($r = 2.1$ to $r = 30$ Å), during the $r$-dependent PDF refinements[12]. The result of such a refinement is given in Figure S7c,d of the SI. It is evident from Figure 2i that for the 7 Å $< r \leq$ 30 Å range, $U_{33} \cong U_{11}$. However, for refinements in the $r <$ 7 Å region, the $U_{33}$ shows an anomalously increasing trend while $U_{11}$ decreases a little with respect to the refined values for $r >$ 7 Å. These refinements thus confirm that the anisotropic ADPs ($U_{11}$ and $U_{33}$) of Mn are '$r$'- dependent. It was verified that the ADPs for Ni are rather small and nearly isotropic, while ADPs for Ga are nearly independent of length scale ($r$) and are significantly lower than the $U_{33}$ of Mn in the SR regime. The drastic increase in $U_{33}$ of Mn in the SR regime gave us the first indication of local structural distortion, similar to that reported in other systems[72-73]. For a better visualization of changes in the SR regime, the anisotropic ADP ratio ($U_{33}/U_{11}$) of Mn at 400 K is shown in the inset of Figure 2i. The anomalous increase in $U_{33}/U_{11}$ in the SR regime suggests the dominant ADP along the $c$-direction as compared to the ADP in the basal plane. A similar behaviour was also observed at other temperatures below $T_{\text{C}}$ and $T_{\text{SRT}}$ (see Figure S7e,f of the SI). The value of thermal parameter ($B$) for Mn, obtained using the relationship $B_{33} = 8\pi^2 U_{33}$[74], comes out to be ~2.22 Å$^2$ in the SR regime for refinements upto $r$ ~ 4.41 Å at 400 K. Such a large values of $B_{33}$ (and $U_{33}$) along the $c$-axis in the SR regime suggest that the Mn atoms may be displaced from their registered Wyckoff positions for the $P6_3/mmc$ SG in the SR regime, as discussed in context of the other systems also[72-73, 75-78].

Since all the atoms in the asymmetric unit of the $P6_3/mmc$ SG occupy special Wyckoff positions, i.e., Mn at 2a (0, 0, 0), Ni at 2d (1/3, 2/3, 3/4), and Ga at 2c (1/3, 2/3, 1/4)[43], any local atomic displacement with respect to these special positions can only be modelled by considering a subgroup of $P6_3/mmc$. Accordingly, we considered all possible hexagonal/primitive trigonal subgroups of $P6_3/mmc$, generated by the ISODISTORT software in the ISOTROPY suite[79-80], given in the SI, one by one for the real-space structure refinement. The subgroups not belonging to the hexagonal/primitive trigonal space groups predict additional peaks in the diffraction pattern. Since all the peaks in the high-$Q$ SXRPD patterns could be indexed using the hexagonal/primitive trigonal system, the lower symmetry subgroups belonging to the orthorhombic, monoclinic and triclinic space groups were not considered for the real-space structure refinements in the SR regime. The observed, calculated and difference PDF patterns obtained by real-space structure refinements in the SR regime ($r \leq 5.25$ Å) for all the hexagonal/primitive trigonal subgroups are compared with that for the $P6_3/mmc$ SG at 300 K in Figure S8 of the SI. The best fit to the experimental PDF was obtained for the noncentrosymmetric $P3$ (153) and $P3m1$ (156) subgroups which gave significantly lower value of $R_w$ ~6.6% as compared to $R_w$ ~11.4% for the $P6_3/mmc$ SG. Between these two subgroups, we consider the higher symmetry SG $P3m1$ as per symmetry lowering criterion. We note that any possible artifacts related to the number of refinable parameters in such PDF data modelling are discarded as discussed in the context of Figure S9 in the SI.

The results of the real-space structure refinements for the SR and MR+LR regimes considering isotropic ADPs with $P6_3/mmc$ and $P3m1$ space groups are compared in Figure 3. It is evident from a comparison of Figure 3a and Figure 3b that the misfits observed at a distance $r \sim$ 2.35 Å, 3.5 Å, 4.25 Å and 4.75 Å for the $P6_3/mmc$ SG have almost disappeared for the $P3m1$ SG, as a result of which the $R_w$ has decreased drastically from 11.8% for the former SG to 6.2% for the latter. This drastic improvement in the fits between the observed and calculated PDF patterns is the result of the emergence of new bond lengths for the $P3m1$ SG, which were otherwise degenerate for the $P6_3/mmc$ SG, as can be seen from Table S4 of the SI. Thus, the results of our PDF analysis presented in Figure 3a,b suggest that the $P3m1$ SG symmetry is the most plausible lower symmetry structure to model the experimental PDF in the SR regime in the PM phase of MnNiGa at 400 K. In sharp contrast to the SR regime, the atomic PDF of MnNiGa in the MR+LR regimes are better modelled by the $P6_3/mmc$ SG as can be seen from a comparison of Figure 3c,d. It is important to mentione here that during MR+LR refinements, the atomic positions were fixed to the values obtained from the refinement in the SR regime (Figure 3c,d), as the displacements in the atoms should not alter on going from the SR to the MR+LR regimes if the $P3m1$ would have the correct LRO structure. There are several published reports, where the atomic positions were fixed corresponding to the SR regime during the refinements for the MR and LR regimes for the same SG[12-13]. The PDF fit in the MR+LR regimes for the $P6_3/mmc$ SG, shown in Figure 3c, leads to a $R_w \sim 13.1\%$ which is lower than that for the $P3m1$ SG, shown in Figure 3d, with $R_w \sim 16.2\%$. The improvement in the fits around some of the PDF peaks is highlighted with arrows in these two figures. Thus, the results of PDF analysis at 400 K reveal that the local structure of MnNiGa in the SR regime is best modeled with the primitive trigonal SG $P3m1$, whereas the structure in the MR+LR regimes corresponds to the average hexagonal structure in the $P6_3/mmc$ SG, obtained by the consideration of the intensity of the Bragg peaks only in the conventional reciprocal space Rietveld refinement using SXRPD data. A similar analysis carried out using the PDF data below the PM to the FM phase transition temperature $T_C \sim 347$ K and below the FM to spin reorientation transition (SRT) at $T_{SRT} \sim 200$ K shows that the local structure in the SR regime and the average structure in the MR+LR regimes continue to correspond to the $P3m1$ and $P6_3/mmc$ space groups, respectively, at these temperatures also (see Figure S10a-S10h and the related discussion in the SI) similar to that at 400 K. The parameters obtained after the PDF refinements in the SR regime at the three selected temperatures 400, 300 and 100 K using the $P3m1$ SG are compared in Table S5 of the SI, which suggests that the $z_{Mn2}$, $z_{Ni2}$ and $z_{Ga2}$ decrease with decreasing temperature and approach the value of 0.5, 0.25 and 0.75, within the esds, expected for the $P6_3/mmc$ SG, at the lowest temperature. However, $z_{Ni1}$ and $z_{Ga1}$ are significantly displaced away from their special Wyckoff positions ($P6_3/mmc$) at all the temperatures elucidating the correctness of the structure modelling using the lower symmetry $P3m1$ SG. Furthermore, to confirm the robustness of our experimental results and data reproducibility, we have further collected high-Q SXRPD data using a higher energy (100 keV) x-ray beam at the P21.1 beamline of PETRA-III, DESY. The corresponding results are included as Figures S14-S17 with related discussion in the SI. These results also confirm the local structure of MnNiGa is trigonal ($P3m1$) while the global structure is hexagonal ($P6_3/mmc$).

In order to understand if the noncentrosymmetric local structure in the $P3m1$ SG is limited to the compound MnNiGa only or if it is a general feature of other isostructural skyrmion-hosting compounds, we performed PDF analysis on the sister skyrmion-hosting compound MnPtGa also. For this, we first confirmed the average hexagonal structure of MnPtGa in the $P6_3/mmc$ SG by the reciprocal space Rietveld refinement using the high-$Q$ SXRPD data (see Figure S11 and the related discussion in the SI). After this, we performed PDF analysis in the SR regime using a procedure similar to that employed for MnNiGa. By taking the Fourier transform of the F($Q$) (given in Figure S12 of the SI), the atomic PDF for MnPtGa is obtained at 300 K as shown in Figure S13 of the SI. The fits between the observed and the calculated PDF patterns in the SR regime for MnPtGa using $P6_3/mmc$ and $P3m1$ space groups are compared in Figure 4a and Figure 4b at 300 K. It reveals an excellent fit for the $P3m1$ SG with considerably lower $R_w \sim 8.7\%$ as compared to the $P6_3/mmc$ SG ($R_w \sim 18.1\%$). This significant improvement in the PDF fitting with the $P3m1$ SG is attributed to the appearance of new bond lengths that were otherwise degenerate for the $P6_3/mmc$ SG, as given in Table S6 of the SI. Further, for the MR+LR regimes of MnPtGa also, the SG $P6_3/mmc$ provides a better fit than the $P3m1$ SG at 300 K (see the fits around the peaks marked with arrows in Figure 4c,d). These results confirm that the local structure in the SR regime is trigonal with the $P3m1$ SG, while the average MR+LR structure remains hexagonal in the $P6_3/mmc$ SG for MnPtGa also. These results obtained from the detail atomic PDF study manifest that the local off-centering of the atoms induces local symmetry breaking which eventually tends towards the local trigonal structure ($P3m1$) embedded in an average hexagonal structure ($P6_3/mmc$).

### 2.4. Local Structure Comparison of MnNiGa and MnPtGa

To get more insight into the off-centeirng, a comparison of local structure is essential. We compare the noncentrosymmetric local structures of MnNiGa and MnPtGa schematically in Figure 5a and Figure 5b, respectively, at 300 K. The local off-center displacements of various atoms in the [0001] direction, with respect to the special positions of the $P6_3/mmc$ SG, are indicated by arrows with approximate values given in nm in Figure 5. From a comparison of Figure 5a with Figure 5b, it is evident that the off-center displacement of the Ga1 atom in MnPtGa (~0.051 nm in Figure 5b) is a little more than twice the value for MnNiGa (~0.024 nm in Figure 5a) at 300 K. This indicates a stronger departure from the centrosymmetric structure and hence stronger local DMI in MnPtGa as compared to MnNiGa for the shortest indirect exchange pathways Mn-Ga-Mn. Apart from the off-centering of the Ga atom, the off-center displacement of the Pt atoms (Pt1 and Pt2) in MnPtGa (~0.018 nm and ~0.009 nm in Figure 5b) is significantly larger than that of Ni atoms (Ni1 and Ni2) in MnNiGa (~0.011 nm and ~0.005 nm in Figure 5a) at 300 K. Thus, stronger DMI in MnPtGa is inferred for the shortest Mn-Pt-Mn indirect exchange pathways also, as compared to the Mn-Ni-Mn pathways in MnNiGa. The direct exchange pathway Mn1-Mn2-Mn1' is also slightly asymmetric as the Mn2 atom is displaced with respect to the special position of the $P6_3/mmc$ SG. Further, the Mn2-Mn1' bond length is smaller than Mn1-Mn2, Mn1-Ga1, Mn2-Ga1, Mn2-Ga2, and Mn1'-Ga2 bond lengths for both the compounds. The height difference between Ni and Ga atoms for the first and the second layers are found to be ~0.019 nm and ~0.002 nm, respectively, for the trigonal $P3m1$ SG of MnNiGa at 300 K. On the other hand, the height difference between the first layer of Pt and Ga atoms (~0.042 nm) is significantly, larger than that (~0.006 nm) for the second layer in the trigonal unit cell of MnPtGa at 300 K. All these observations suggest that the asymmetry in the structure due to the absence of inversion symmetry in the SR regime, and hence the local DMI, are expected to be significantly larger in MnPtGa than in MnNiGa. This is also expected from the stronger spin-orbit coupling (SOC) for Pt as compared to Ni[81]. It is worth mentioning here that these values for the height differences (~0.042 nm and ~0.006 nm) are slightly different from those reported (~0.033 nm and ~0.017 nm) for single-crystals of MnPtGa[47]. This small difference might be due to the fact that the PDF technique used in the present study captures the truly local trigonal structure in the SR regime, whereas it was averaged over the entire crystal in the previous single-crystal study[47]. The important point to emphasize is that the essential features in the present study for the SR regime are similar to the previous study using single-crystal[47] but are fundamentally different in the MR+LR regimes, for which our PDF study does not favour the average primitive trigonal structure proposed in Ref.[47]. Our findings for the MR+LR regimes are in better agreement with the average LR hexagonal structure reported through the reciprocal space Rietveld refinement study of MnPtGa[48] including the more recent thin film and single crystalline studies on MnPtGa[49-51]. One of the consequences of the presence of non-zero DMI is that the FM phase resulting from the PM phase can adopt non-collinear magnetic structure. In fact, even in the PM region close to $T_C$, the fluctuating spin-clusters with FM correlations can acquire non-collinearity. The fact that even a modest magnetic field can stabilize skyrmionic texture in MnNiGa even at 340 K suggests that the non-zero DMI is facilitating such a texture against thermal fluctuations. Moreover, it is worth to discuss here another important fact in context of comparing the size of topological spin-textures. Usually the size of topological spin-textures in the centrosymmetric bulk magnets with no net DMI has been found significantly smaller (by almost one order of magnitude) compared to the noncentrosymmetric materials in which finite DMI is assumed as a primary factor behind stabilizing the topological spin-textures[23, 39]. For instance, the diameter of spin-textures lies in the 2-5 nm range in centrosymmetric Gd-based materials ($Gd_2PdSi_3$, $Gd_2Ru_2Si_2$)[32, 38] while much larger diameter has been observed in the noncentrosymmetric materials for example MnSi (18 nm)[28], $Co_8Zn_8Mn_4$ (125 nm)[82], $Mn_{1.4}Pt_{0.9}Pd_{0.1}Sn$ (150 nm)[83], $(Fe_{0.5}Co_{0.5})_5GeTe_2$ (200 nm)[84]. Contrarily, in MnNiGa and MnPtGa (which were previously assumed to be average centrosymmetric materials with $Ni_2In$-type hexagonal structure in the $P6_3/mmc$ space group[40-41, 43-44, 47-51] average diameter of spin-textures has been reported to be around 90-200 nm and 190 nm, respectively, which is in marked contrast and significantly bigger compared to the Gd-based centrosymmetric materials[32, 38]. Such larger size of spin-textures in MnNiGa and MnPtGa are comparable to those observed in noncentrosymmetric materials. This reflects the major type of interaction involved behind the stabilizing spin-textures in such materials (MnNiGa and MnPtGa) is analogous to DMI that is nothing but hidden DMI locally in the average centrosymmetric structure. Our experimental PDF results discoles such hidden DMI in both $Ni_2In$-type compounds (MnNiGa and MnPtGa) and suggest that the local DMI is a key factor behind stabilization of topological spin-textures in centrosymmetric compounds as poposed recently[52-57].

## 2.5. Theoretical Calculations

In addition to the present experimental results, we have also performed theoretical first principle calculations to get more insight about the local and global structure of both the compounds (MnNiGa and MnPtGa). The details of the present calculations is described in section X of the SI. Since the PDF provides the local trigonal ($P3m1$) and global hexagonal ($P6_3/mmc$), a mixed structure model is considered for the calculation. In this particular case, to simulate such mixed structure, we consider a supercell that is formed by a $3 \times 3 \times 3$ $P6_3/mmc$ supercell and the unit cell at the center was replaced by one-unit cell of trigonal $P3m1$ structure (the light green region in the mixed structure shown in Figure 6). For the calculation with the pure structures ($P6_3/mmc$ and $P3m1$), experimentally obtained lattice parameters as well as atomic positions have been used. The result of the calculation is shown in Figure 6. The results for both MnNiGa and MnPtGa structures have similar traits. In both cases i.e., MnNiGa and MnPtGa, the mixed structure has a lower ground state energy in comparison with the $P3m1$ and $P6_3/mmc$ structures. The introduction of spin canting (non-collinear spin structure) around the centered $P3m1$ cell further lowers the ground state energy of the mixed structures. Using the collinear $P3m1$ arrangement as the zero of energy, the collinear $P6_3/mmc$ polymorph is lower by 92 meV/f.u. in MnPtGa and 25.665 meV/f.u. in MnNiGa (see Figure 6a-6b). Switching to the mixed structure (without spin canting) already yields a substantial stabilization, lowering the energy to –163 meV/f.u. for MnPtGa and –34.916 meV/f.u. for MnNiGa. We found similar results using the larger 5 × 5 × 5 $P6_3/mmc$ supercell as shown in Figure S18 of the SI. Allowing the moments within this same lattice to adopt their fully non-collinear ground state drives the energies down further to –936 meV/f.u. and –38.282 meV/f.u., respectively with a net gain of about 773 meV/f.u. for MnPtGa and 3.37 meV/f.u. for MnNiGa (see Figure 6). This consideration of spin canting (non-collinearity) leads to stabilize the vortex-type spin structure within the P3*m*1 unit cell as highlighted in Figure 6a,b. These stabilizations lie within the range expected for Dzyaloshinskii-Moriya (DM) interactions: the pronounced reduction in MnPtGa reflects the strong SOC introduced by Pt[81], whereas the smaller but finite gain in MnNiGa still shows an operative local chiral exchange. The *P*3*m*1 local distortion therefore eliminates the inversion centre at Mn–Mn mid-points while retaining the 3m polar axis, so Moriya's rules permit a DM vector mainly along the c-axis for the Mn–Mn pairs traversing the trigonal nanodomain, exactly the symmetry that promotes Neel-type skyrmions. Thus, the mixed structure is more stable than the global *P*3*m*1 and $P6_3/mmc$ symmetry structures in both MnNiGa and MnPtGa. Meanwhile, the presence of spin canting (non-collinearity) further lowers the ground state energy of the mixed structures. It is essential to emphasize here that the changes in the ground state energy of MnNiGa mix structure compared to MnPtGa mix structure are relatively small may be due to the less local distortion in the MnNiGa's *P*3*m*1 structure as observed experimentally (see Figure 5 and related discussion). Such a stark differences between the Pt and Ni containing compounds almost certainly originates from the vastly different SOC strengths of 5d-Pt versus 3d-Ni[81]. In a non-collinear framework the total energy contains three SOC-derived terms that can lower the energy when the spin moments are allowed to rotate: (i) the single-ion magnetocrystalline anisotropy (MCA), which scales roughly as $\lambda^2/\Delta$ (where, λ is the atomic SOC constant and $\Delta$ is the crystal-field splitting), (ii) anisotropic exchange and (iii) the antisymmetric DMI. For Pt, $\lambda$ is much higher than that for Ni (it is an order of magnitude smaller for Ni)[81]. So, all three contributions are intrinsically two orders of magnitude larger in the MnPtGa's lattice. Once chemical disorder is introduced in the $3 \times 3 \times 3$ supercell, local inversion symmetry is broken on many Mn pairs, enabling sizable DMI vectors that favour the canted non-collinear spin pattern visualized in Figure 6a. The resulting 773 meV/f.u. gain via spin canting in MnPtGa system is therefore consistent with a dominant SOC stabilization, whereas the relatively small 3.37 meV/f.u. gain in MnNiGa is roughly what one would expect from the SOC scale of Ni. However, it is important to note that in both compounds we see the similar trends. Moreover since the DFT calculation reveals only the trends with change in structure (or related parameters), information about the absolute energy scale is not expected. Nevertheless, these calculations further support our experimental findings of local trigonal and global hexagonal structure in these compounds. Our results from synchrotron PDF analysis show that such local distortions ($P3m1$) are confined to the SR regime (~5-6 Å, as evident by Figure 2i), and are not correlated across long distances. The MR and LR structure remain well described by the average $P6_3/mmc$ symmetry. This is consistent with the model of uncorrelated or randomly oriented $P3m1$-type nanodomains embedded within a globally centrosymmetric matrix i.e., the $P3m1$ distortions do not propagate coherently throughout the material and remain confined to local domains or clusters.

## 3. Conclusion

In summary, using a detailed analysis of the local structure by atomic PDF technique, we have for the first time confirmed that the structure of both MnNiGa and MnPtGa in the SR regime corresponds to the noncentrosymmetric trigonal SG $P3m1$ which is different from the MR+LR regimes structure, obtained by the conventional reciprocal space Rietveld refinement, which correspond to the centrosymmetric SG *P*$6_3$/*mmc*. The DMI induced by local inversion symmetry breaking in an otherwise average centrosymmetric structure has been identified as a key factor in recent theoretical calculations for the stabilization of the skyrmionic textures[52-57] as well as impacting macroscopic properties such as electronic band structure with explaining magnetism of Weyl semimetal $Co_3Sn_2S_2$[85] and inducing gapped topological magnons in $Mn_5Ge_3$[86]. The compelling evidence presented here on the locally broken inversion symmetry, which can induce local DMI, elucidating as an additional key factor behind the stabilization of the skyrmionic textures in the centrosymmetric $Ni_2In$-type skyrmion-hosting materials. The experimental result is also supported by theoretical DFT calculations. The results of the present study thus acquire broader significance in the light of the theoretical predictions for the non-zero DMI induced by local symmetry breaking for explaining a host of phenomena of immense topical interest[52-57]. We believe that our work would encourage a systematic search for local symmetry breaking ins the other so-called "centrosymmetric crystals" for understanding the genesis of several topological phenomena.

**Acknowledgements**

S.S. is thankful to the Science and Engineering Research Board of India for financial support through the award of Ramanujan Fellowship (grant no: SB/S2/RJN-015/2017) and Core Research Grant (grant no. CRG/2021/003256) and UGC-DAE CSR, Indore for financial support through its "CRS" Scheme. Portions of this research were conducted at the light source PETRA III of DESY, a member of the Helmholtz Association (HGF) and at the ESRF, France. We acknowledge the FRM-II for neutron diffraction experimensts. Financial support from the Department of Science and Technology (DST), Government of India within the framework of the India@DESY is gratefully acknowledged. We would like to thank Dr. Sanjay Kumar Mishra, Solid State Physics

Division, Bhabha Atomic Research Centre, Mumbai, for helpful discussions on the isotropy subgroups. We thank Dr. Keshav Kumar for helpful discussion about subgroup and Rietveld analysis.

## Conflict of Interest

The authors declare no conflict of interest.

## Data Availability Statement

The data that support the findings of this study are available from the corresponding author upon reasonable request.

## References

1. Du, L.; Hasan, T.; Castellanos-Gomez, A.; Liu, G.-B.; Yao, Y.; Lau, C. N.; Sun, Z., Engineering symmetry breaking in 2D layered materials. *Nature Reviews Physics* **2021,** *3* (3), 193-206.

2. Feynman, R. P., Superfluidity and Superconductivity. *Reviews of Modern Physics* **1957,** *29* (2), 205.

3. Ikegami, H.; Tsutsumi, Y.; Kono, K., Chiral symmetry breaking in superfluid 3He-A. *Science* **2013,** *341* (6141), 59-62.

4. Lines, M. E.; Glass, A. M., *Principles and Applications of Ferroelectrics and related Materials*. Oxford University Press: 2001.

5. Coey, J. M., *Magnetism and Magnetic Materials*. Cambridge University Press: 2010.

6. Feng, X.-Y.; Zhang, G.-M.; Xiang, T., Topological characterization of quantum phase transitions in a spin-1/2 model. *Physical Review Letters* **2007,** *98* (8), 087204.

7. Shi, Q.-Q.; Zhou, H.-Q.; Batchelor, M. T., Universal order parameters and quantum phase transitions: a finite-size approach. *Scientific Reports* **2015,** *5* (1), 7673.

8. Di Matteo, S.; Joly, Y.; Bombardi, A.; Paolasini, L.; de Bergevin, F.; Natoli, C., Local chiral-symmetry breaking in globally centrosymmetric crystals. *Physical Review Letters* **2003,** *91* (25), 257402.

9. Lu, L.; Song, M.; Liu, W.; Reyes, A.; Kuhns, P.; Lee, H.; Fisher, I.; Mitrović, V., Magnetism and local symmetry breaking in a Mott insulator with strong spin orbit interactions. *Nature Communications* **2017,** *8* (1), 14407.

10. Zhao, K.; Bruinsma, R.; Mason, T. G., Local chiral symmetry breaking in triatic liquid crystals. *Nature Communications* **2012,** *3* (1), 801.

11. Hu, L.; Qin, F.; Sanson, A.; Huang, L.-F.; Pan, Z.; Li, Q.; Sun, Q.; Wang, L.; Guo, F.; Aydemir, U., Localized symmetry breaking for tuning thermal expansion in ScF3 nanoscale frameworks. *Journal of the American Chemical Society* **2018,** *140* (13), 4477-4480.

12. Koch, R. J.; Sinclair, R.; McDonnell, M. T.; Yu, R.; Abeykoon, M.; Tucker, M. G.; Tsvelik, A.; Billinge, S.; Zhou, H.; Yin, W.-G., Dual Orbital Degeneracy Lifting in a Strongly Correlated Electron System. *Physical Review Letters* **2021,** *126* (18), 186402.

13. Bozin, E. S.; Yin, W.; Koch, R.; Abeykoon, M.; Hor, Y. S.; Zheng, H.; Lei, H.; Petrovic, C.; Mitchell, J.; Billinge, S., Local orbital degeneracy lifting as a precursor to an orbital-selective Peierls transition. *Nature Communications* **2019,** *10* (1), 3638.
14. Khim, S.; Landaeta, J.; Banda, J.; Bannor, N.; Brando, M.; Brydon, P.; Hafner, D.; Küchler, R.; Cardoso-Gil, R.; Stockert, U., Field-induced transition within the superconducting state of $CeRh_2As_2$. *Science* **2021,** *373* (6558), 1012-1016.
15. Pourret, A.; Knebel, G., Driving multiphase superconductivity. *Science* **2021,** *373* (6558), 962-963.
16. Jeong, I.-K.; Darling, T.; Lee, J.; Proffen, T.; Heffner, R.; Park, J.; Hong, K.; Dmowski, W.; Egami, T., Direct observation of the formation of polar nanoregions in $Pb(Mg_{1/3}Nb_{2/3})O_3$ using neutron pair distribution function analysis. *Physical Review Letters* **2005,** *94* (14), 147602.
17. Dmowski, W.; Vakhrushev, S.; Jeong, I.-K.; Hehlen, M.; Trouw, F.; Egami, T., Local lattice dynamics and the origin of the relaxor ferroelectric behavior. *Physical Review Letters* **2008,** *100* (13), 137602.
18. Zhou, Y.; Xue, D.; Tian, Y.; Ding, X.; Guo, S.; Otsuka, K.; Sun, J.; Ren, X., Direct evidence for local symmetry breaking during a strain glass transition. *Physical Review Letters* **2014,** *112* (2), 025701.
19. Jona, F.; Shirane, G., *Ferroelectric Crystals, International Series of Monographs on Solid State Physics*. Pergamon press Oxford, UK:: 1962.
20. Senn, M.; Keen, D.; Lucas, T.; Hriljac, J.; Goodwin, A., Emergence of long-range order in $BaTiO_3$ from local symmetry-breaking distortions. *Physical review letters* **2016,** *116* (20), 207602.
21. Yoneda, Y.; Ohara, K.; Nagata, H., Local structure and phase transitions of $KNbO_3$. *Japanese Journal of Applied Physics* **2018,** *57* (11S), 11UB07.
22. Fert, A.; Reyren, N.; Cros, V., Magnetic skyrmions: advances in physics and potential applications. *Nature Reviews Materials* **2017,** *2* (7), 17031.
23. Nagaosa, N.; Tokura, Y., Topological properties and dynamics of magnetic skyrmions. *Nature Nanotechnology* **2013,** *8* (12), 899.
24. Bogdanov, A. N.; Panagopoulos, C., Physical foundations and basic properties of magnetic skyrmions. *Nature Reviews Physics* **2020,** *2* (9), 492-498.
25. Xu, T.; Guo, X. Y.; Liu, Y.; Du, H.; Shao, D. F., Thin-Film Magnetic Skyrmions for Spintronic Devices. *Advanced Functional Materials* **2025**, 2504100.
26. Dzyaloshinsky, I., A thermodynamic theory of "weak" ferromagnetism of antiferromagnetics. *Journal of Physics and Chemistry of Solids* **1958,** *4* (4), 241-255.
27. Moriya, T., Anisotropic superexchange interaction and weak ferromagnetism. *Physical Review* **1960,** *120* (1), 91.
28. Mühlbauer, S.; Binz, B.; Jonietz, F.; Pfleiderer, C.; Rosch, A.; Neubauer, A.; Georgii, R.; Böni, P., Skyrmion lattice in a chiral magnet. *Science* **2009,** *323* (5916), 915-919.
29. Yu, X.; Kanazawa, N.; Onose, Y.; Kimoto, K.; Zhang, W.; Ishiwata, S.; Matsui, Y.; Tokura, Y., Near room-temperature formation of a skyrmion crystal in thin-films of the helimagnet FeGe. *Nature Materials* **2011,** *10* (2), 106-109.
30. Jamaluddin, S.; Manna, S. K.; Giri, B.; Madduri, P. P.; Parkin, S. S.; Nayak, A. K., Robust Antiskyrmion Phase in Bulk Tetragonal Mn–Pt (Pd)–Sn Heusler System Probed by Magnetic Entropy Change and AC-Susceptibility Measurements. *Advanced Functional Materials* **2019**, 1901776.

31. Yu, X.; Tokunaga, Y.; Kaneko, Y.; Zhang, W.; Kimoto, K.; Matsui, Y.; Taguchi, Y.; Tokura, Y., Biskyrmion states and their current-driven motion in a layered manganite. *Nature Communications* **2014,** *5*, 3198.
32. Khanh, N. D.; Nakajima, T.; Yu, X.; Gao, S.; Shibata, K.; Hirschberger, M.; Yamasaki, Y.; Sagayama, H.; Nakao, H.; Peng, L., Nanometric square skyrmion lattice in a centrosymmetric tetragonal magnet. *Nature Nanotechnology* **2020,** *15* (6), 444-449.
33. Yang, M.; Li, Q.; Chopdekar, R.; Dhall, R.; Turner, J.; Carlström, J.; Ophus, C.; Klewe, C.; Shafer, P.; N'Diaye, A., Creation of skyrmions in van der Waals ferromagnet $Fe_3GeTe_2$ on $(Co/Pd)_n$ superlattice. *Science Advances* **2020,** *6* (36), eabb5157.
34. Zuo, S.; Qiao, K.; Zhang, Y.; Zhao, T.; Jiang, C.; Shen, B., Spontaneous Biskyrmion Lattice in a Centrosymmetric Rhombohedral Rare-Earth Magnet with Easy-Plane Anisotropy. *Nano Letters* **2023,** *23* (2), 550-557.
35. Hayami, S.; Motome, Y., Square skyrmion crystal in centrosymmetric itinerant magnets. *Physical Review B* **2021,** *103* (2), 024439.
36. Lin, S.-Z.; Hayami, S., Ginzburg-Landau theory for skyrmions in inversion-symmetric magnets with competing interactions. *Physical Review B* **2016,** *93* (6), 064430.
37. Capic, D.; Garanin, D. A.; Chudnovsky, E. M., Biskyrmion lattices in centrosymmetric magnetic films. *Physical Review Research* **2019,** *1* (3), 033011.
38. Gomilšek, M.; Hicken, T.; Wilson, M.; Franke, K.; Huddart, B.; Štefančič, A.; Holt, S.; Balakrishnan, G.; Mayoh, D.; Birch, M., Anisotropic skyrmion and multi-q spin dynamics in centrosymmetric Gd 2 PdSi 3. *Physical Review Letters* **2025,** *134* (4), 046702.
39. Takagi, R.; Matsuyama, N.; Ukleev, V.; Yu, L.; White, J. S.; Francoual, S.; Mardegan, J. R.; Hayami, S.; Saito, H.; Kaneko, K., Square and rhombic lattices of magnetic skyrmions in a centrosymmetric binary compound. *Nature communications* **2022,** *13* (1), 1472.
40. Buschow, K.; van Engen, P., Note on the magnetic and magneto-optical properties of $Ni_2In$ type 3d transition metal compounds. *Phys. Status Solidi (a)* **1983,** *76* (2), 615-620.
41. Buschow, K.; De Mooij, D., Crystal structure and magnetic properties of PtMnGa and PtMnAl. *Journal of the Less Common Metals* **1984,** *99* (1), 125-130.
42. Xiao, X.; Peng, L.; Zhao, X.; Zhang, Y.; Dai, Y.; Guo, J.; Tong, M.; Li, J.; Li, B.; Liu, W., Low-field formation of room-temperature biskyrmions in centrosymmetric MnPdGa magnet. *Applied Physics Letters* **2019,** *114* (14), 142404.
43. Wang, W.; Zhang, Y.; Xu, G.; Peng, L.; Ding, B.; Wang, Y.; Hou, Z.; Zhang, X.; Li, X.; Liu, E., A centrosymmetric hexagonal magnet with superstable biskyrmion magnetic nanodomains in a wide temperature range of 100–340 K. *Adv. Mater.* **2016,** *28* (32), 6887-6893.
44. Peng, L.; Zhang, Y.; Wang, W.; He, M.; Li, L.; Ding, B.; Li, J.; Sun, Y.; Zhang, X.-G.; Cai, J., Real-space observation of nonvolatile zero-field biskyrmion lattice generation in MnNiGa magnet. *Nano Letters* **2017,** *17* (11), 7075-7079.
45. Xu, G.; You, Y.; Tang, J.; Zhang, H.; Li, H.; Miao, X.; Gong, Y.; Hou, Z.; Cheng, Z.; Wang, J., Simultaneous tuning of magnetocrystalline anisotropy and spin reorientation transition via Cu substitution in Mn-Ni-Ga magnets for nanoscale biskyrmion formation. *Physical Review B* **2019,** *100* (5), 054416.
46. Li, X.; Zhang, S.; Li, H.; Venero, D. A.; White, J. S.; Cubitt, R.; Huang, Q.; Chen, J.; He, L.; van der Laan, G., Oriented 3D Magnetic Biskyrmions in MnNiGa Bulk Crystals. *Advanced Materials* **2019,** *31* (17), 1900264.

47. Srivastava, A. K.; Devi, P.; Sharma, A. K.; Ma, T.; Deniz, H.; Meyerheim, H. L.; Felser, C.; Parkin, S. S., Observation of Robust Néel Skyrmions in Metallic PtMnGa. *Advanced Materials* **2020,** *32* (7), 1904327.
48. Cooley, J. A.; Bocarsly, J. D.; Schueller, E. C.; Levin, E. E.; Rodriguez, E. E.; Huq, A.; Lapidus, S. H.; Wilson, S. D.; Seshadri, R., Evolution of noncollinear magnetism in magnetocaloric MnPtGa. *Physical Review Materials* **2020,** *4* (4), 044405.
49. Ibarra, R.; Lesne, E.; Sabir, B.; Gayles, J.; Felser, C.; Markou, A., Anomalous Hall Effect in Epitaxial Thin Films of the Hexagonal Heusler MnPtGa Noncollinear Hard Magnet. *Advanced Materials Interfaces* **2022**, 2201562.
50. Ibarra, R.; Lesne, E.; Ouladdiaf, B.; Beauvois, K.; Sukhanov, A.; Wawrzyńczak, R.; Schnelle, W.; Devishvili, A.; Inosov, D.; Felser, C., Noncollinear magnetic order in epitaxial thin films of the centrosymmetric MnPtGa hard magnet. *Applied Physics Letters* **2022,** *120* (17), 172403.
51. Dwari, G.; Dan, S.; Maity, B. B.; Ramakrishnan, S.; Lakshan, A.; Kulkarni, R.; Sharma, V.; Nandi, S.; Jana, P. P.; Ptok, A., Unveiling the interplay of magnetic order and electronic band structure in the evolution of the anomalous Hall effect in single crystalline MnPtGa. *Physical Review B* **2024,** *110* (4), 045111.
52. Hayami, S., Skyrmion crystal and spiral phases in centrosymmetric bilayer magnets with staggered Dzyaloshinskii-Moriya interaction. *Physical Review B* **2022,** *105* (1), 014408.
53. Lin, S.-Z., Skyrmion lattice in centrosymmetric magnets with local Dzyaloshinsky–Moriya interaction. *Materials Today Quantum* **2024,** *2*, 100006.
54. Rastogi, S.; Shahi, N.; Kumar, V.; Shukla, G. K.; Bhattacharjee, S.; Singh, S., Revealing the origin of the topological Hall effect in the centrosymmetric shape memory Heusler alloy $Mn_2NiGa$: A combined experimental and theoretical investigation. *Physical Review B* **2023,** *108* (22), 224108.
55. Cui, Q.; Zhu, Y.; Jiang, J.; Cui, P.; Yang, H.; Chang, K.; Wang, K., Anatomy of Hidden Dzyaloshinskii–Moriya Interactions and Topological Spin Textures in Centrosymmetric Crystals. *Nano Letters* **2024,** *24* (24), 7358-7365.
56. Wang, Z.; Ji, J.; Yu, H.; Xu, C., Skyrmionic states induced by local Dzyaloshinskii-Moriya interaction. *Physical Review B* **2025,** *111* (5), 054417.
57. Moody, S.; Bereciartua, P.; Francoual, S.; Littlehales, M.; Wilson, M.; Gomilšek, M.; Birch, M.; Mayoh, D.; Balakrishnan, G.; Hatton, P., Local Dzyaloshinskii-Moriya Interactions Driving Quasi-2D Magnetism in a Centrosymmetric Nanoskyrmion Material. *Physical Review Letters* **2025,** *135* (7), 076706.
58. Kaduk, J. A., Billinge, S.J.L., Dinnebier, R.E. et al. , Powder diffraction. *Nature Reviews Methods Primers* **2021,** *1*, 77.
59. Jagodzinski, H.; Frey, F., Disorder Diffuse Scattering of X-rays and Neutrons, Ch. 2. IUCr: 1993; Vol. B, , pp 407-442.
60. Egami, T.; Billinge, S. J., *Underneath the Bragg Peaks: Structural Analysis of Complex Materials*. Elsevier: 2003.
61. Petkov, V., Pair distribution functions analysis. *Characterization of Materials* **2002**, 1-14.
62. Billinge, S. J., The rise of the X-ray atomic pair distribution function method: a series of fortunate events. *Philosophical Transactions of the Royal Society A* **2019,** *377* (2147), 20180413.
63. Terban, M. W.; Billinge, S. J., Structural analysis of molecular materials using the pair distribution function. *Chemical Reviews* **2021,** *122*, 1208-1272.

64. Jeong, I.-K.; Ahn, J.; Kim, B.; Yoon, S.; Singh, S. P.; Pandey, D., Short-and medium-range structure of multiferroic $Pb(Fe_{1/2}Nb_{1/2})O_3$ studied using neutron total scattering analysis. *Physical Review B* **2011,** *83* (6), 064108.
65. Singh, A. K.; Singh, S.; Pandey, D., Pair distribution function study of $Ni_2MnGa$ magnetic shape memory alloy: Evidence for the precursor state of the premartensite phase. *Physical Review B* **2021,** *104* (6), 064110.
66. Zhu, H.; Huang, Y.; Ren, J.; Zhang, B.; Ke, Y.; Jen, A. K. Y.; Zhang, Q.; Wang, X. L.; Liu, Q., Bridging Structural Inhomogeneity to Functionality: Pair Distribution Function Methods for Functional Materials Development. *Advanced Science* **2021,** *8* (6), 2003534.
67. O'Quinn, E. C.; Sickafus, K. E.; Ewing, R. C.; Baldinozzi, G.; Neuefeind, J. C.; Tucker, M. G.; Fuentes, A. F.; Drey, D.; Lang, M. K., Predicting short-range order and correlated phenomena in disordered crystalline materials. *Science Advances* **2020,** *6* (35), eabc2758.
68. Keen, D. A.; Goodwin, A. L., The crystallography of correlated disorder. *Nature* **2015,** *521* (7552), 303-309.
69. Thygesen, P. M.; Paddison, J. A.; Zhang, R.; Beyer, K. A.; Chapman, K. W.; Playford, H. Y.; Tucker, M. G.; Keen, D. A.; Hayward, M. A.; Goodwin, A. L., Orbital dimer model for the spin-glass state in $Y_2Mo_2O_7$. *Physical Review Letters* **2017,** *118* (6), 067201.
70. Greedan, J.; Gout, D.; Lozano-Gorrin, A.; Derahkshan, S.; Proffen, T.; Kim, H.-J.; Božin, E.; Billinge, S., Local and average structures of the spin-glass pyrochlore $Y_2Mo_2O_7$ from neutron diffraction and neutron pair distribution function analysis. *Physical Review B* **2009,** *79* (1), 014427.
71. Lee, S.; Xu, H., Using powder XRD and pair distribution function to determine anisotropic atomic displacement parameters of orthorhombic tridymite and tetragonal cristobalite. *Acta Crystallographica Section B: Structural Science, Crystal Engineering and Materials* **2019,** *75* (2), 160-167.
72. Masadeh, A. S.; Shatnawi, M. T.; Adawi, G.; Ren, Y., Total-scattering pair-distribution function analysis of zinc from high-energy synchrotron data. *Modern Physics Letters B* **2019,** *33* (33), 1950410.
73. Masadeh, A.; Božin, E.; Farrow, C.; Paglia, G.; Juhas, P.; Billinge, S.; Karkamkar, A.; Kanatzidis, M., Quantitative size-dependent structure and strain determination of CdSe nanoparticles using atomic pair distribution function analysis. *Physical Review B* **2007,** *76* (11), 115413.
74. Trueblood, K.; Bürgi, H.-B.; Burzlaff, H.; Dunitz, J.; Gramaccioli, C.; Schulz, H.; Shmueli, U.; Abrahams, S., Atomic dispacement parameter nomenclature. Report of a subcommittee on atomic displacement parameter nomenclature. *Acta Crystallographica Section A: Foundations of Crystallography* **1996,** *52* (5), 770-781.
75. Knox, K.; Bozin, E.; Malliakas, C.; Kanatzidis, M.; Billinge, S., Local off-centering symmetry breaking in the high-temperature regime of SnTe. *Physical Review B* **2014,** *89* (1), 014102.
76. Dutta, M.; Pal, K.; Etter, M.; Waghmare, U. V.; Biswas, K., Emphanisis in Cubic (SnSe) 0.5 ($AgSbSe_2$) 0.5: Dynamical Off-Centering of Anion Leads to Low Thermal Conductivity and High Thermoelectric Performance. *Journal of the American Chemical Society* **2021,** *143* (40), 16839-16848.
77. Vasdev, A.; Dutta, M.; Mishra, S.; Kaur, V.; Kaur, H.; Biswas, K.; Sheet, G., Local ferroelectric polarization switching driven by nanoscale distortions in thermoelectric $Sn_{0.7}Ge_{0.3}Te$. *Scientific Reports* **2021,** *11* (1), 17190.
78. Oliveira, G.; Pereira, A.; Lopes, A.; Amaral, J.; Dos Santos, A.; Ren, Y.; Mendonça, T.; Sousa, C.; Amaral, V.; Correia, J., Dynamic off-centering of $Cr^{3+}$ ions and short-range magneto-electric clusters in $CdCr_2S_4$. *Physical Review B* **2012,** *86* (22), 224418.

79. H. T. Stokes, D. M. H., and B. J. Campbell, ISOTROPY Software Suite, iso.byu.edu.
80. Haines, C. R.; Howard, C. J.; Harrison, R. J.; Carpenter, M. A., Group-theoretical analysis of structural instability, vacancy ordering and magnetic transitions in the system troilite (FeS)–pyrrhotite ($Fe_{1-x}S$). *Acta Crystallographica Section B: Structural Science, Crystal Engineering and Materials* **2019,** *75* (6), 1208-1224.
81. Jo, D.; Go, D.; Lee, H.-W., Gigantic intrinsic orbital Hall effects in weakly spin-orbit coupled metals. *Physical Review B* **2018,** *98* (21), 214405.
82. Karube, K.; White, J.; Reynolds, N.; Gavilano, J.; Oike, H.; Kikkawa, A.; Kagawa, F.; Tokunaga, Y.; Rønnow, H. M.; Tokura, Y., Robust metastable skyrmions and their triangular–square lattice structural transition in a high-temperature chiral magnet. *Nature materials* **2016,** *15* (12), 1237-1242.
83. Nayak, A. K.; Kumar, V.; Ma, T.; Werner, P.; Pippel, E.; Sahoo, R.; Damay, F.; Rößler, U. K.; Felser, C.; Parkin, S. S., Magnetic antiskyrmions above room temperature in tetragonal Heusler materials. *Nature* **2017,** *548* (7669), 561-566.
84. Zhang, H.; Raftrey, D.; Chan, Y.-T.; Shao, Y.-T.; Chen, R.; Chen, X.; Huang, X.; Reichanadter, J. T.; Dong, K.; Susarla, S., Room-temperature skyrmion lattice in a layered magnet (Fe0. 5Co0. 5) 5GeTe2. *Science advances* **2022,** *8* (12), eabm7103.
85. Zhang, Q.; Zhang, Y.; Matsuda, M.; Garlea, V. O.; Yan, J.; McGuire, M. A.; Tennant, D. A.; Okamoto, S., Hidden Local Symmetry Breaking in a Kagome-Lattice Magnetic Weyl Semimetal. *Journal of the American Chemical Society* **2022,** *144* (31), 14339-14350.
86. dos Santos Dias, M.; Biniskos, N.; Dos Santos, F.; Schmalzl, K.; Persson, J.; Bourdarot, F.; Marzari, N.; Blügel, S.; Brückel, T.; Lounis, S., Topological magnons driven by the Dzyaloshinskii-Moriya interaction in the centrosymmetric ferromagnet Mn5Ge3. *Nature Communications* **2023,** *14* (1), 7321.

## Figures

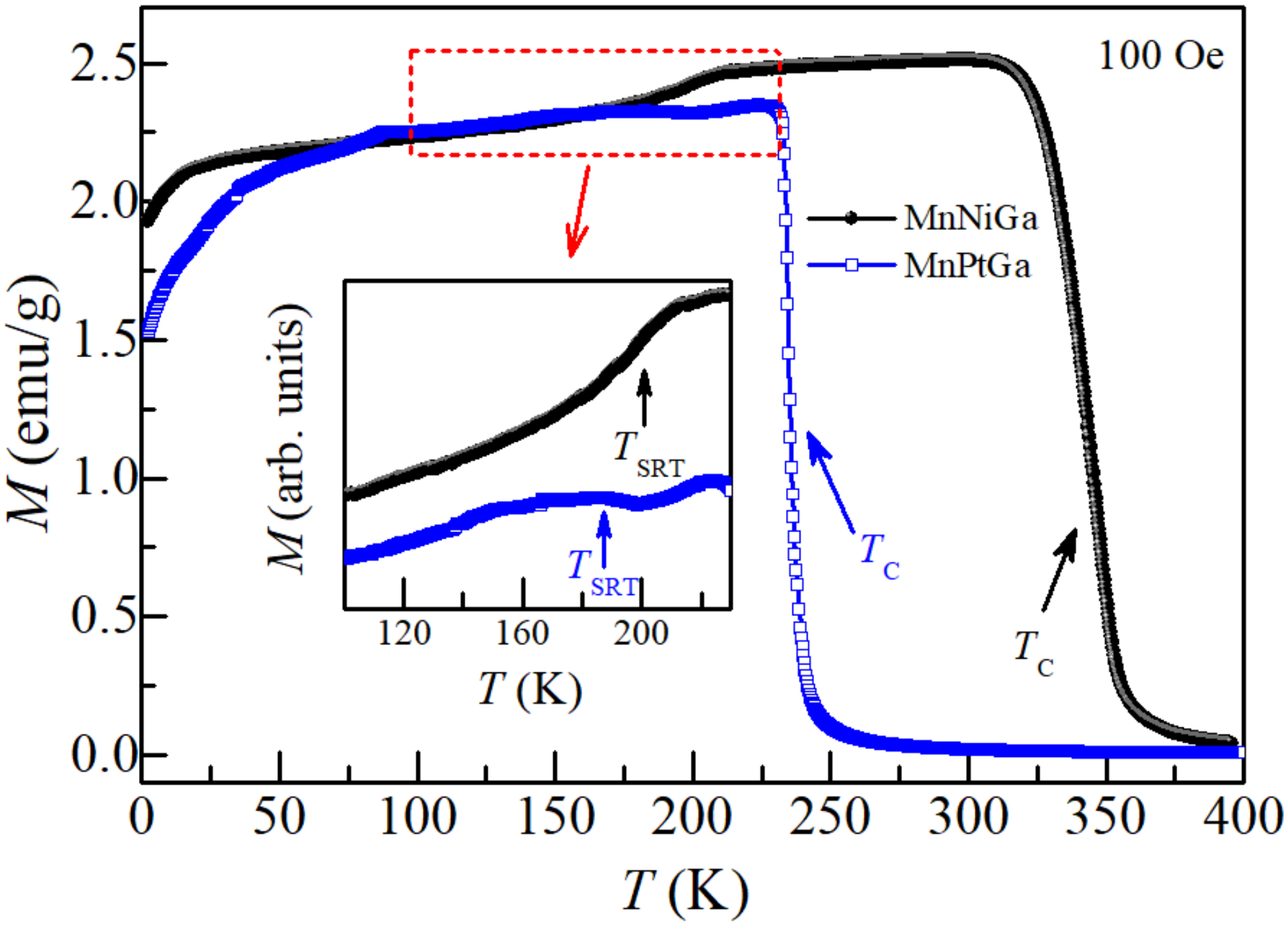

**Figure 1:** Temperature-dependence of magnetization (*M*), measured at 100 Oe magnetic field on zero-field cooled sample in the warming cycle of MnNiGa (black coloured circles) and MnPtGa (blue coloured squares), showing the anomalies at the ferromagnetic $T_C$ and spin reorientation transition $T_{SRT}$ temperatures. The inset shows an enalarged view of *M* from 100 to 230 K (red colored bounded region).

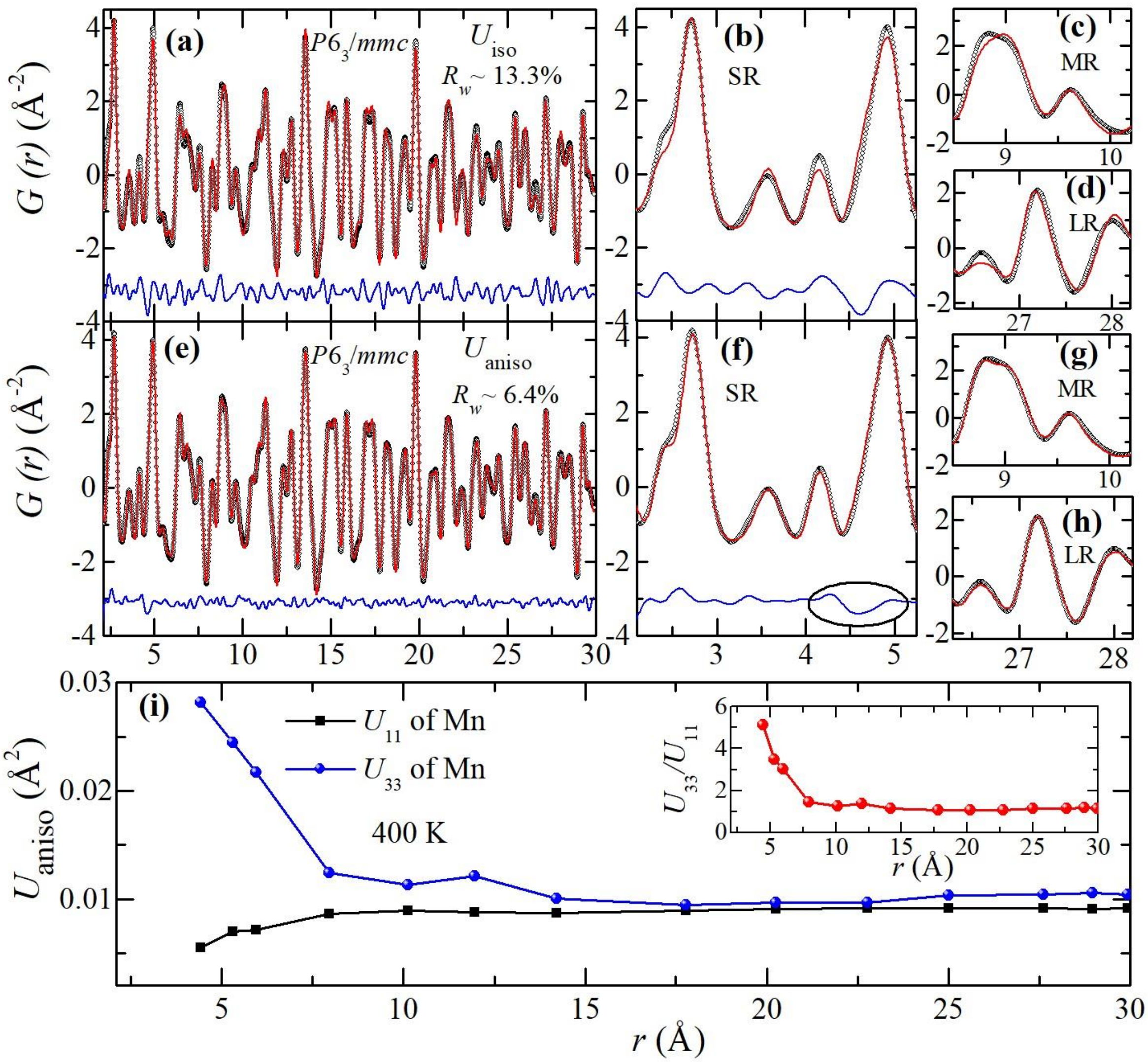

**Figure 2:** **(a)** The experimental atomic PDF (black circles), calculated PDF (red line), their difference (blue line at the bottom), and weighted agreement factor ($R_w$) obtained by real-space structure refinements of MnNiGa, using observed PDF up to 30 Å at 400 K, for the $P6_3/mmc$ SG with isotropic atomic displacement parameters (ADPs; $U_{iso}$), **(b)**, **(c)** and **(d)** depict the enlarged views of **(a)** for $r$ ~ 2.1 to 5.25 Å, $r$ ~ 8.5 to 10.2 Å and $r$ ~ 26.3 to 28.2 Å regions, representative of SR, MR and LR regimes, respectively. **(e)** Depicts the results for anisotropic ADPs ($U_{aniso}$) with enlarged views of the SR, MR and LR regimes in **(f)**, **(g)** and **(h)**, respectively. The encircled region in the difference PDF of **(f)** shows some misfit just below 5 Å peak. **(i)** Variation of the anisotropic atomic displacement parameters $U_{11}$ (black squares) and $U_{33}$ (blue spheres) of Mn atom with the range ($r$) upto which the experimental PDF was used for the refinements (i.e., $r$ = 4.41, 5.26, …, and 30 Å) at 400 K. The inset of **(i)** depicts the variation of the ratio $U_{33}/U_{11}$ with $r$.

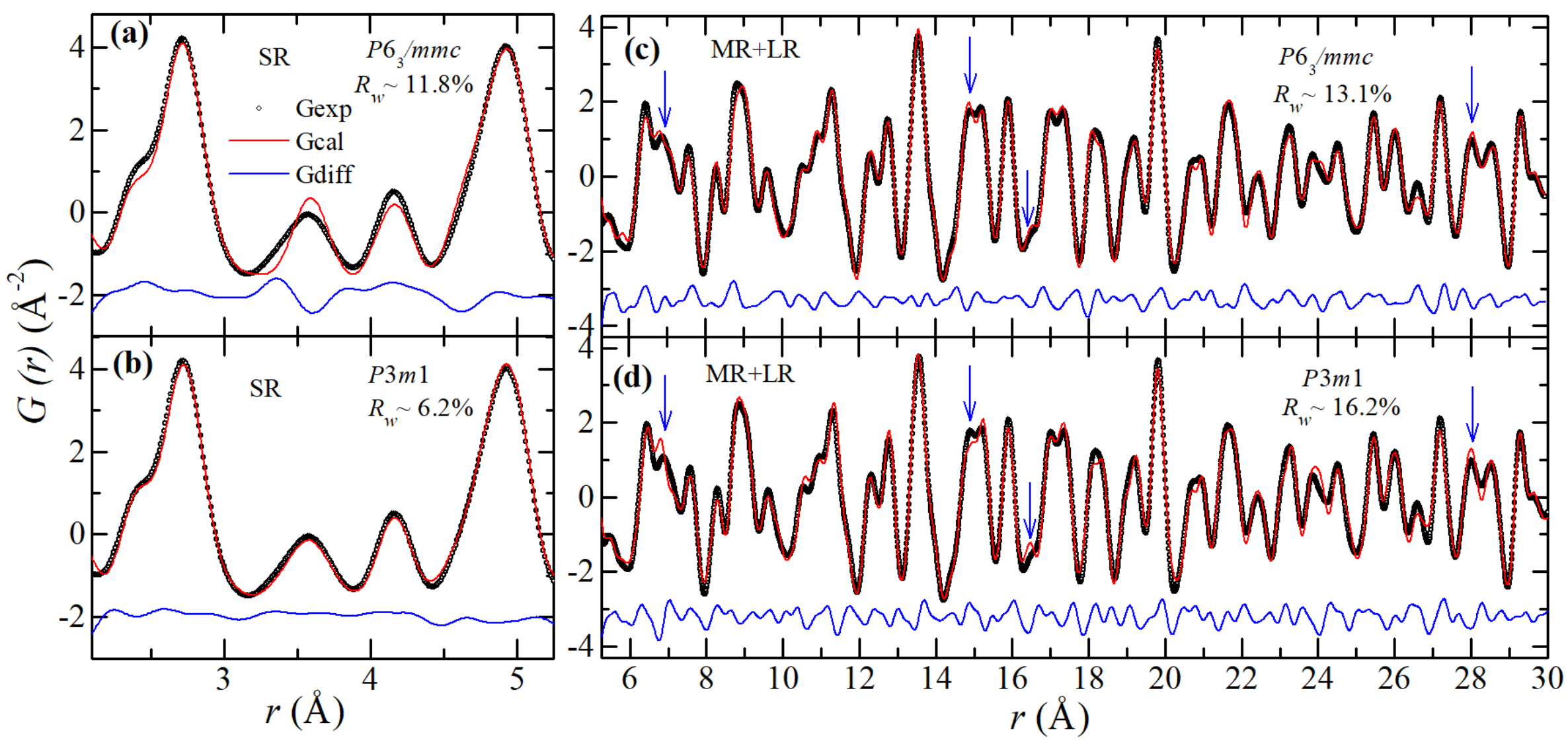

**Figure 3:** The experimental atomic PDF (black circles), calculated PDF (red line), their difference (blue line at the bottom), and weighted agreement factor ($R_w$) obtained by real-space structure refinements of MnNiGa at 400 K in the SR regime using the space groups **(a)** $P6_3/mmc$ and **(b)** $P3m1$; the corresponding fits for the MR+LR regimes without altering the atomic positions obtained from the SR regime refinements, are shown in **(c)** and **(d)** for $P6_3/mmc$ and $P3m1$ space groups. The arrow marks in **(c)** and **(d)** highlight some of the regions where the fit has improved significantly for the $P6_3/mmc$ SG as compared to that for the $P3m1$ SG in the MR+LR regimes.

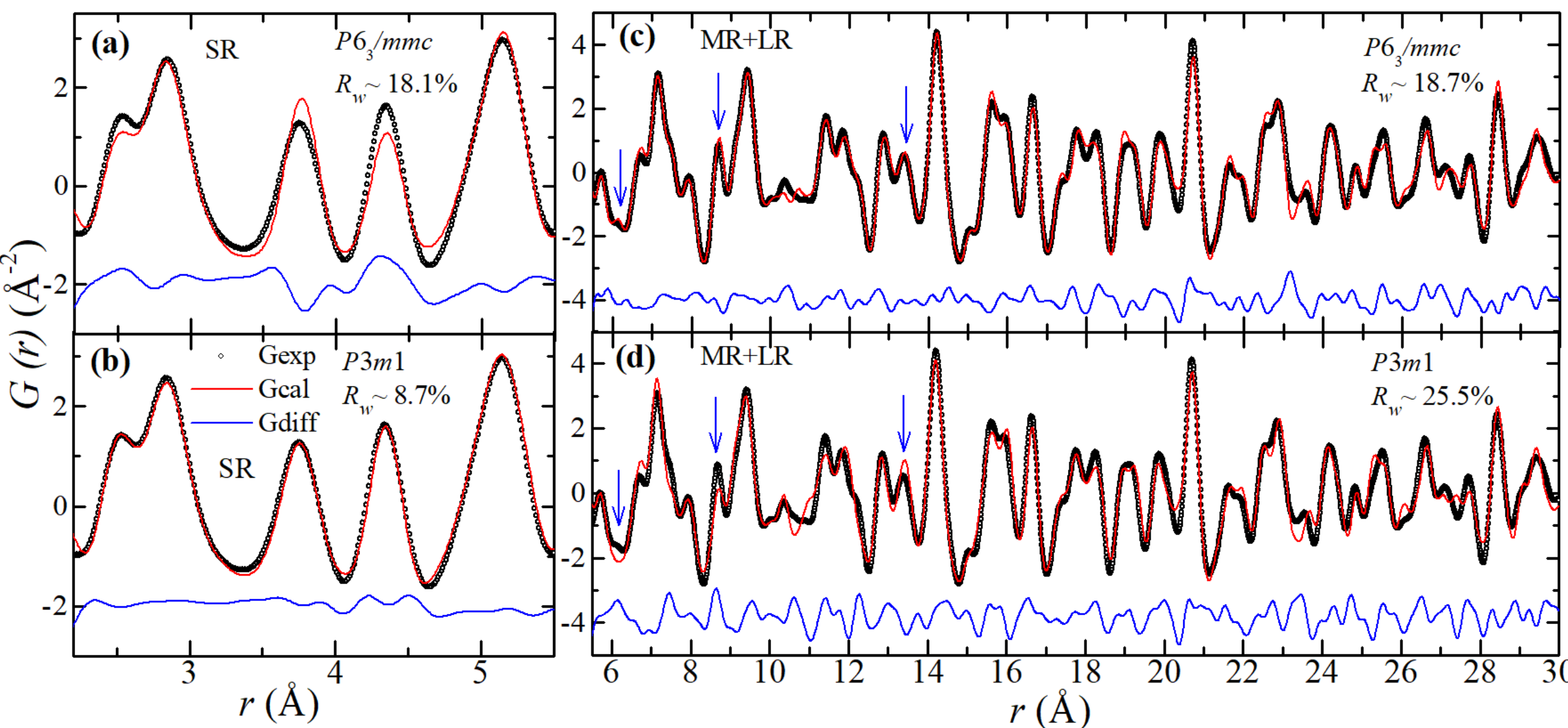

**Figure 4:** The experimental atomic PDF (black circles), calculated PDF (red line), their difference (blue line at the bottom), and weighted agreement factor ($R_w$) obtained by real-space structure refinements of MnPtGa at 300 K in the SR regime using the space groups **(a)** $P6_3/mmc$ and **(b)** $P3m1$; the corresponding fits for the MR+LR regimes without altering the atomic positions obtained from the SR regime refinements, are shown in **(c)** and **(d)** for $P6_3/mmc$ and $P3m1$ space groups. The arrow marks in **(c)** and **(d)** highlight some of the regions where the fit has improved significantly for the $P6_3/mmc$ SG as compared to that for the $P3m1$ SG in the MR+LR regimes.

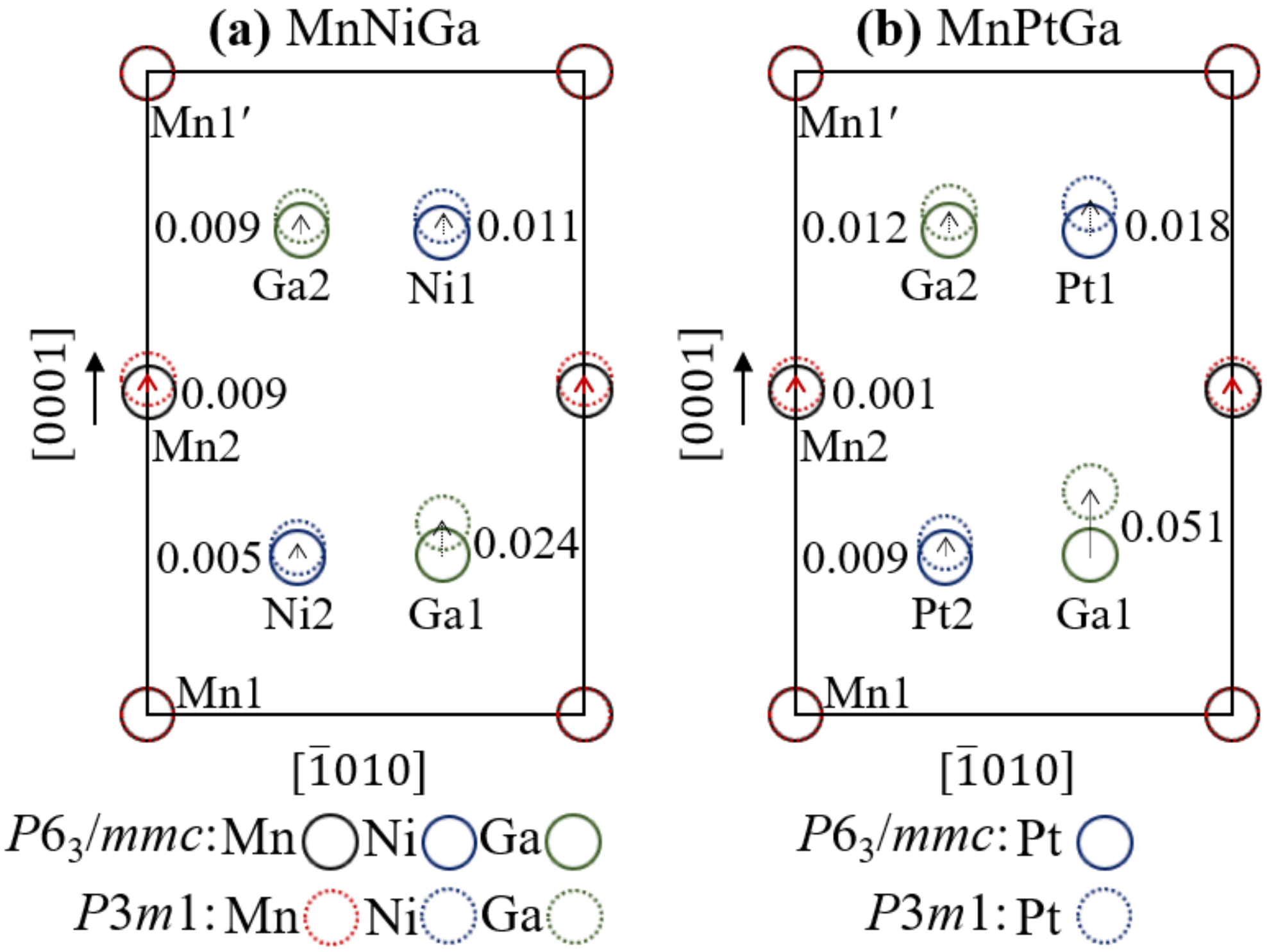

**Figure 5:** A schematic diagram of primitive hexagonal ($P6_3/mmc$) and trigonal ($P3m1$) unit cells (the axes are not to the scale) at 300 K of **(a)** MnNiGa and **(b)** MnPtGa. The arrows within the circles represent the off-centering of atoms. The length of arrows is scaled such that they are equivalent to the approximate magnitude of the value of off-centered distances in nm. The indices of the *c* and *b*-axis are indicated. From the corner bottom Mn1 atoms, considering the Ni-Ga (or Pt-Ga) layers only, the first layer contains Ni2 and Ga1 in **(a)** (or Pt2 and Ga1 in **(b)**), whereas the second layer contains the Ga2 and Ni1 in **(a)** [or Ga2 and Pt1 in **(b)**].

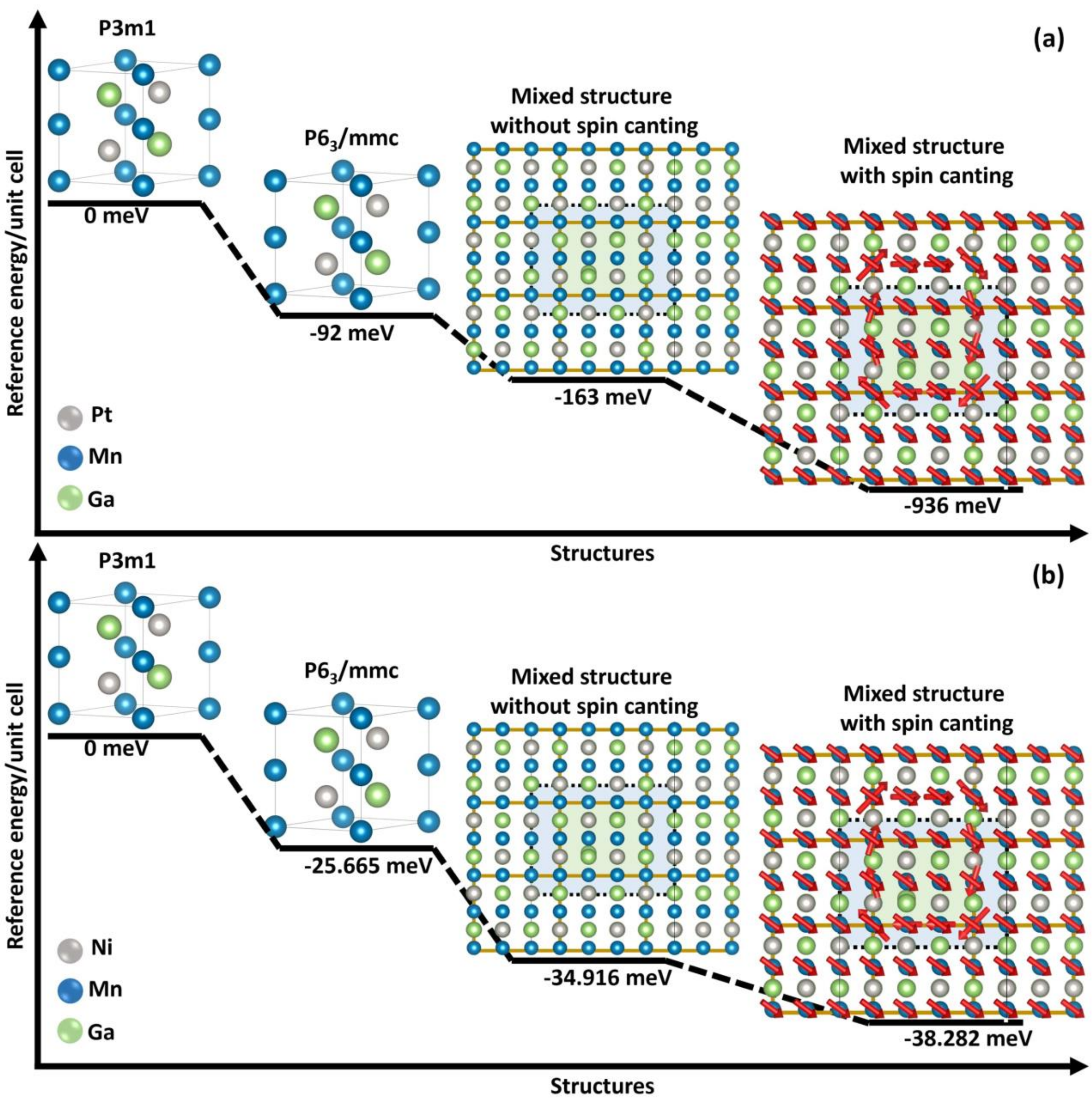

**Figure 6:** Schematic diagram for the comparison of ground state energy per unit cell for different structures (see text for more detail) for **(a)** MnPtGa and **(b)** MnNiGa. The *P*3*m*1 and *P*6$_3$/*mmc* structures are considered in a single unit cell, and the mixed structure is considered as 3 × 3 × 3 supercell, wherein the light green region shows one-unit cell of trigonal *P*3*m*1 structure replaced in place of the unit cell at the center of the *P*6$_3$/*mmc* supercell. The light gray region in the mixed structure shows one outer layer of atoms which is allowed to relax with P3*m*1 (see text for more detail). Blue, green, and white balls correspond to Mn, Ga, and Pt/Ni, respectively. Consideration of spin canting (non-collinearity) leads to stabilize the vortex-type spin structure within the P3*m*1 unit cell as highlighted in the right-most figures.

# Supporting Information

## Local Symmetry Breaking in Skyrmion-Hosting Centrosymmetric Hexagonal Compounds

*Anupam K. Singh, Krishna K. Dubey, Parul Devi, Pritam Das, Martin Etter, Ola. G. Grendal, Catherine Dejoie, Andrew Fitch, Anatoliy Senyshyn, Seung-Cheol Lee, Satadeep Bhattacharjee*, Sanjay Singh* and Dhananjai Pandey*

## I. Experimental Methods

Details of the preparation of the MnNiGa polycrystalline samples and characterization are described elsewhere [1]. The conventional arc-melting technique was employed to prepare the polycrystalline ingot with nominal composition MnPtGa under the argon atmosphere [2]. The appropriate quantity of each constituent element with at least 99.99% purity were melted several times to get good homogeneity. The melt-cast ingot was annealed in vacuum-sealed quartz ampoule at 800ºC for 6 days to achieve further homogeneity [3]. The chemical composition was checked using the energy dispersive analysis of x-rays (EDAX) technique. The EDAX analysis was performed on several regions of the samples using EVO-Scanning Electron Microscope MA15/18 (ZEISS) in the backscattered electron mode. The average composition obtained by EDAX is found to be $Mn_{0.95}Pt_{0.98}Ga_{1.07}$, which corresponds to MnPtGa. The EDAX-determined composition has typically around 2-3% error depending on the atomic number and the overlap of the peaks vis-a-vis instrumental resolution of the employed EDAX system. The temperature dependent magnetization measurements on MnNiGa and MnPtGa samples were carried out at an applied magnetic field of 100 Oe during the warming cycle on zero-field cooled samples using the VSM module attached with a Physical Properties Measurement System (PPMS, Quantum Design). The temperature-dependent (100-400 K) synchrotron x-ray powder diffraction (SXRPD) measurements of MnNiGa were carried out in high-$Q$ and high-resolution mode using high flux and high energy (60 keV) x-rays with a wavelength ($\lambda$) of 0.20742 Å at the P02.1 beamline of PETRA-III, DESY, Germany [4]. Liquid nitrogen-based Oxford cryostreamer was used for temperature dependent (100 to 400 K) SXRPD measurements, where sample capillary was cooled through continuous flow of liquid-$N_2$ by cryostreamer. For measurements above 300 K, continuous flow of hot air was used to heat the sample capillary in the cryostreamer. The temperature variation *w.r.t* the set temperature was around $\pm$0.2 K during measurements. The sample was kept in a borosilicate capillary for temperature dependent SXRPD measurements. The high-$Q$ measurements were performed with a maximum instrumental $Q$-value $Q_{maxinst}$ ~ 23 Å$^{-1}$. The high-$Q$ SXRPD data were also collected on empty borosilicate capillary required for the background subtraction. The high-$Q$ SXRPD data for MnPtGa were recorded using a wavelength of $\lambda$ ~ 0.207 Å with $Q_{maxinst}$ ~ 24.5 Å$^{-1}$ at ID22 beamline of ESRF, Grenoble, France [5]. To check the robustness and reproducibility of the features in the experimental data, high-Q SXRPD data were also collected using a 100 keV x-ray beam providing $\lambda$ ~ 0.1204 Å with $Q_{maxinst}$ ~ 29 Å$^{-1}$ at the P21.1 beamline of PETRA-III DESY [6]. Moreover, for the detailed average long-range ordered (LRO) structure studies, the powder neutron diffraction data for MnPtGa was collected using neutron beam with λ ~ 1.548 Å at SPODI diffractometer of FRM II, Munich, Germany [7].

## II. Confirmation of the average LRO crystal structure of MnNiGa and MnPtGa

High-resolution SXRPD data with excellent signal-to-noise ratio were used for the confirmation of the average LRO structure of both MnNiGa and MnPtGa compounds by Rietveld technique [8]. The refinement was first carried out using the FULLPROF package [9] for the hexagonal structure in the $P6_3/mmc$ space group, considering all the atoms at the special Wyckoff positions, i.e., Mn at 2a (0, 0, 0), Ni at 2d (1/3, 2/3, 3/4) and Ga at 2c (1/3, 2/3, 1/4) [10,11]. Figure S1a,b show the excellent fits between the observed and calculated peak profiles by accounting for all the Bragg peaks confirming the average LR hexagonal structure ($P6_3/mmc$) of both MnNiGa and MnPtGa compounds. Moreover, considering the refinement using the trigonal structure ($P3m1$) for both MnNiGa (Figure S1c) and MnPtGa (Figure S1d) compounds, the major characteristic reflections of the $P3m1$ space group (marked by "*" in Figure S1c and S1d) are absent as can be seen by the insets (i) and (ii) in Figure S1c,d. These extra reflections are forbidden in the $P6_3/mmc$ space group. All these results confirm that the average LRO structure of MnNiGa and MnPtGa is hexagonal with $P6_3/mmc$ space group. For comparison, the results of Rietveld refinements using high-resolution SXRPD data at 300 K for the hexagonal structure ($P6_3/mmc$) and trigonal structure ($P3m1$) for both compounds are shown in Figure S1a-S1d.

## III. Powder neutron diffraction of MnPtGa

To further confirm the average LR structure, we also collected powder neutron diffraction data for the MnPtGa. Since the atomic form factor varies randomly with atomic weight for neutrons, contrary to the gradually decreasing atomic form factor in case of x-rays with decreasing atomic number, neutron diffraction is more sensitive technique than x-ray diffraction to capture the signatures of site-disorder which is one of the common sources of diffuse scattering in alloys [12]. Due to widely different form factors, neutrons can easily distinguish the two nearby elements, e.g., Mn and Ga in the present case and hence capture any significant site-disorder [12]. Rietveld refinement using the powder neutron diffraction data on MnPtGa at 300 K shows an excellent fit for the $P6_3/mmc$ space group, as can be seen from Figure S2. It is worth to note here that the major characteristic reflections expected for the $P3m1$ space group, which are otherwise forbidden in the $P6_3/mmc$ space group are marked by "*" in Figure S2 and its inset. The absence of these $P3m1$ reflections in powder neutron diffraction data of MnPtGa further confirms the average LRO hexagonal structure in the $P6_3/mmc$ space group of MnPtGa compound in agreement with the results of analysis using SXRPD data. We note that the presence of any possible any anti-site disorder is discarded based on the Rietveld analysis using neutron diffraction data. Our observations are in agreement with the LRO structure of MnPtGa reported by several other researchers [13,14]. A recent combined high-resolution SXRPD and neutron powder diffraction study on the MnPtGa also did not find any evidence of the LRO trigonal structure ($P3m1$) and supports the LRO hexagonal structure ($P6_3/mmc$) [15]. More recently, the LRO hexagonal structure ($P6_3/mmc$) of MnPtGa has also been confirmed in a single crystalline thin film specimen [16,17] as well as bulk crystal [18]. Similar to MnPtGa, the average LRO structure of MnNiGa has also been reported to be hexagonal ($P6_3/mmc$) by the high-resolution SXRPD and neutron diffraction studies [11].

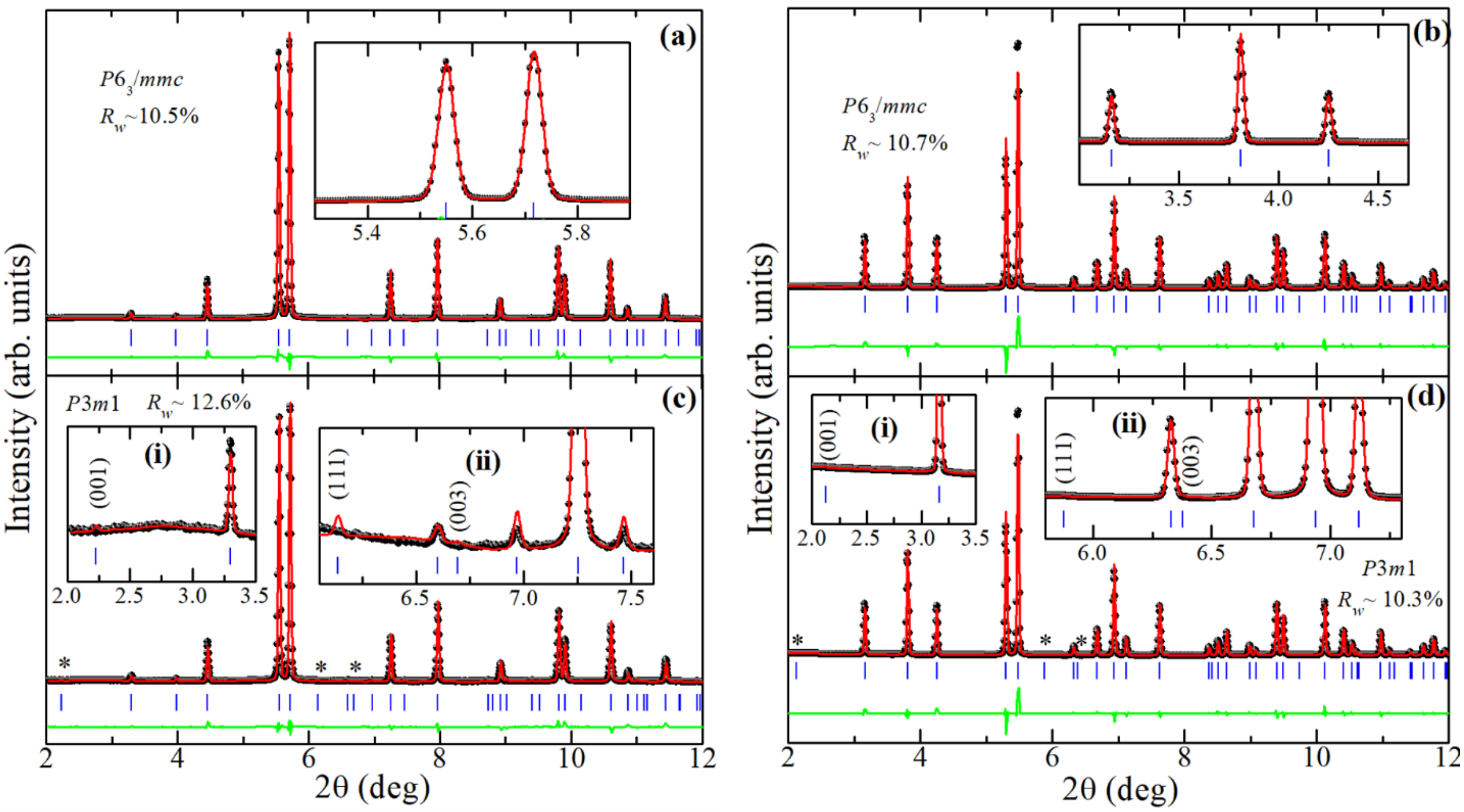

**Figure S1:** The observed profile (black spheres), calculated profile (red line), difference profile (green line), Bragg peak positions (blue ticks) and weighted agreement factor ($R_w$) obtained after Rietveld refinement for the hexagonal structure in the $P6_3/mmc$ space group using high-resolution SXRPD data at 300 K of **(a)** MnNiGa and **(b)** MnPtGa. The results of similar refinement for the trigonal structure in the $P3m1$ space group using high-resolution SXRPD data at 300 K of **(c)** MnNiGa and **(d)** MnPtGa. The insets in **(a)** and **(b)** show enlarged view of fits around a few Bragg peaks. The major extra reflections of $P3m1$ (which are forbidden in the $P6_3/mmc$) are marked by "*" in **(c)** and **(d)**. The insets **(i)** and **(ii)** in **(c)** and **(d)** show enlarged view to check the presence of the hexagonal $P6_3/mmc$ space group forbidden reflections.

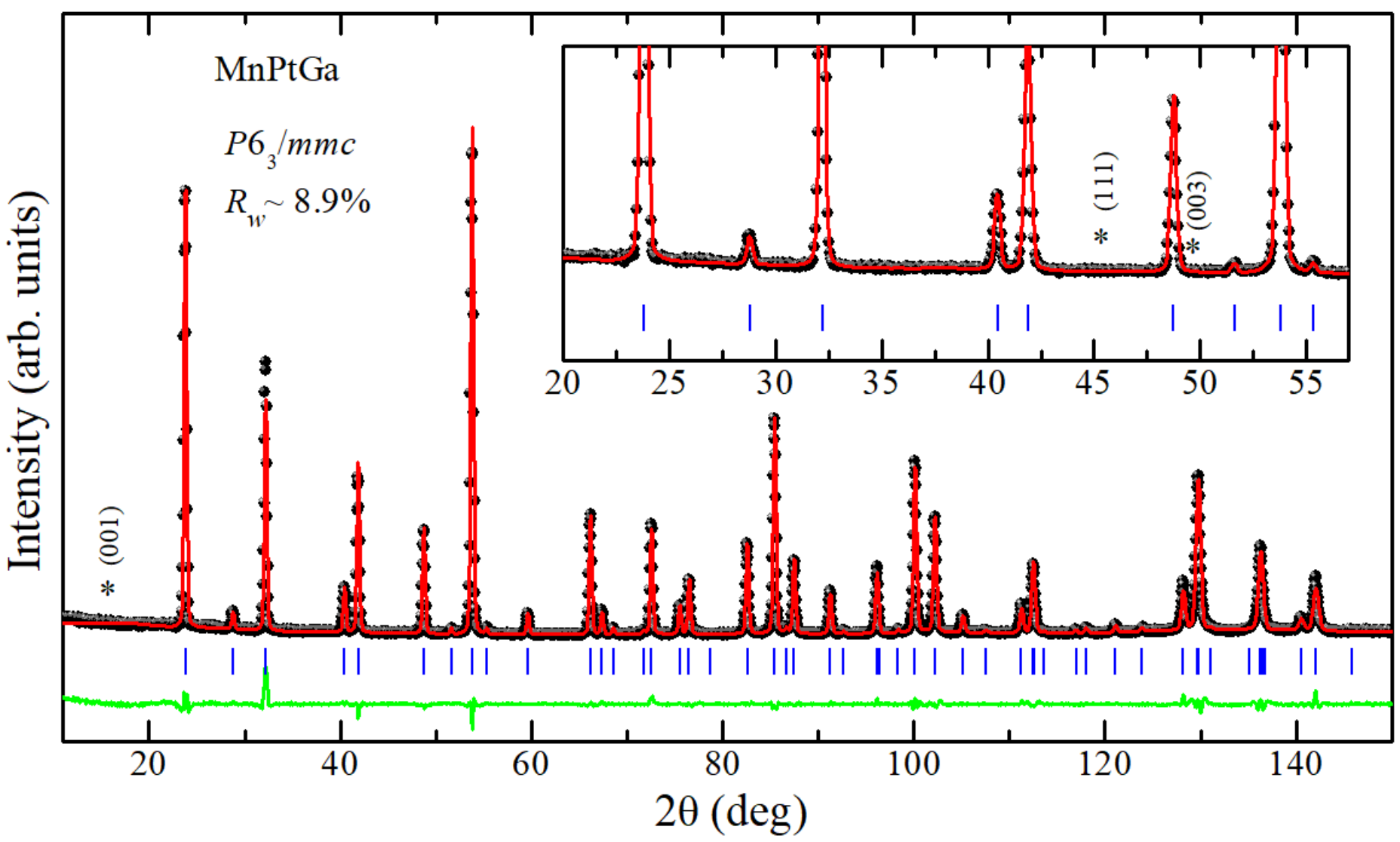

**Figure S2:** The observed profile (black spheres), calculated profile (red line), difference profile (green line), Bragg peak positions (blue ticks) and weighted agreement factor ($R_w$) obtained after Rietveld refinement for the hexagonal structure in the $P6_3/mmc$ space group using powder neutron diffraction data of MnPtGa at 300 K. The inset show enlarged view to check the characteristic reflections (miller indices are labelled) of $P3m1$, which is forbidden in the $P6_3/mmc$. The "*" indicate the reflections which are forbidden for $P6_3/mmc$ and allowed in the $P3m1$ space group.

## IV. Atomic pair distribution function

The atomic pair distribution function (PDF) is a powerful method for investigating the local structure of materials in the real-space [19-21]. It captures any deviation in the local short-range regime from the average long-range structure and relies on the Fourier transformation method [19-21]. The PDF considers the contribution of both Bragg and diffuse scatterings to the total scattering structure function S($Q$). It is, therefore, also known as the total scattering method [19-21]. The S($Q$) were obtained from the high-$Q$ ($Q_{max}$ = 22.7 Å$^{-1}$ for MnNiGa and $Q_{max}$ = 24.5 Å$^{-1}$ for MnPtGa) SXRPD data after applying the standard normalization and background corrections to raw synchrotron x-rays diffraction data using the PDFgetX3 program [22]. After that, reduced structure function F($Q$), was determined using the relation F($Q$) = $Q$[S($Q$)-1]. Finally, the experimental reduced atomic PDF ($G(r)$) was obtained using the PDFgetX3 program [22] by taking the Fourier transform of F($Q$) using the following equation [19-21]:

$$G(r) = 4\pi r\,[\rho(r) - \rho_0] = \frac{2}{\pi}\int_{Q_{min}}^{Q_{max}} F(Q)\, sinQr\, dQ \ldots (1),$$

where $Q$ is the magnitude of the scattering vector $\mathbf{Q} = \mathbf{k}_i - \mathbf{k}_f$ while $\mathbf{k}_i$ and $\mathbf{k}_f$ are the incident and reflected wave vectors, respectively. The $Q_{min}$ and $Q_{max}$ are the minimum and maximum cut-offs, respectively, for $Q$, whereas $\rho(r)$ is the atomic number density and $\rho_0$ is the average atomic number density [19-21]. It is important to mention here that since the present PDF data were collected using high-energy (60 keV and 100 keV) synchrotron x-rays in the transmission mode, the x-rays could penetrate fully through the powder sample in the sample holder and provide information about the average bulk behavior.

## V. Temperature-dependent high-$Q$ synchrotron x-ray diffraction study of MnNiGa

Although Rietveld refinements using high-resolution SXRPD data in the reciprocal space revealed the average hexagonal structure in the $P6_3/mmc$ space group for both compounds, we further confirmed the average structure using high-$Q$ SXRPD data. The Rietveld refinements were carried out using high-$Q$ SXRPD patterns collected at the three selected temperatures (400, 300 and 100 K), covering both magnetic transitions ($T_C$ ~ 347 K and $T_{SRT}$ ~ 200 K) observed in the magnetization (see Figure 1 of the main text). The results of the Rietveld refinement at 400, 300 and 100 K are shown in Figure S3a, Figure S3b and Figure S3c, respectively. Figures S3a-S3c show excellent fits between the observed and calculated profiles by accounting for all the Bragg peaks. This suggests that the average hexagonal structure ($P6_3/mmc$) of MnNiGa is stable in the 400-100 K range. The parameters obtained after the Rietveld refinements at the three selected temperatures (400, 300 and 100 K) are given in Table S1. It is worth to mention here that we did not find any appealing signature of site-disorder effect during Rietveld refinement using high-$Q$ (Figure S3) as well as high-resolution SXRPD data (Figure S1) which agrees with the fact of absence of any site-disorder inferred from the neutron diffraction data (Figure S2). We would like to add here that considering the anisotropy ADPs in the Rietveld refinement**,** we find usual information about the anisotropy effect apparent in the ADPs values related to the hexagonal structure as can be seen in Figure S3d, where equivalent ADPs ($U_{equiv}$) lies in good agreement with the isotropic ADPs listed in Table S1. However, slightly higher ADPs values for Ni and Ga is indicating some extent of hidden disorder. This aspect is comprehensively discussed using real space atomic PDF analysis in the next section.

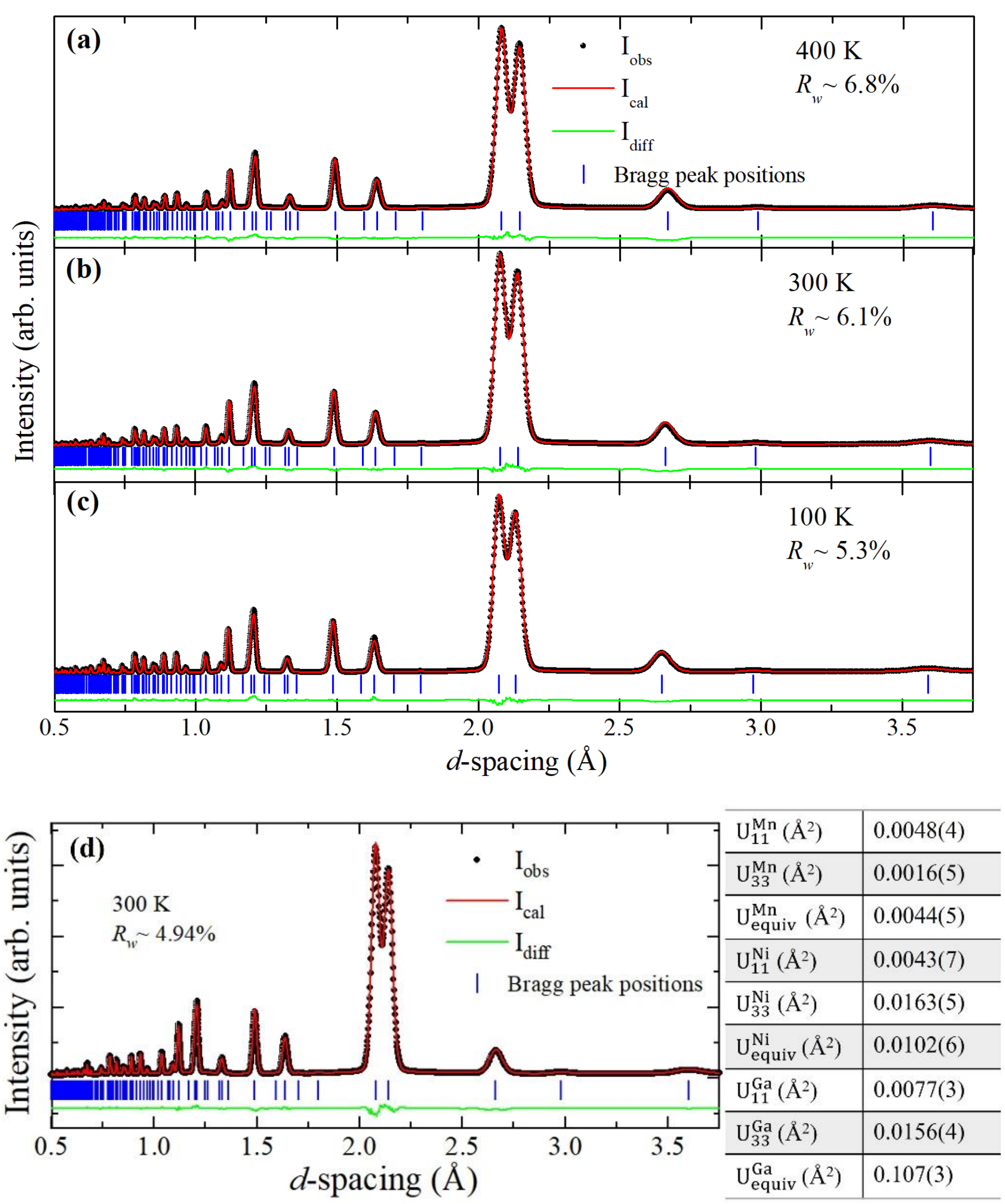

| $U_{11}^{Mn}$ (Å$^2$) | 0.0048(4) |
|---|---|
| $U_{33}^{Mn}$ (Å$^2$) | 0.0016(5) |
| $U_{equiv}^{Mn}$ (Å$^2$) | 0.0044(5) |
| $U_{11}^{Ni}$ (Å$^2$) | 0.0043(7) |
| $U_{33}^{Ni}$ (Å$^2$) | 0.0163(5) |
| $U_{equiv}^{Ni}$ (Å$^2$) | 0.0102(6) |
| $U_{11}^{Ga}$ (Å$^2$) | 0.0077(3) |
| $U_{33}^{Ga}$ (Å$^2$) | 0.0156(4) |
| $U_{equiv}^{Ga}$ (Å$^2$) | 0.107(3) |

**Figure S3:** The observed (black spheres), calculated (red line) and the difference profiles (green line), Bragg peak positions (blue ticks) and weighted agreement factor ($R_w$) obtained after Rietveld refinement for the hexagonal structure in the *P*6$_3$/*mmc* space group using the high-*Q* synchrotron x-ray powder diffraction data for MnNiGa compound at **(a)** 400 K, **(b)** 300 K and **(c)** 100 K. **(d)** Result of after the Rietveld refinement considering anisotropic ADPs for the hexagonal structure (*P*6$_3$/*mmc*) using high-*Q* SXRPD data for MnNiGa compound at 300 K. The anisotropic ADPs obtained after the refinement is listed in the table on the right side of figure **(d)**.

**TABLE S1**. The lattice parameters (*a* and *c*), isotropic atomic displacements parameters ($U_{iso}$) and agreement factors obtained after Rietveld refinements for the hexagonal structure (*P*6$_3$/*mmc)* space group using temperature dependent high-*Q* SXRPD data of MnNiGa at the selected temperatures.

| **Temperature →** / **Parameters ↓** | **400 K** | **300 K** | **100 K** |
|---|---|---|---|
| *a* (Å) | 4.16220(5) | 4.15552(4) | 4.14569(2) |
| *c* (Å) | 5.33731(9) | 5.32476(8) | 5.29734(6) |
| $U_{iso}^{Mn}$ (Å$^2$) | 0.0064(3) | 0.0051(2) | 0.0027(3) |
| $U_{iso}^{Ni}$ (Å$^2$) | 0.0104(2) | 0.0080(1) | 0.0047(1) |
| $U_{iso}^{Ga}$ (Å$^2$) | 0.0128(1) | 0.0093(1) | 0.0033(1) |
| $R_p$ (%) | 6.34 | 5.67 | 4.79 |
| $R_{wp}$ (%) | 6.85 | 6.13 | 5.32 |

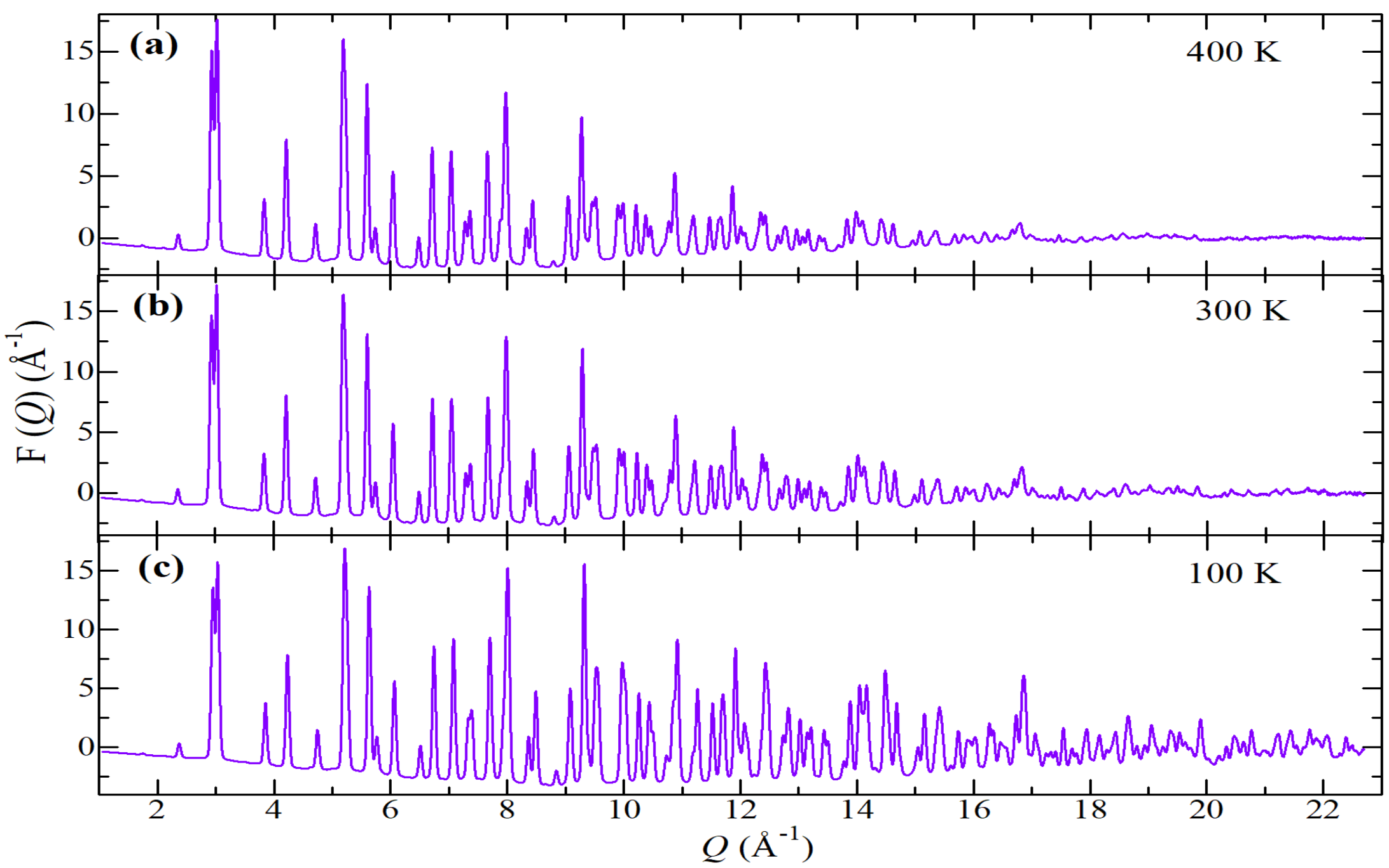

**Figure S4**: The reduced structure function F(*Q*) versus *Q* plot for MnNiGa at **(a)** 400 K, **(b)** 300 K and **(c)** 100 K.

The reduced structure function F(*Q*), obtained after the standard corrections and normalization of the high-*Q* SXRPD patterns (shown in Figure S3a-S3c) are depicted in Figure S4a, S4b and S4c at 400, 300 and 100 K, respectively. We note that the intensity of the peaks in the F(*Q*) diminishes significantly towards higher *Q* values suggesting the dominance of the diffuse scattering [19].

Also, the intensity of the peaks in the F(*Q*) diminishes quickly with *Q* at 400 K compared to 100 K. This is attributable to thermal broadening [19]. As a refinement result given in Table S1, the value of thermal parameter (B) with relation $B = 8\pi^2 U_{iso}$ is estimated to be $B_{Mn}\approx0.50$ Å$^2$, $B_{Ni}\approx0.82$ Å$^2$, $B_{Ga}\approx1.0$ Å$^2$, respectively, at 400 K (where, $U_{iso}$ is atomic displacement parameter, ADP). These values of thermal parameters lie within physically allowed limit and does not provide any direct indication of possible static disorders as reported for other systems [23,24].

## VI. Temperature-dependent atomic pair distribution function analysis of MnNiGa

The atomic PDF *G(r)* in the short-range (SR), medium-range (MR) and long-range (LR) regimes at the three selected temperatures (400, 300 and 100 K) are shown in Figure S5a, S5b and S5c, respectively. Fourier ripples are clearly visible below the first physical atomic PDF distance ~2.38 Å in Figure S5a. It is evident from Figure S5 that the PDF peaks around 4.6 Å, 8.9 Å, 18.2 Å, 20.8 Å, 27.3 Å etc. split at 100 K on lowering the temperature from 400 K as indicated by arrow marks in Figure S5. These splittings indicate the appearance of new interatomic distances (or new phase) [19]. However, a careful analysis reveals (discussed later) that these splitting in the PDF peaks are related to the effect of temperature only, which can cause PDF peaks to broaden at higher temperatures [19]. Apart from the thermal effects, the appearance of any new PDF peak with the change in temperature was not observed. This is further attributed to the absence of any structural phase transition in the temperature range of 400 to 100 K.

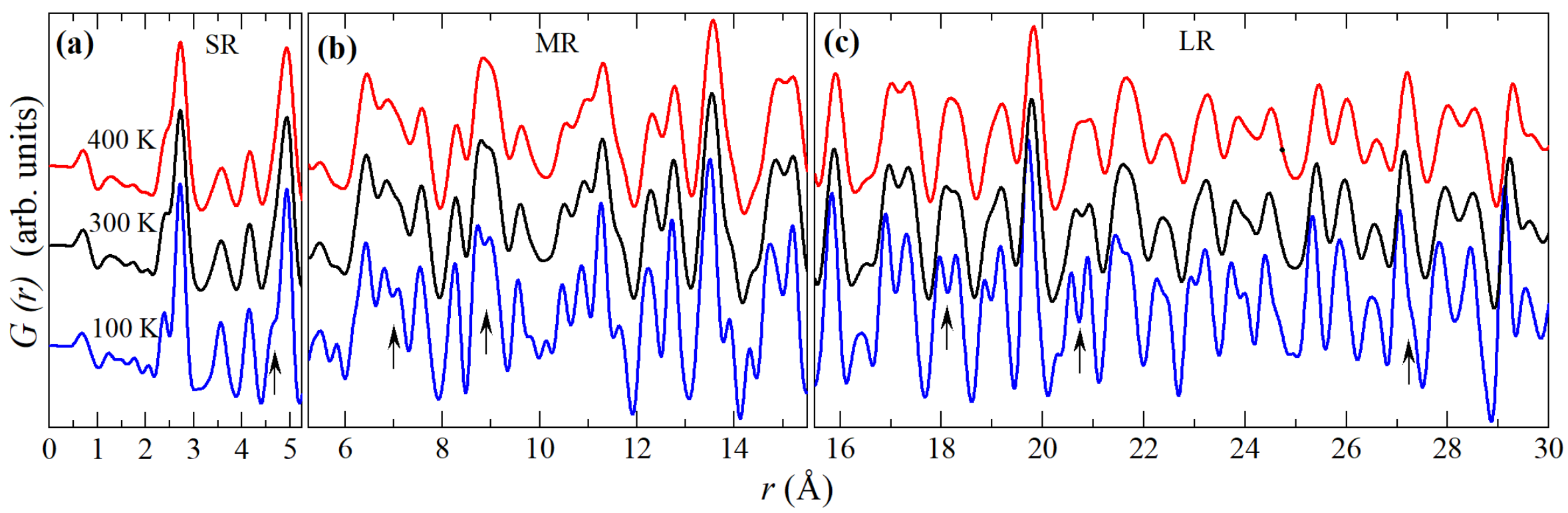

**Figure S5:** Experimental atomic PDF *G(r)* of MnNiGa on the vertically-spaced scale at the three selected temperatures (400, 300 and 100 K) in the regimes **(a)** short-range (SR), **(b)** medium-range (MR) and **(c)** long-range (LR). The arrow marks in **(a)**-**(c)** highlight some of the peaks, which gets sharpen/splitted at the low temperature.

It is essential to note that in principle, the systematic errors (additive and multiplicative errors introduced via incoherent Compton scattering, background scattering, absorption, polarization of beam, etc.) may be present in the atomic PDF [19]. These systematic errors can, however, be minimized (or eliminated) by careful data processing protocols as per the established practices [19]. The PDFgetX3 program use the polynomial fitting to overcome the systematic errors introduced in the atomic PDF during data processing [22]. There are a few criteria has been proposed for testing the quality of PDF data in the literature [25]. The two most robust criteria for good quality PDF data are:

(i) The high-$Q$ portion of structure function S($Q$) should approach unity, i.e., $\lim_{Q\to\infty} S(Q) = 1$ and

(ii) $\Delta G_{low} = \frac{\int_0^{r_{low}} [rG(r)+4\pi r^2 \rho_{fit}]^2 dr}{\int_0^{r_{low}} [4\pi r^2 \rho_{fit}]^2 dr}$ … (2) should be minimum,

where $r_{low}$ and $\rho_{fit}$ are the distances below the first interatomic distance and average atomic number density obtained after fitting the PDF at low-$r$ region, respectively. We took $r_{low}$~2.1 Å for MnNiGa as it is below the first interatomic distance ~2.38 Å for this compound, whereas $r_{low}$~ 2.2 Å was taken for MnPtGa as the first interatomic distance ~ 2.48 Å in this case. We used these two scale independent criteria to judge the quality of the PDF data. The S($Q$) for both compounds, given in Figure S6a-S6b, clearly shows that S($Q$) approaches unity in the high-$Q$ region.

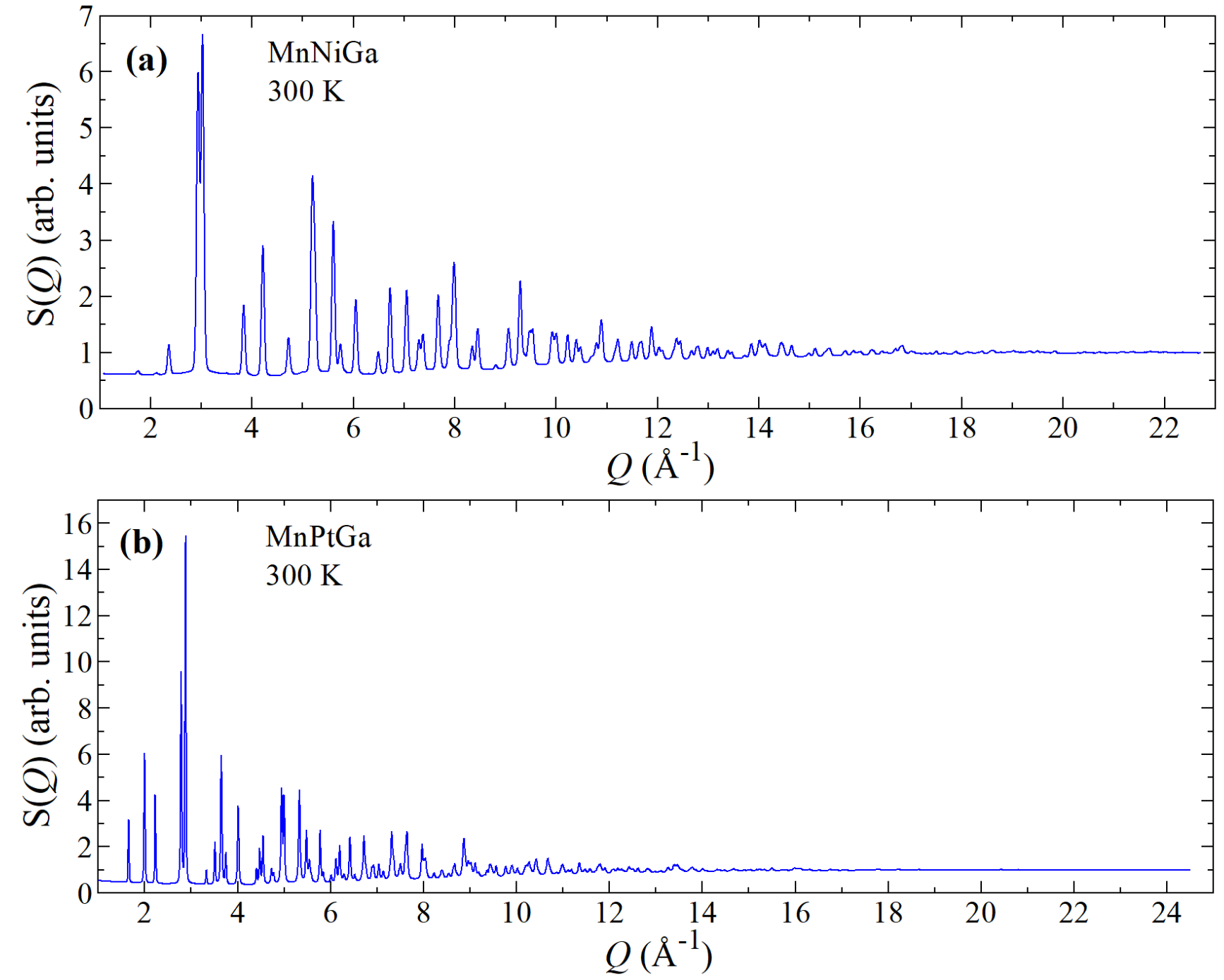

**Figure S6:** The structure function S($Q$) at 300 K for **(a)** MnNiGa and **(b)** MnPtGa.

Now to verify the second criteria, we first need the value of $\rho_{fit}$. Following the equation (1), $G(r)=4\pi r[\rho(r)-\rho_0]$. For low-$r$ region, i.e., $r \rightarrow 0$, this relationship becomes,

$$G(r) = -4\pi r\rho_0 \ldots (3).$$

In the above equation, $\rho_0$ is equivalent to $\rho_{fit}$ [25]. Now, following the above equation (3), we obtained the value of $\rho_0$~0.043 atoms/Å$^3$ for MnNiGa and $\rho_0$~0.038 atoms/Å$^3$ for MnPtGa by fitting the low-$r$ PDF using the equation of a straight line. The value of $\rho_0$~0.043 atoms/Å$^3$, so obtained, was used in conjunction with the selected value of $r_{low}$~2.1 Å to determine the $\Delta G_{low}$ for MnNiGa. After putting these values in equation (2), we obtained $\Delta G_{low}$~0.01, which is a very small value and in good agreement with the recommendations in the literature for obtaining reliable quality PDF data [25]. Similarly, we obtained the $\Delta G_{low}$~0.02, which is also very small, for MnPtGa. Thus, our S($Q$) and $G(r)$ satisfy the two main criteria used for PDF data quality test. This suggest that the present atomic PDFs are characteristics of the sample only and not the experimental artefacts. Further, since there is no scale-dependent analysis (e.g., integration of PDF peak intensities) is carried in the present study, accurate data normalization is less important provided the sample scale factor is a variable parameter in the structure modelling [25].

For a detailed investigation of the peak splitting observed in the experimental PDFs at low temperature (Figure S5), the PDF refinements were carried out for the hexagonal structure ($P6_3/mmc$) at the three selected temperatures (400, 300 and 100 K). All the PDF refinements were carried in the real-space using the program PDFgui [26]. The initial value of lattice parameters, atomic displacement parameters (ADPs) and atomic positions were taken from the reciprocal space Rietveld refined crystallographic structure model at the respective temperatures given in Table S1. To get the instrument resolution parameters ($Q_{damp}$ and $Q_{broad}$), the high-$Q$ SXRPD data was also collected for standard $LaB_6$ sample. The PDFs obtained from this standard sample was fitted and value of $Q_{damp}$=0.030 Å$^{-1}$ and $Q_{broad}$=0.008 Å$^{-1}$ were obtained which is used for the PDF refinements of MnNiGa compound. The lattice parameters, ADPs, scale factor and atomic correlation parameter were refined during the PDF refinements (taking care of correlation between different parameters during the refinement), as all the atoms of MnNiGa in the asymmetric unit for $P6_3/mmc$ space group occupy the special Wyckoff positions only [10,11]. The results of the PDF refinements with the isotropic ADPs consideration in the 2.1-30 Å range at 300 K ($< T_C$) and 100 K ($< T_{SRT}$) are shown in Figure S7a and Figure S7b, respectively, which reveal that the fittings are not satisfactory (evidenced by the significant difference PDFs in Figure S7a and S7b). In contrast, considering the anisotropic ADPs along the basal plane and out-of-plane (*c*-direction) in the Mn, Ni and Ga atoms during the PDF refinements, the PDF fits look satisfactory, as evident by the reduced value of $R_w$ and difference PDFs in Figure S7c and S7d for 300 K and 100 K, respectively (compared to Figure S7a and S7b). This suggests the presence of strong anisotropy, which has been reported for the hexagonal MnNiGa [11]. The parameters obtained after the PDF refinements for $r$ = 2.1-30 Å at the three selected temperatures (400, 300 and 100 K), covering both magnetic transitions ($T_C$ ~ 347 K and $T_{SRT}$ ~ 200 K), are given in Table S2a and Table S3a for isotropic ADPs and anisotropic ADPs, respectively. We note that to avoid any further possible correlation effect during the anisotropic ADPs consideration in the PDF refinement [27], the value of atomic correlation parameter ($\delta_2$) is kept fixed to the value obtained from the isotropic ADPs consideration in the refinement as evident by Table S2a and Table S3a. Also, it is important to note here that although the isotropic ADPs obtained from full-range PDF refinements (shown in Table S2a) are comparable to those obtained from reciprocal space Rietveld refinements (shown in Table S1), the ADPs from the full-range PDF refinements are more reliable as experimental background is subtracted during the atomic PDF conversion. A careful inspection of the anisotropic ADPs values (Table S3a) shows that the $U_{33}$ of Ga is anomalously large and is hinting at the possibility of hidden disorder (similar to Figure S3d) as discussed in more detail hereafter.

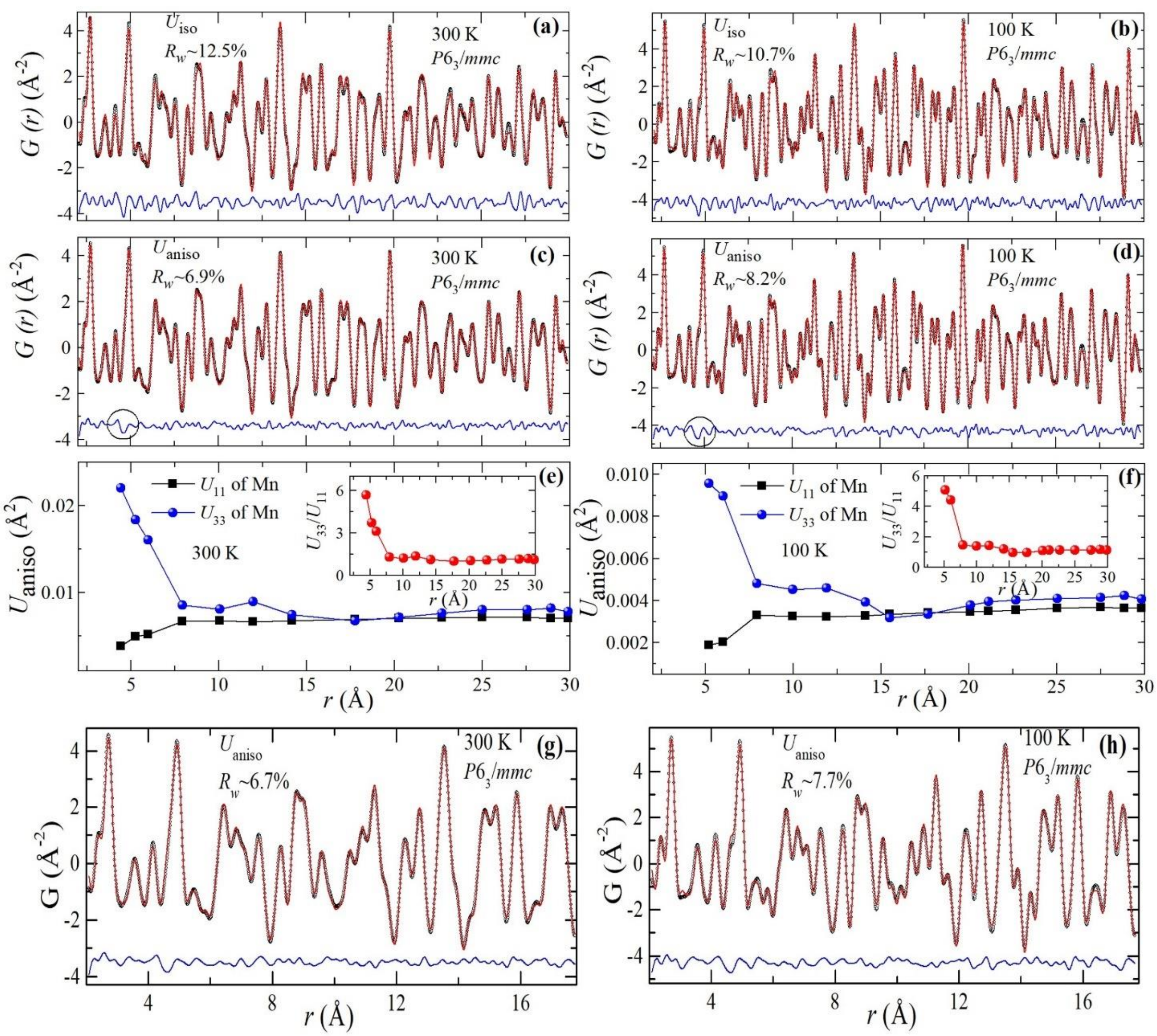

**Figure S7:** The experimental PDF (black circles), calculated PDF (red line), their difference (blue line at the bottom) and weighted agreement factor ($R_w$) obtained by real-space structure refinements of MnNiGa, using observed PDF up to 30 Å for the $P6_3/mmc$ space group with isotropic atomic displacement parameters (ADPs; $U_{iso}$) at **(a)** 300 K and **(b)** 100 K. Results of similar refinements using the observed PDF up to 30 Å with anisotropic ADPs ($U_{aniso}$) at **(c)** 300 K and **(d)** 100 K. The encircled regions in the difference PDF of **(c)** and **(d)** show some misfit just below ~5 Å peak. Variation of anisotropic in-plane $U_{11}$ (black squares) and the out-of-plane $U_{33}$ (blue spheres) of Mn atom with the range ($r$) upto which the observed PDF was used for refinements (i.e., $r = 4.41, 5.26, \ldots$ and 30 Å) at **(e)** 300 K and **(f)** 100 K. The insets of **(e)** and **(f)** depict the ratio $U_{33}/U_{11}$ with $r$ at 300 K and 100 K, respectively. PDF fits upto $r$=17.75 Å with $U_{aniso}$ for the $P6_3/mmc$ space group at **(g)** 300 K and **(h)** 100 K.

A careful visual inspection reveals the slight misfit in the short-range (SR) regime in comparison to the medium-range and long-range (MR+LR) regimes (see the encircled region in the difference PDF just below ~5 Å PDF peak in Figure S7c,d). In order to get more insight into these misfits, the PDF refinements were carried out by varying the maximum distance from $r_{max}$ = 4.41 to 30 Å (i.e., $r$-dependent PDF refinement) at the respective temperatures. We note that during the $r$-dependent PDF refinements, the lattice parameters and correlation parameter ($\delta_2$) were kept fixed to the value obtained from the full-range PDF refinement (2.1-30 Å) [27,28]. The anisotropic ADPs for Mn atoms obtained after $r$-dependent refinements are shown in Figure S7e and Figure S7f, while the result of such a refinement up to $r_{max}$ = 17.75 Å is shown in Figure S7g and Figure S7h for 300 K and 100 K, respectively. It is evident from Figure S7e,f that anisotropic ADPs corresponding to Mn changes drastically in the SR regime, which is an indication of the local structure distortion as reported in other systems [29,30]. For better visualization of changes in the SR regime, the ratio of anisotropic ADPs ($U_{33}/U_{11}$) of Mn atom at 300 K and 100 K are shown in the inset of Figure S7e and Figure S7f, respectively.

**TABLE S2(a).** The lattice parameters (*a* and *c*), isotropic atomic displacements parameters ($U_{iso}$), atomic correlation parameter ($\delta_2$) and weighted agreement factor ($R_w$) obtained from the PDF refinements upto 30 Å for the hexagonal structure in the *P*6$_3$/*mmc* space group using the experimental atomic PDF data of MnNiGa at the three selected temperatures.

| **Temperature** → **Parameters** ↓ | **400 K** | **300 K** | **100 K** |
|---|---|---|---|
| *a* (Å) | 4.1612(8) | 4.1547(7) | 4.1452(4) |
| *c* (Å) | 5.334(1) | 5.321(1) | 5.295(1) |
| $U_{iso}^{Mn}$ (Å$^2$) | 0.0089(4) | 0.0072(3) | 0.0036(2) |
| $U_{iso}^{Ni}$ (Å$^2$) | 0.0091(5) | 0.0073(4) | 0.0038(2) |
| $U_{iso}^{Ga}$ (Å$^2$) | 0.0105(5) | 0.0080(4) | 0.0041(2) |
| $\delta_2$ (Å$^2$) | 2.4(5) | 2.3(4) | 1.9(5) |
| $R_w$ (%) | 13.3 | 12.5 | 10.7 |

**TABLE S2(b).** Result of the similar refinements in the SR (upto ~5.25 Å) for the hexagonal structure (*P*6$_3$/*mmc*) using the atomic PDF data of MnNiGa. These refinements are done with keeping the $\delta_2$ fixed to the value obtained from the full range refinements given in Table S2(a).

| **Temperature** → **Parameters** ↓ | **400 K** | **300 K** | **100 K** |
|---|---|---|---|
| *a* (Å) | 4.171(6) | 4.162(6) | 4.149(4) |
| *c* (Å) | 5.34(2) | 5.33(1) | 5.29(1) |
| $U_{iso}^{Mn}$ (Å$^2$) | 0.013(1) | 0.010(1) | 0.0048(6) |
| $U_{iso}^{Ni}$ (Å$^2$) | 0.007(1) | 0.005(1) | 0.0031(6) |
| $U_{iso}^{Ga}$ (Å$^2$) | 0.008(1) | 0.006(1) | 0.0037(6) |
| $R_w$ (%) | 11.8 | 11.4 | 10.7 |

**TABLE S3(a)**. The lattice parameters ($a$ and $c$), anisotropic atomic displacements parameters ($U_{11} = U_{22}$ and $U_{33}$), atomic correlation parameter $\delta_2$ (kept fixed to the value obtained from the full range refinements using isotropic ADPs given in Table S2a) and $R_w$ obtained from the PDF refinements upto 30 Å for the hexagonal structure ($P6_3/mmc$) using the atomic PDF data of MnNiGa.

| **Temperature →** / **Parameters ↓** | **400 K** | **300 K** | **100 K** |
|---|---|---|---|
| $a$ (Å) | 4.1623(6) | 4.1553(5) | 4.1451(3) |
| $c$ (Å) | 5.336(2) | 5.323(2) | 5.296(1) |
| $U_{11}^{Mn}$ (Å$^2$) | 0.0090(6) | 0.0070(4) | 0.0036(2) |
| $U_{33}^{Mn}$ (Å$^2$) | 0.010(2) | 0.008(1) | 0.0042(5) |
| $U_{11}^{Ni}$ (Å$^2$) | 0.0080(7) | 0.0064(4) | 0.0031(2) |
| $U_{33}^{Ni}$ (Å$^2$) | 0.013(2) | 0.010(1) | 0.0070(8) |
| $U_{11}^{Ga}$ (Å$^2$) | 0.0064(6) | 0.0052(4) | 0.0036(2) |
| $U_{33}^{Ga}$ (Å$^2$) | 0.024(3) | 0.018(2) | 0.0055(7) |
| $\delta_2$ (Å$^2$) | 2.4(5) | 2.3(4) | 1.9(5) |
| $R_w$ (%) | 6.4 | 6.9 | 8.2 |

**TABLE S3(b)**. Result of the similar refinements in the SR (upto ~5.25 Å) for the hexagonal structure ($P6_3/mmc$) for MnNiGa. These refinements are done while keeping the $\delta_2$ and lattice parameters fixed to the value obtained from the full range refinements given in Table S3(a).

| **Temperature →** / **Parameters ↓** | **400 K** | **300 K** | **100 K** |
|---|---|---|---|
| $U_{11}^{Mn}$ (Å$^2$) | 0.006(1) | 0.005(1) | 0.0019(6) |
| $U_{33}^{Mn}$ (Å$^2$) | 0.024(5) | 0.019(4) | 0.009(3) |
| $U_{11}^{Ni}$ (Å$^2$) | 0.009(1) | 0.007(1) | 0.0046(8) |
| $U_{33}^{Ni}$ (Å$^2$) | 0.011(4) | 0.008(4) | 0.003(2) |
| $U_{11}^{Ga}$ (Å$^2$) | 0.006(1) | 0.005(1) | 0.0033(7) |
| $U_{33}^{Ga}$ (Å$^2$) | 0.022(6) | 0.017(5) | 0.007(3) |
| $R_w$ (%) | 7.4 | 7.3 | 8.1 |

A large increase in $U_{33}/U_{11}$ in the SR regime suggests the dominating ADP along the *c*-direction as compared to the ADP along the basal plane. Such a drastic change in the ADPs manifests that the atoms deviate from their Wyckoff positions assigned for the SR regime during the refinement [29,31-34]. The value thermal parameter (*B*), obtained using the relation $B_{33} = 8\pi^2 U_{33}$ [35], comes out ~1.74 Å$^2$ and ~0.75 Å$^2$ at 300 K and 100 K, respectively, for the PDF refinements in the SR regime. Such a large value of $B_{33}$ and the $U_{33}$ along the *c*-axis indicate the off-centering of the Mn atom from its original atomic positions in the SR regime [29,31-34]. In addition, the interatomic distances also vary drastically in the SR regime. These observations suggest that the local SR structure of MnNiGa can differ from the average LRO crystal structure ($P6_3/mmc$) of MnNiGa. We note that although anisotropic ADPs value are found to be nearly same for the full-range for the Mn atom, we find a considerable difference between $U_{11}$ and $U_{33}$ anisotropic ADPs for the Ni and Ga atoms (well separated beyond the estimated standard deviations) as evident by Table S3a. This leads to significant reduction in the fitting agreement factor ($R_w$) compared to the case considering the isotropic ADPs. Furthermore, we find the value of anisotropic ADPs for Ni and Ga remained independent of *r*-space after the *r*-dependent PDF analysis, and only deviation is found in the ADPs of Mn in the SR as given in Figure 2i of the main text and Figure S7e,f. The structural anisotropy of Mn disappears at the higher length-scale and leading to abrupt drop in $U_{33}$ of Mn as function of $r_{max}$ (Figure 2i of the main text).

## VII. Identification of the most suitable lower symmetric structure of MnNiGa

The different subgroups of $P6_3/mmc$ (194) were considered one by one in the PDF refinements to determine a better local structure. The subgroups were obtained using the ISODISTORT software in the ISOTROPY suite [36,37]. The subgroups of $P6_3/mmc$ were found to be hexagonal, trigonal, orthorhombic, monoclinic and triclinic structures. Out of them, the subgroups with the hexagonal structure are $P\bar{6}2c$ (190), $P\bar{6}m2$ (187), $P6_3mc$ (186), $P6_322$ (182), $P6_3/m$ (176), $P\bar{6}$ (174) and $P6_3$ (173). On the other hand, the subgroups with the trigonal structure are $P\bar{3}m1$ (164), $P\bar{3}1c$ (163), $P31c$ (159), $P3m1$ (156), $P321$ (150), $P312$ (149), $P\bar{3}$ (147) and $P3$ (143). The subgroups with the orthorhombic structure are $Cmcm$ (63), $Ama2$ (40), $C222_1$ (20), $Amm2$ (38) and $Cmc2_1$ (36). The subgroups with the monoclinic structure are $C2/c$ (15), $C2/m$ (12), $P2_1/m$ (11), $Cc$ (9), $Cm$ (8), $Pm$ (6), $C2$ (5) and $P2_1$ (4), while the subgroups with the triclinic structure are $P\bar{1}$ (2) and $P1$ (1). Out of all these subgroups, the orthorhombic, monoclinic and triclinic structural subgroups are discarded as no extra peak, or peak splitting is observed in the present temperature dependent experimental PDFs of MnNiGa, i.e., it does not show any signs of phase transition (see Figure S5 and S7). The asymmetric unit of the remaining subgroups were obtained using composition and space group [38].

In the parent $P6_3/mmc$, all the atoms are occupied their special Wyckoff positions, i.e., Mn at 2a (0, 0, 0), Ni at 2d (1/3, 2/3, 3/4) and Ga at 2c (1/3, 2/3, 1/4) with total six number of atoms per formula unit (or asymmetric unit) of MnNiGa [10,11]. The PDF fits using this space group in the SR regime at 300 K is shown in Figure S8a, which reveals a significant misfit in the peaks at $r \sim$ 2.35 Å, 3.5 Å, 4.25 Å and 4.75 Å with $R_w$ ~11.4%. In order to get a better structural model, the PDF refinements were carried out using each hexagonal and trigonal subgroup of $P6_3/mmc$ and the result of the refinements are shown in Figure S8b-S8p in the SR regime at 300 K.

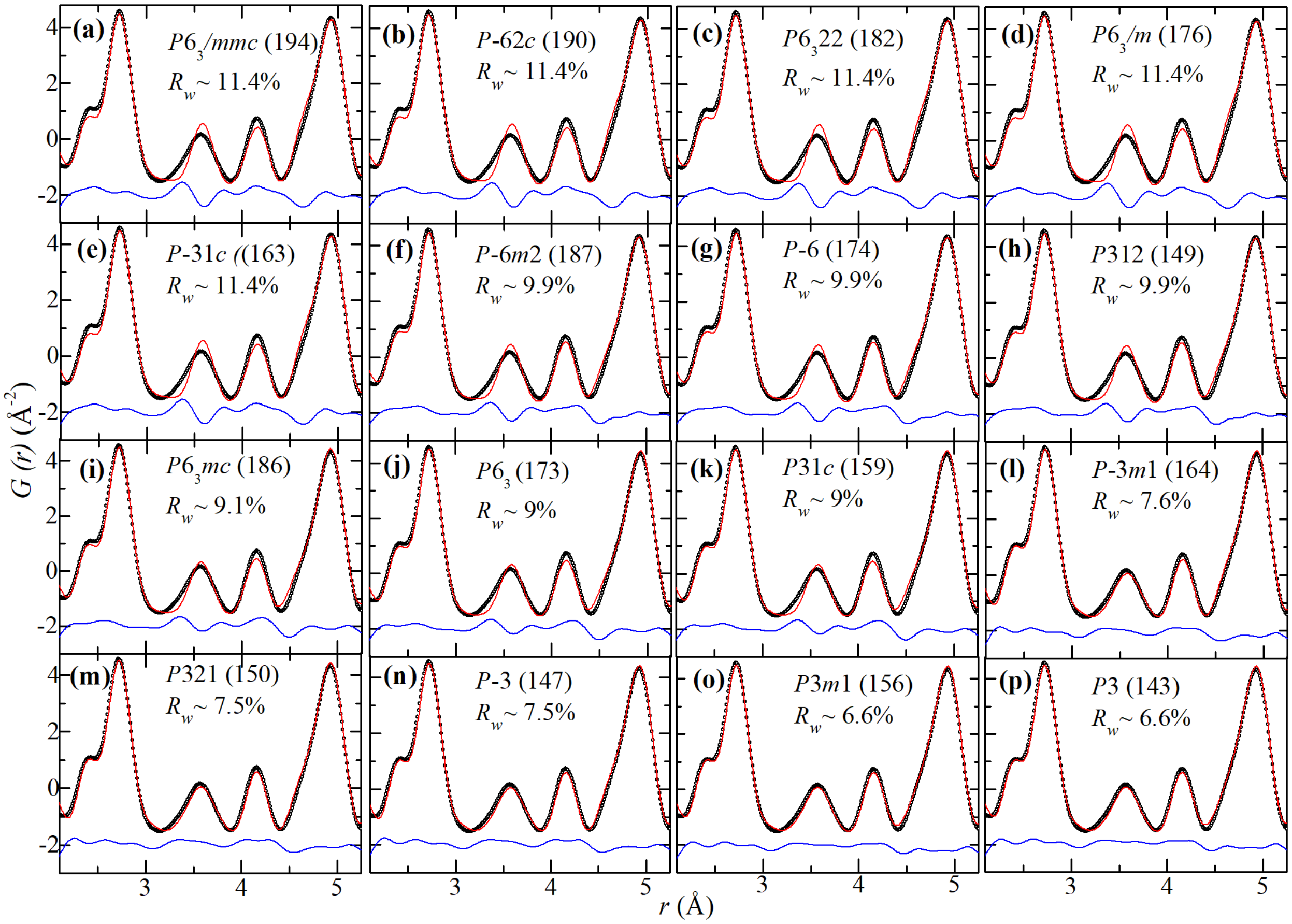

**Figure S8:** The experimental PDF (black circles), calculated PDF (red line), their difference (blue line at the bottom) and weighted agreement factor ($R_w$) obtained by real-space structure refinements of MnNiGa at 300 K in the SR regime using space groups **(a)** $P6_3/mmc$, **(b)** $P\bar{6}2c$, **(c)** $P6_322$, **(d)** $P6_3$/m, **(e)** $P\bar{3}1c$, **(f)** $P\bar{6}m2$, **(g)** $P\bar{6}$, **(h)** $P312$, **(i)** $P6_3mc$, **(j)** $P6_3$, **(k)** $P31c$, **(l)** $P\bar{3}m1$, **(m)** $P321$, **(n)** $P\bar{3}$, **(o)** $P3m1$ and **(p)** $P3$.

In order to get a better structural model, the PDF refinements were carried out using each hexagonal and trigonal subgroup of $P6_3/mmc$ and the result of the refinements are shown in Figure S8b-S8p in the SR regime at 300 K. The lattice parameters, isotropic ADPs and suitable atomic positions were varied during the PDF refinements. Since the asymmetric unit of $P\bar{6}2c$, $P6_322$, $P6_3/m$ and $P\bar{3}1c$ contains Wyckoff positions similar to $P6_3/mmc$; they provide similar fittings to that observed for $P6_3/mmc$ (see Figure S8a-S8e). This suggests that $P\bar{6}2c$, $P6_322$, $P6_3/m$ and $P\bar{3}1c$ do not belong to a suitable lower symmetry space group. Since the asymmetric unit of $P\bar{6}m2$, $P\bar{6}$, $P312$, $P6_3mc$, $P6_3$ and $P31c$ contain refinable Wyckoff positions; they provide a slightly better fit compared to $P6_3/mmc$ (see Figure S8a and Figure S8f-S8k). Although the fitting from these subgroups ($P\bar{6}m2$, $P\bar{6}$, $P312$, $P6_3mc$, $P6_3$ and $P31c$) look good, slight misfits are still present in the peaks at $r$ ~ 2.35 Å, 3.5 Å and 4.25 Å. Therefore, a further possibility was investigated to achieve the best model structure. The possibility of *Pnma* space group, which is an orthorhombic martensite phase, appears via transformation from the hexagonal austenite phase observed in *MnM'X* class of materials [39], was discarded as this space group (*Pnma*) provided a bad fit to the experimental PDF.

Among all the subgroups, it is found that the $P\bar{3}m1$, P321 and $P\bar{3}$ provide excellent fits to the experimental PDF, as shown in Figure S8l, Figure S8m and Figure S8n, respectively. We note that these three subgroups ($P\bar{3}m1$, $P321$ and $P\bar{3}$) provide almost similar fitting (with $R_w$ ~ 7.6%) to each other as they have similar kinds of Wyckoff positions. Therefore, among these three subgroups ($P\bar{3}m1$, $P321$ and $P\bar{3}$), the highest symmetric subgroup, which is $P\bar{3}m1$ is considered as the model structure. Interestingly, the PDF refinements using the subgroups $P3m1$ and $P3$ provide an even better fit compared to $P\bar{3}m1$. The PDF fits using the subgroups $P3m1$ and $P3$ are shown in Figure S8o and Figure S8p, respectively. We note that the $P3m1$ and $P3$ provide similar fitting (with $R_w$ ~ 6.6%) to each other as they have similar Wyckoff positions. Therefore, among these two subgroups ($P3m1$ and $P3$), the higher symmetric subgroup, which is $P3m1$, is considered as the structural model. It is important to mention here that the z-coordinate of Mn1 was considered at zero in the 1a (0, 0, z) Wyckoff position to fix the origin at (0, 0, 0) during the PDF refinements using the $P3m1$ and $P3$ space groups. It is evident from Figure S8o that tiny misfits in the peaks at $r$ ~ 3.5 Å and 4.25 Å using $P\bar{3}m1$ (see Figure S8l) almost disappears using the $P3m1$ space group. This is also reflected in the value of $R_w$, which decreased from 7.6% (for $P\bar{3}m1$) to 6.6% (for $P3m1$). Although the reduction in the $R_w$ is not very significant (unit magnitude), results have been reported on the basis of such a small reduction in the value of $R_w$ in the literature [32,40]. Further, the average $P3m1$ space group has been reported for a sister compound (MnPtGa) [41]. All these results indicate that the trigonal $P3m1$ is the most suitable lower symmetry space group to model the present experimental PDF. Thus, the local structure of MnNiGa in the SR regime corresponds to the primitive trigonal in the $P3m1$ space group at 300 K. It is important to mention here that the Rietveld and PDF fits can be easily improved by reducing the crystal symmetry and thereby introducing extra refinable parameters, which may not always have physical meaning, but it is not always true. For example, the fits in the MR+LR regimes do not improve, but rather deteriorate, by lowering the symmetry to trigonal space group with fixed atomic positions of SR regimes (see Figure 3 and Figure 4 of the main text). The main point is that if there are features coming from a particular symmetry, they would not be successfully modelled even after using additional parameters by rejecting those parameters to values within the estimated standard deviations (esd). Thus, if there is a certain feature which can be modelled only by reducing the crystal symmetry then that lower symmetry structure is considered to be the correct model structure of the material. In the present study, the features in the atomic PDF (e.g., misfits at $r \sim 2.35$ Å, 3.5 Å, 4.25 Å and 4.75 Å in the SR regime) could not be captured using standard hexagonal structure. Consideration of a lower symmetry model (trigonal) was the logical next step to interpret those features in the atomic PDF. In the absence of the lower symmetry trigonal structure, the PDF in the SR regime cannot be successfully interpreted in terms of the hexagonal symmetry.

Moreover, for a better interpretation of the lower symmetry structural model, we performed the refinements using atomic PDF data at 300 K in the SR regime keeping equal number of parameters in the both hexagonal ($P6_3/mmc$) and trigonal ($P3m1$) structural models. There are 6 parameters (two lattice parameters, three isotropic atomic displacement parameters, and one correlation parameter) for the hexagonal $P6_3/mmc$ space group, which leads to $R_w$~11.4 % as given in Figure S8a. In case of the trigonal $P3m1$ space group, 11 parameters (5 atomic positions on lowering the symmetry and the 5 other structural parameters like two lattice parameters, three isotropic atomic displacement parameters and one correlation parameter) were refined giving $R_w$~6.6 %. However, even if we fix the lattice and atomic displacement parameters to the values equal to those obtained after the PDF refinements using $P6_3/mmc$ symmetry leaving out only 6 refinable parameters making the same number of refinable parameters to the hexagonal space group with 6 parameters, refinement using $P3m1$ space group led to a value of agreement factor $R_w$~7.3 %, which is a significantly lower value over the $R_w$~11.4 % for the hexagonal symmetry with a far better fit as can be seen from a comparison of Figure S9 and Figure S8a. In fact, the agreement factor so obtained by the constrained refinement for the trigonal symmetry is quite close to the value of $R_w$~6.6 % obtained after refining all the 11 parameters (see Figure S8o). This suggests that our results on the presence of lower symmetry structure with trigonal symmetry ($P3m1$) in the SR regime are not an artefact of increased number of parameters but represent the real feature of the atomic PDF.

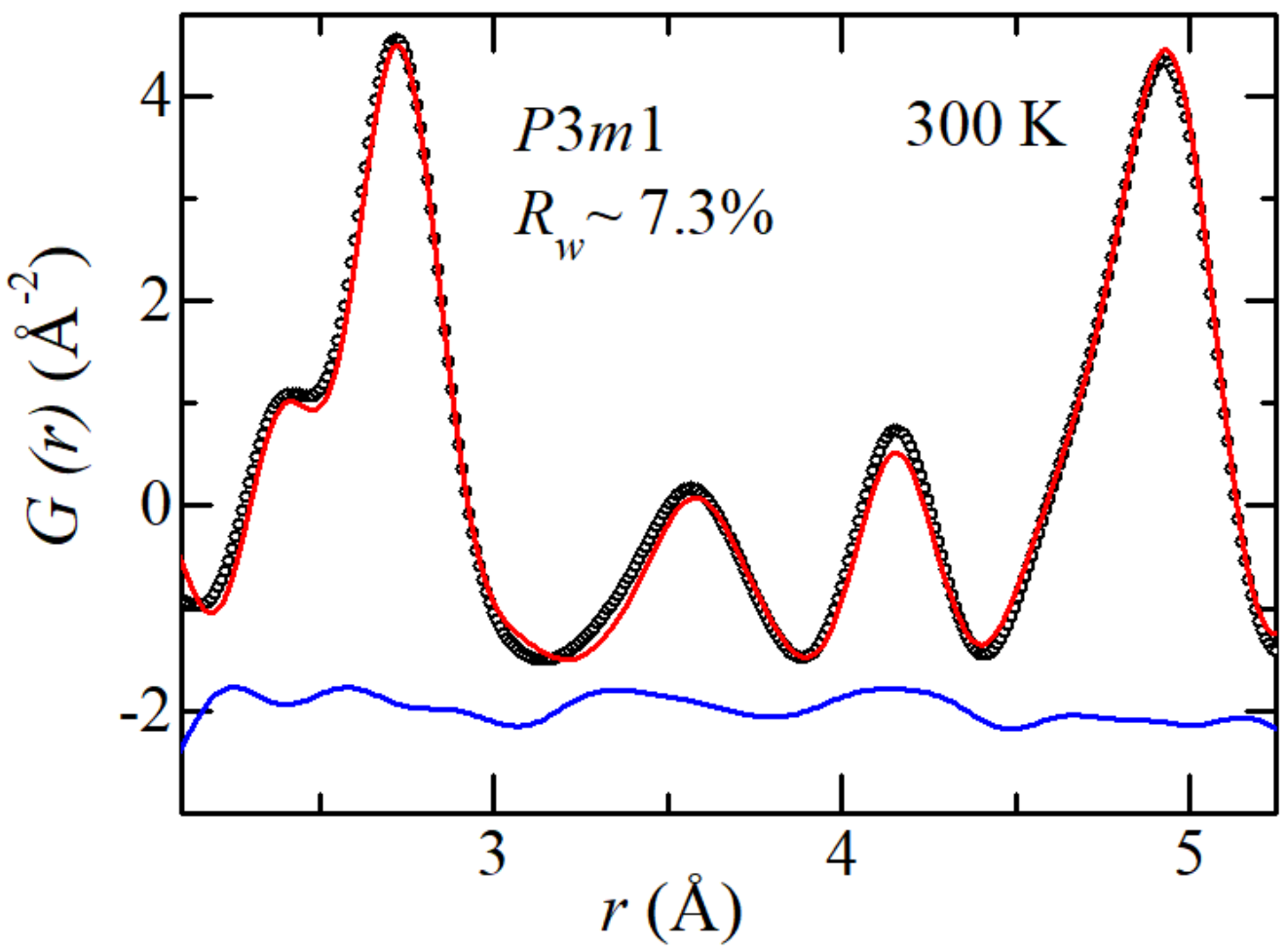

**Figure S9:** The experimental PDF (black circles), calculated PDF (red line), their difference (blue line at the bottom), and weighted agreement factor ($R_w$) obtained by PDF refinement in the SR regime using the $P3m1$ model for MnNiGa at 300 K, where the lattice parameters and isotropic ADPs were fixed at the values obtained by PDF refinement using hexagonal structure ($P6_3/mmc$).

As MnNiGa exhibits paramagnetic to ferromagnetic (FM) phase transition at $T_C$ ~ 347 K and FM to spin reorientation transition (SRT) transition at $T_{SRT}$ ~ 200 K (see Figure 1 of the main text), it is interesting to see the stability of the present local structure ($P3m1$) of MnNiGa with change in temperature. The local (SR) trigonal structure in the $P3m1$ space group has been confirmed in the paramagnetic phase ($T_C < 400$ K $> T_{SRT}$) (see Figure 3b and related discussion in the main text). For comparison, the bond distances obtained from the PDF refinements in the SR regime with the $P6_3/mmc$ and $P3m1$ space groups using the experimental atomic PDF data of MnNiGa at 400 K are given in Table S4. The comparison of PDF refinements using the $P6_3/mmc$ and $P3m1$space groups in the SR regime at 300 K (FM phase) and 100 K (SRT phase) are shown in Figure S10a- S10d. The parameters obtained from the PDF refinements in the SR regime using the $P3m1$ space group at the three selected temperatures are given in Table S5. The PDF refinement in the FM phase at 300 K ($T_C$ > 300 K) provides better fit with $P3m1$ (see Figure S10b with $R_w$ ~ 6.6%) in comparison to $P6_3/mmc$ (see Figure S10a with $R_w$ ~ 11.4%). This suggests that the local structure of MnNiGa at 300 K in the SR regime corresponds to the primitive trigonal ($P3m1$).

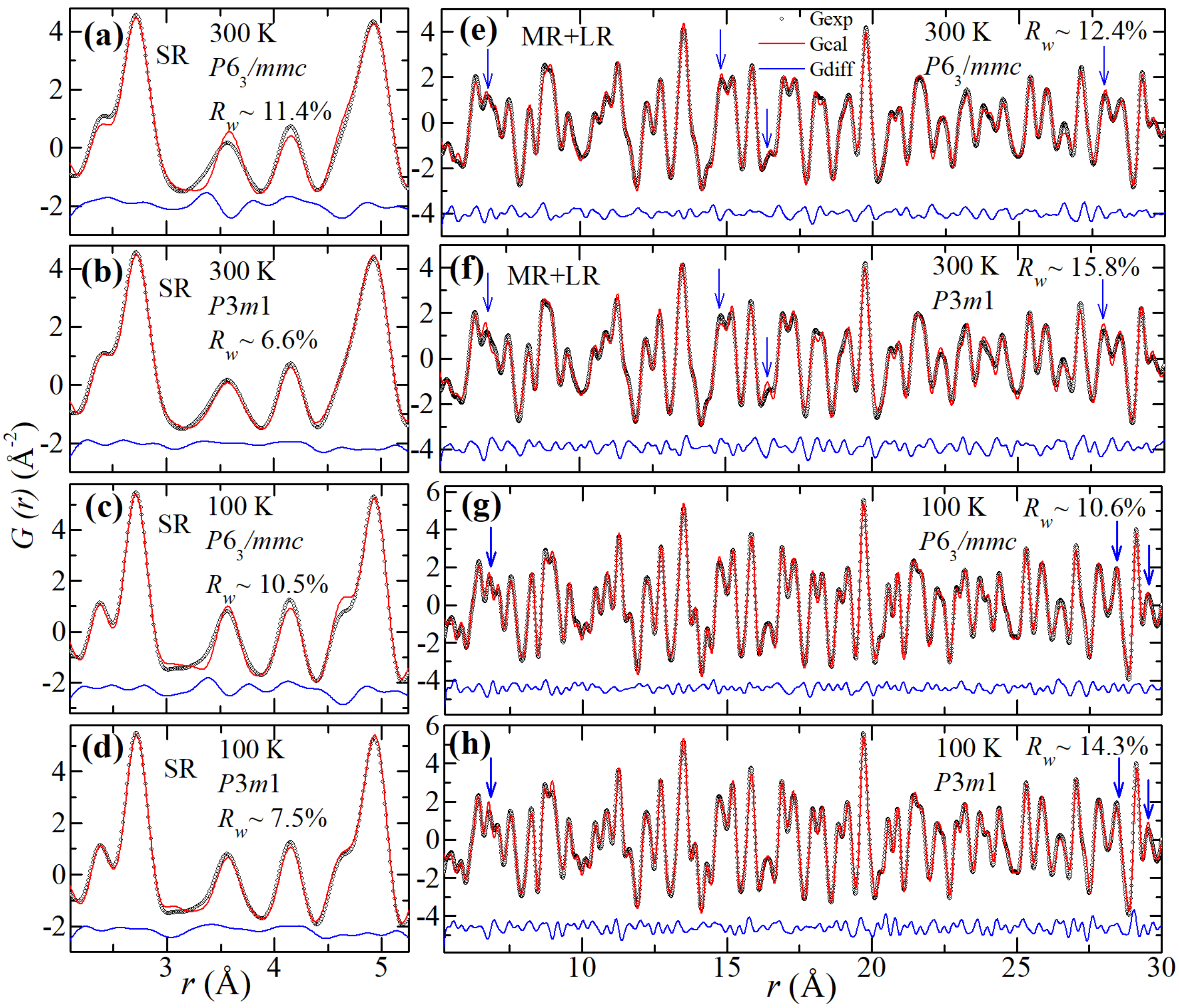

**Figure S10:** The experimental PDF (black circles), calculated PDF (red line), their difference (blue line at the bottom) and weighted agreement factor ($R_w$) obtained by real-space structure refinements of MnNiGa at 300 K in the SR regime using the space groups **(a)** $P6_3/mmc$ and **(b)** $P3m1$; while at 100 K using the space groups **(c)** $P6_3/mmc$ and **(d)** $P3m1$. The corresponding fits in the MR+LR regimes without altering the atomic positions obtained from the SR regime refinements at 300 K using **(e)** $P6_3/mmc$ and **(f)** $P3m1$ space groups; while at 100 K using **(g)** $P6_3/mmc$ and **(h)** $P3m1$ space groups. The arrow marks in **(e)** to **(h)** highlight some of the regions where the fit has improved significantly for the $P6_3/mmc$ space group as compared to that for the $P3m1$ space group in the MR+LR regimes.

Further, the PDF refinement in the SRT phase at 100 K ($T_{SRT} > 100$ K) provides a better fit with $P3m1$ (see Figure S10d with $R_w \sim 7.5\%$) compare to the $P6_3/mmc$ space group (see Figure S10c with $R_w \sim 10.5\%$). This suggests that the trigonal structure in the $P3m1$space group is present in the SR regime even at 100 K. All these results suggest that the local SR structure of MnNiGa corresponds to primitive trigonal in the $P3m1$ space group and stable down to 100 K from 400 K.

Moreover, we have also compared the MR+LR regimes structure of MnNiGa with the $P3m1$ and $P6_3/mmc$ space groups in the FM and SRT phase (covering both the magnetic phase transition, i.e., $T_C$ and $T_{SRT}$). The MR+LR regimes structure in the paramagnetic phase ($T_C < 400$ K $> T_{SRT}$) is compared in the main text on the basis of PDF refinements given in Figure 3c and 3d of the main text. The comparison of PDF refinements using $P6_3/mmc$ and $P3m1$space groups in the MR+LR regimes at 300 K (FM phase) and 100 K (SRT phase) are shown in Figure S10(e)-S10(h). A better fit between experimental and calculated PDFs is observed for $P6_3/mmc$ (see Figure S10e with $R_w \sim 12.4\%$) in comparison to $P3m1$ space group (see Figure S10f with $R_w \sim 15.8\%$) at 300 K in the MR+LR regimes. The improvement in the fits around some of the PDF peaks is highlighted with arrows in these two figures. Further, such a better fit is observed down to 100 K for $P6_3/mmc$ (see Figure S10g with $R_w \sim 10.6\%$) compared to the $P3m1$ space group (see Figure S10h with $R_w \sim 14.3\%$) in the MR+LR regimes. In addition, the Rietveld refinements (shown in Figure S1 and S3) clearly reveal that the average LR crystal structure of MnNiGa is hexagonal in the $P6_3/mmc$ space group. Thus, the local structure in the SR regime corresponds to the primitive trigonal in the $P3m1$ space group, whereas the structure in the MR+LR regimes corresponds to the average hexagonal in the $P6_3/mmc$ space group of MnNiGa compound in the 400-100 K range.

It is important to emphasis here that fits to the MR+LR regimes should be effectively same using either $P6_3/mmc$ or $P3m1$ with special Wyckoff positions for both space groups. Moreover, in the refinements for the $P3m1$ space group for MR+LR regimes, presented in Figure 3c and 3d of the main text and Figure S10f and S10h, the atomic positions were fixed to the values obtained from refinements in the SR regime, as the displacements in the atoms should not alter on going from the SR to the MR+LR regimes if this ($P3m1$) is the correct LRO structure. There are several previous reports, where the atomic positions were fixed corresponding to the SR regime during the refinements for the MR and LR regimes for the same space group [28,42]. Therefore, the structure of MnNiGa belongs to the hexagonal in the MR+LR regime.

**TABLE S4.** Comparison of bond distances obtained from the real-space refinements for the hexagonal ($P6_3/mmc$) and trigonal ($P3m1$) structures in the short-range regime using the experimental PDF data of MnNiGa at 400 K.

| **S. N.** | **Pairs** | **Bond distances for $P6_3/mmc$ (Å)** | **Pairs** | **Bond distances for $P3m1$ (Å)** |
|---|---|---|---|---|
| 1 | Ni-Ga<br>Ni-Ga | 2.384(3)<br>2.420(5) | Ni-Ga | 2.380(2), 2.387(2), 2.416(5), 2.423(5) |
| 2 | Ni-Ga= Mn-Mn<br>Mn-Ni=Mn-Ga<br>Mn-Ni=Mn-Ga | 2.674(7)<br>2.733(4)<br>2.765(6) | Ni-Ga<br>Mn-Mn<br>Mn-Ga<br>Mn-Ga<br>Mn-Ni<br>Mn-Ni<br>Mn-Ni | 2.51(1), 2.65(1), 2.68(2), 2.81(1)<br>2.54(1), 2.79(1)<br>2.646(8), 2.665(7), 2.679(8), 2.697(5),<br>2.731(4), 2.762(6), 2.885(7), 2.914(8)<br>2.659(8), 2.692(5), 2.731(6), 2.737(8)<br>2.762(4), 2.762(7), 2.768(8), 2.786(5)<br>2.817(3) |
| 3 | Ga-Ga=Ni-Ni<br>Ga-Ga-Ni-Ni | 3.58(1)<br>3.61(1) | Ga-Ga<br>Ni-Ni | 3.45(3), 3.47(1), 3.70(1), 3.72(2)<br>3.55(2), 3.57(1), 3.62(3) |
| 4 | Mn-Mn=Ni-Ni=Ga-Ga | 4.171(4) | Mn-Mn=Ni-Ni=Ga-Ga | 4.164(4) |
| 5 | Mn-Ga= Mn-Ni<br>Mn-Ga= Mn-Ni<br>Ni-Ga<br>Ni-Ga | 4.67(1)<br>4.68(2)<br>4.81(1)<br>4.841(5) | Mn-Ga<br>Mn-Ga<br>Mn-Ni<br>Mn-Ni<br>Ni-Ga<br>Ni-Ga | 4.39(2), 4.41(1), 4.64(2), 4.66(1),<br>4.76(1), 4.78(2), 4.80(1), 4.82(2)<br>4.55(1), 4.57(1), 4.63(2), 4.64(1)<br>4.65(1), 4.66(1), 4.77(2), 4.79(1)<br>4.796(6), 4.799(4), 4.832(3), 4.836(4)<br>4.86(1) |
| 6 | Mn-Mn=Ni-Ga<br>Mn-Ga= Mn-Ni<br>Mn-Ga= Mn-Ni | 4.96(1)<br>4.987(6)<br>5.02(1) | Mn-Mn<br>Ni-Ga<br>Mn-Ga<br>Mn-Ga<br>Mn-Ni<br>Mn-Ni | 4.88(1), 5.01(1)<br>4.93(1), 4.95(2), 5.02(1)<br>4.934(6), 4.94(4), 4.97(1), 4.979(3)<br>4.98(1), 5.01(2), 5.066(5), 5.10(1)<br>4.941(3), 4.976(6), 4.980(4), 4.983(2)<br>5.01(3), 5.02(1), 5.04(2) |
| 7 | Mn-Mn=Ni-Ni=Ga-Ga | 5.347(2) | Mn-Mn=Ni-Ni=Ga-Ga | 5.32(1) |

**TABLE S5.** The lattice parameters (*a* and *c*), isotropic atomic displacements parameters ($U_{iso}$), atomic correlation parameter ($\delta_2$), atomic positions (z) and weighted agreement factor ($R_w$) obtained by the PDF refinements in short-range regime for the trigonal structure in the *P*3*m*1 space group using the experimental atomic PDF of MnNiGa at the three selected temperatures.

| **Temperature** → / **Parameters** ↓ | **400 K** | **300 K** | **100 K** |
|---|---|---|---|
| *a* (Å) | 4.164(6) | 4.159(5) | 4.149(4) |
| *c* (Å) | 5.32(2) | 5.31(2) | 5.29(1) |
| $U_{iso}^{Mn}$ (Å$^2$) | 0.005(2) | 0.004(1) | 0.0035(5) |
| $U_{iso}^{Ni}$ (Å$^2$) | 0.010(2) | 0.007(1) | 0.0037(7) |
| $U_{iso}^{Ga}$ (Å$^2$) | 0.010(1) | 0.006(1) | 0.0030(6) |
| $z_{Mn2}$ | 0.523(7) | 0.517(8) | 0.509(8) |
| $z_{Ni1}$ | 0.777(8) | 0.771(7) | 0.767(47) |
| $z_{Ni2}$ | 0.27(1) | 0.26(2) | 0.25(1) |
| $z_{Ga1}$ | 0.306(5) | 0.296(5) | 0.277(4) |
| $z_{Ga2}$ | 0.775(8) | 0.768(7) | 0.760(6) |
| $\delta_2$ (Å$^2$) | 3.4(5) | 2.8(5) | 2.1(5) |
| $R_w$ (%) | 6.2 | 6.6 | 7.5 |

We note that the first signature of possible static atomic displacements *w.r.t.* the special atomic positions of the hexagonal structure ($P6_3/mmc$) was provided by the anomalous thermal parameters in the SR regime (see Figure 2i of the main text and Figure S7e and S7f). This motivated us to consider static atomic displacements corresponding to the lower symmetry trigonal structure (*P*3*m*1) in the real space refinement using the PDF data which eventually led to normal thermal parameter values. Since there is a systematic variation in the isotropic ADP ($U_{iso}$) with temperature (see Table S5), the dynamic displacements seem to have been modelled correctly using the thermal parameters. We note that the misfits between the experimental and calculated atomic PDF in the short-range PDFs may be related to correlated atomic motion i.e., can be dynamic in character [43]. Therefore, we have considered the atomic correlation parameter ($\delta_2$) related to correlated atomic motion during the entire PDF refinements and related obtained value of $\delta_2$ is given in Tables

S2, S3, S5. Therefore, we infer that the misfits between the experimental and calculated atomic PDF in the SR regime are related to usual atomic displacement and not related to the dynamical correlated atomic motion. The additional displacements in the atomic positions, modelled by the lower symmetry trigonal structure may therefore be regarded as static. However to confirm this, time-resolved PDF data (e.g., variable-shutter atomic PDF [44]) above the picosecond level is required, which is the matter of future investigation.

## VIII. High-*Q* synchrotron x-ray powder diffraction analysis of MnPtGa

First of all, the average long-range ordered crystal structure is determined by the Rietveld refinement of high-*Q* data in the reciprocal space [8]. The refinement was carried out using the hexagonal structure in the $P6_3/mmc$ space group, considering all the atoms at the special Wyckoff positions, i.e., Mn at 2a (0, 0, 0), Pt at 2d (1/3, 2/3, 3/4) and Ga at 2c (1/3, 2/3, 1/4) [15]. The result of the refinement at 300 K is shown in Figure S11, which shows an excellent fit between the observed and calculated peak profiles by accounting for all the Bragg peaks. This suggests the average hexagonal structure ($P6_3/mmc$) of MnPtGa at 300 K. The lattice parameters obtained after the refinements are $a = b = 4.33765(7)$ Å and $c = 5.59143(9)$ Å, which is in good agreement with the previous reports [15,41].

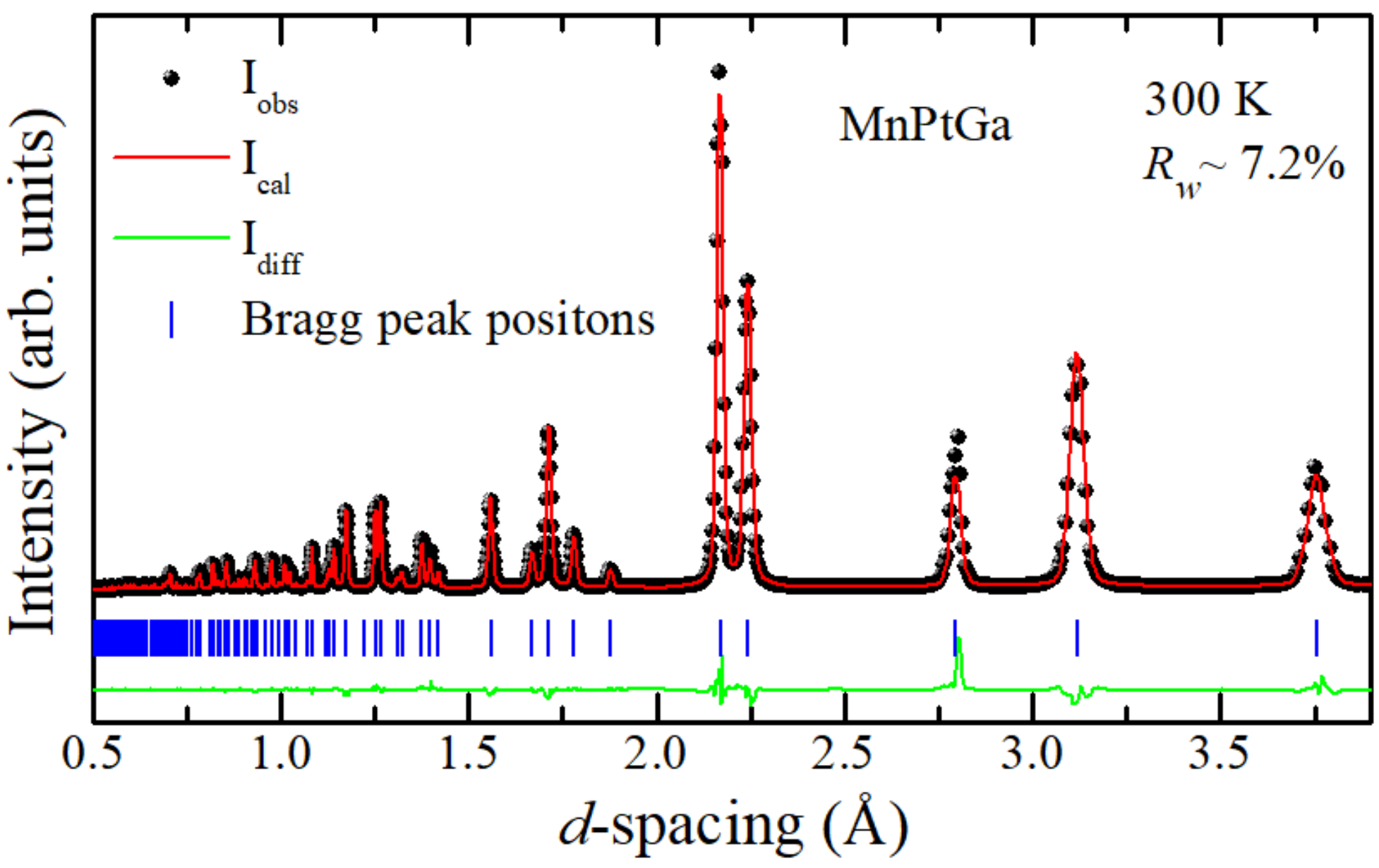

**Figure S11:** The observed profile (black spheres), calculated profile (red line), difference profile (green line), Bragg peak positions (blue ticks) and weighted agreement factor ($R_w$) obtained after Rietveld refinement for the hexagonal structure in the $P6_3/mmc$ space group using high-*Q* synchrotron x-ray powder diffraction data of MnPtGa at 300 K.

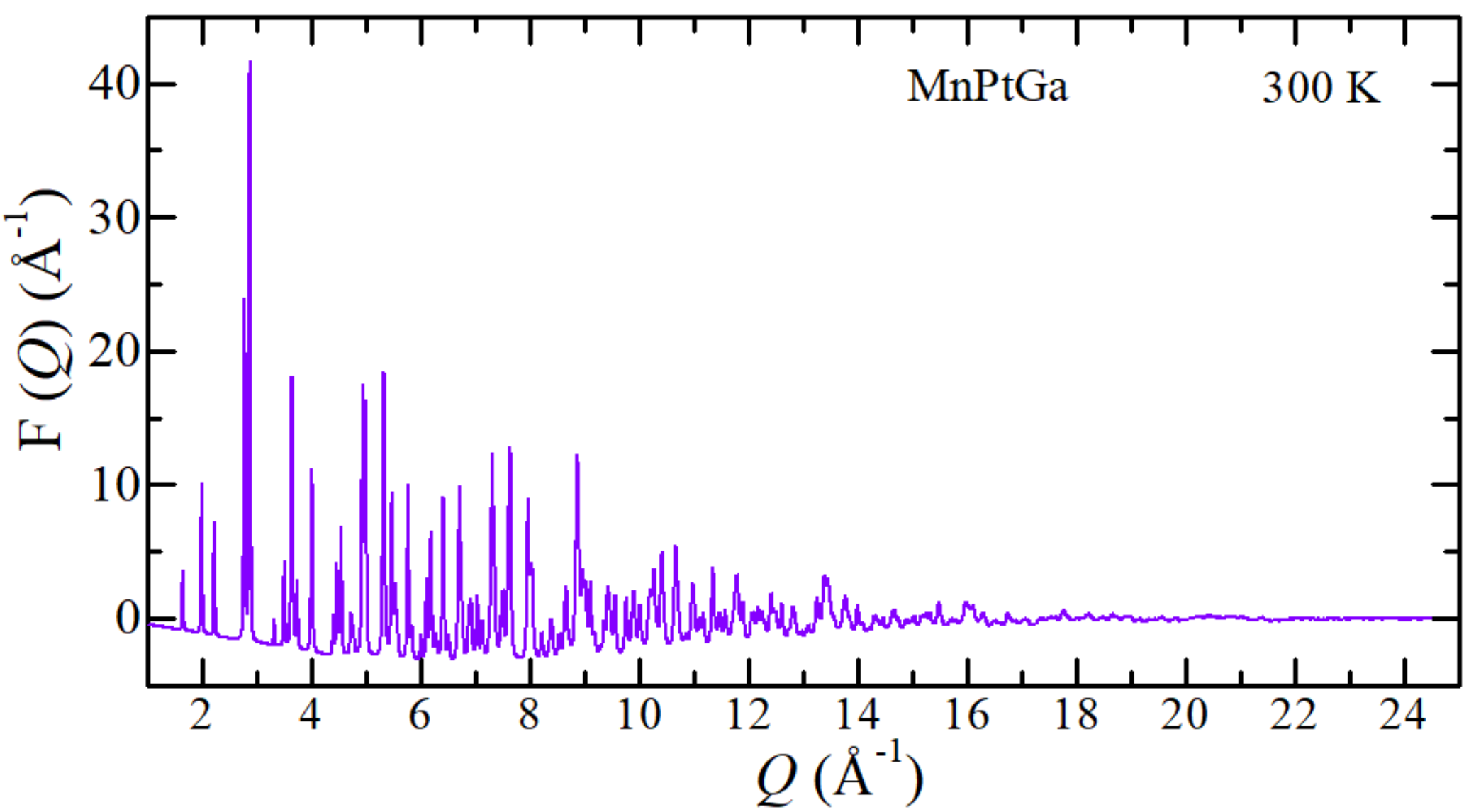

**Figure S12:** The reduced structure function F(*Q*) versus *Q* of MnPtGa at 300 K.

The reduced structure function F(*Q*) is obtained from the high-*Q* SXRPD patterns (shown in Figure S11) and is depicted in Figure S12 (see section IV for more detail about conversion). The intensity of the peaks in the F(*Q*) diminishes significantly towards higher *Q* values suggesting the dominance of the diffuse scattering. After that, following the similar procedure as described for MnNiGa in section IV, the atomic PDF in the SR, MR and LR regimes at room temperature is shown in Figure S13a-S13c. Fourier ripples are clearly visible below the first interatomic distance at $r$~2.48 Å in the Figure S13a. All the PDF peaks at $r\geq 2.48$ Å (first interatomic distance) are related to sample only as can be seen in the Table S6.

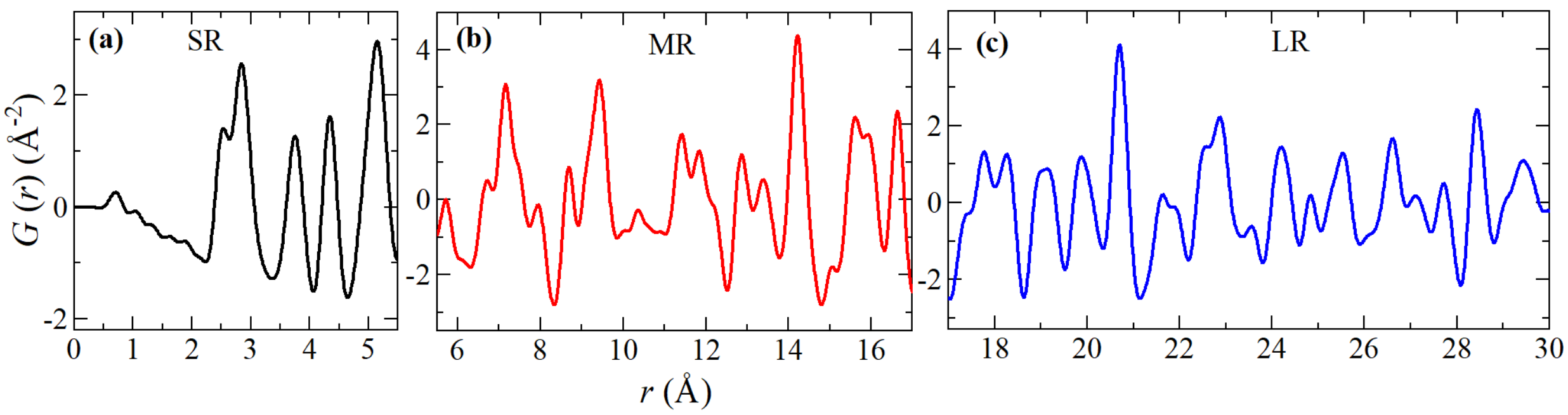

**Figure S13:** The experimental atomic PDF *G(r)* of MnPtGa at room temperature in the regimes (a) short-range (SR), (b) medium-range (MR) and (c) long-range (LR).

**TABLE S6.** Comparison of bond distances obtained from the real-space PDF refinements for the hexagonal ($P6_3/mmc$) and trigonal ($P3m1$) structures in the short-range regime using the experimental PDF data of MnPtGa at 300 K.

| S. N. | Pairs | Bond distances for $P6_3/mmc$ (Å) | Pairs | Bond distances for $P3m1$ (Å) |
|---|---|---|---|---|
| 1 | Pt-Ga<br>Pt-Ga | 2.487(3)<br>2.525(8) | Pt-Ga | 2.474(2), 2.485(2), 2.519(5), 2.523(5) |
| 2 | Pt-Ga= Mn-Mn<br>Mn-Pt=Mn-Ga<br>Mn-Pt=Mn-Ga | 2.805(4)<br>2.85(1)<br>2.88(1) | Pt-Ga<br>Mn-Mn<br>Mn-Ga<br>Mn-Ga<br>Mn-Pt<br>Mn-Pt<br>Mn-Pt | 2.556(1), 2.778(4), 3.13(1)<br>2.78(1),<br>2.642(1), 2.794(2), 2.828(2)<br>2.907(3), 2.939(3), 3.137(5), 3.167(5)<br>2.765(2), 2.799(2), 2.811(2), 2.814(2)<br>2.844(2), 2.901(3), 2.933(3), 2.941(3),<br>2.973(3) |
| 3 | Ga-Ga=Pt-Pt<br>Ga-Ga-Pt-Pt | 3.75(1)<br>3.77(1) | Ga-Ga<br>Pt-Pt | 3.460(7), 3.487(7), 4.04(1), 4.07(1)<br>3.678(9), 3.704(8), 3.812(9), 3.836(9) |
| 4 | Mn-Mn=Pt-Pt=Ga-Ga | 4.352(8) | Mn-Mn=Pt-Pt=Ga-Ga | 4.346(3) |
| 5 | Mn-Ga= Mn-Pt<br>Mn-Ga= Mn-Pt<br>Pt-Ga<br>Pt-Ga | 4.88(2)<br>4.90(2)<br>5.01(1)<br>5.050(6) | Mn-Ga<br>Mn-Ga<br>Mn-Pt<br>Mn-Pt<br>Pt-Ga<br>Pt-Ga | 4.44(1), 4.46(1), 4.78(1), 4.80(1)<br>4.98(1), 5.006(1), 5.086(1), 5.123(1)<br>4.73(1), 4.75(1), 4.79(1), 4.81(1)<br>4.95(1), 4.97(1), 5.04(1), 5.151(1)<br>5.001(1), 5.007(1), 5.023(1), 5.044(1)<br>5.061(1), 5.158(6) |
| 6 | Mn-Mn=Pt-Ga<br>Mn-Ga= Mn-Pt<br>Mn-Ga= Mn-Pt | 5.17(1)<br>5.205(7)<br>5.24(1) | Mn-Mn<br>Pt-Ga<br>Mn-Ga<br>Mn-Ga<br>Mn-Pt<br>Mn-Pt<br>Ga-Ga | 5.164(6), 5.178(6)<br>5.184(6), 5.356(8)<br>5.167(1), 5.204(1), 5.229(1), 5.265(1)<br>5.32(1), 5.33(2), 5.360(3), 5.395(3)<br>5.176(1), 5.188(1), 5.213(1), 5.226(1)<br>5.248(2), 5.262(1), 5.284(2)<br>5.555(4), 5.589(4) |
| 7 | Mn-Mn=Pt-Pt=Ga-Ga | 5.61(2) | Mn-Mn=Pt-Pt=Ga-Ga | 5.60(2) |

## IX. Analysis of high-$Q$ SXRPD data collected at P21.1 beamline of PETRA-III

We note that although present results of local trigonal structure ($P3m1$) for MnNiGa and MnPtGa compounds are determined using the high-quality atomic PDF data analysis, any trivial unintentional experimental errors might significantly modify the outcomes due to the fact of highly-sensitivity of such atomic PDF relying on the experimental data quality [19,25]. Therefore in order to further confirm the robustness of our experimental results and data reproducibility, the high-$Q$ SXRPD data were also collected using a 100 keV x-ray beam (with much shorter $\lambda \sim 0.1204$ Å and higher $Q_{maxinst} \sim 29$ Å$^{-1}$) at the P21.1 beamline of PETRA-III, DESY [6]. This provides better peak-to-peak PDF resolution in the real-space compared to the data collected using 60 keV x-ray beam at P02.1. In the first step, reciprocal space Rietveld refinements is carried out using such high-$Q$ SXRPD data. The result of such refinement at 300 K is shown in Figure S14 which shows excellent fit between the observed and calculated profiles by accounting for all the Bragg peaks and confirms the average hexagonal structure ($P6_3/mmc$) of MnNiGa. Overall behavior of this data looks in excellent match with data collected at P02.1 (see Figure S3) albeit, reciprocal space resolution is compromised due to availability of higher $Q_{max}$-value for the data recorded at P21.1, which is expected.

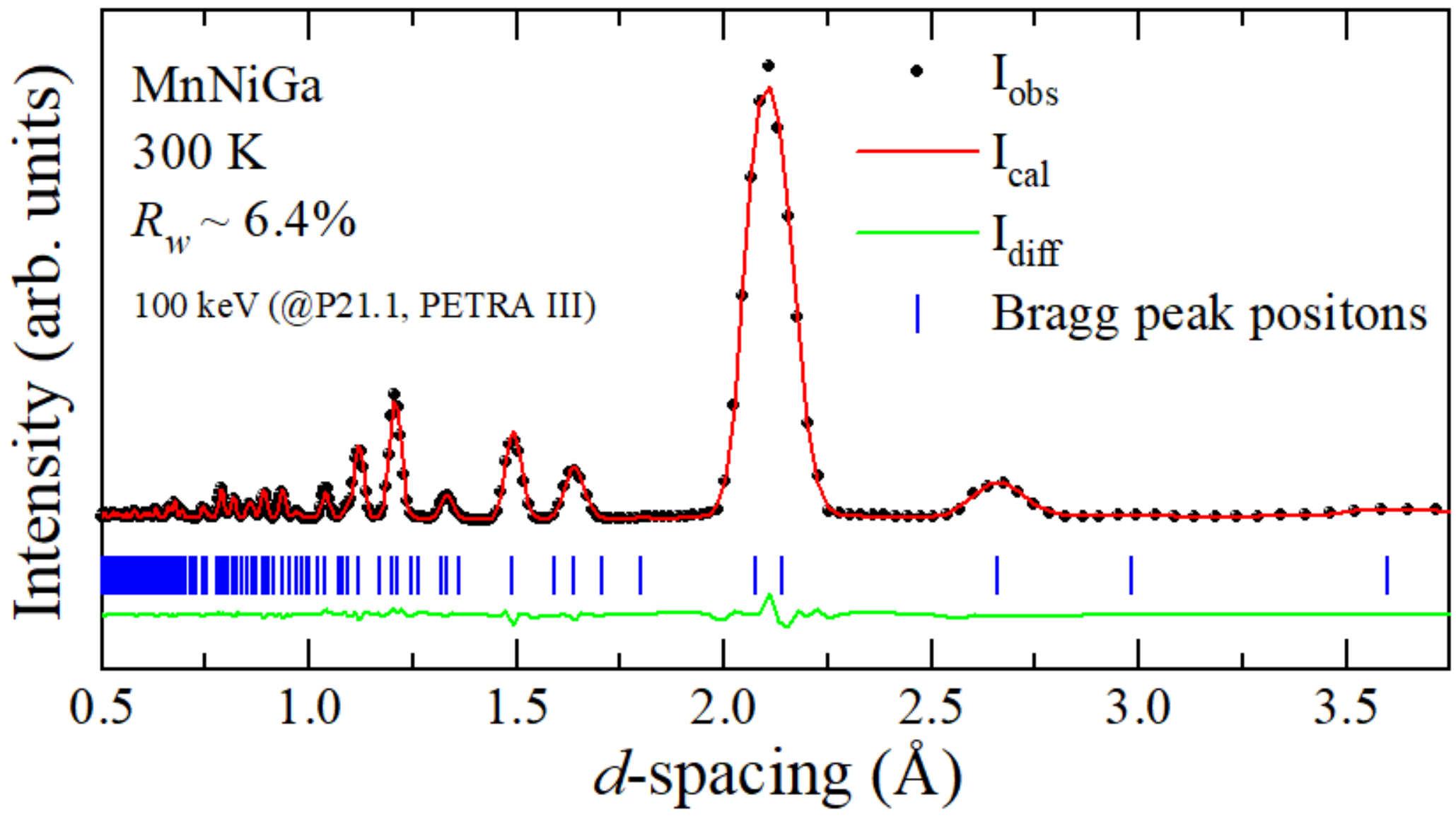

**Figure S14:** The observed profile (black spheres), calculated profile (red line), difference profile (green line), Bragg peak positions (blue ticks) and weighted agreement factor ($R_w$) obtained after Rietveld refinement for the hexagonal structure in the $P6_3/mmc$ space group using room temperature high-$Q$ SXRPD data collected at P21.1 beamline of PETRA-III.

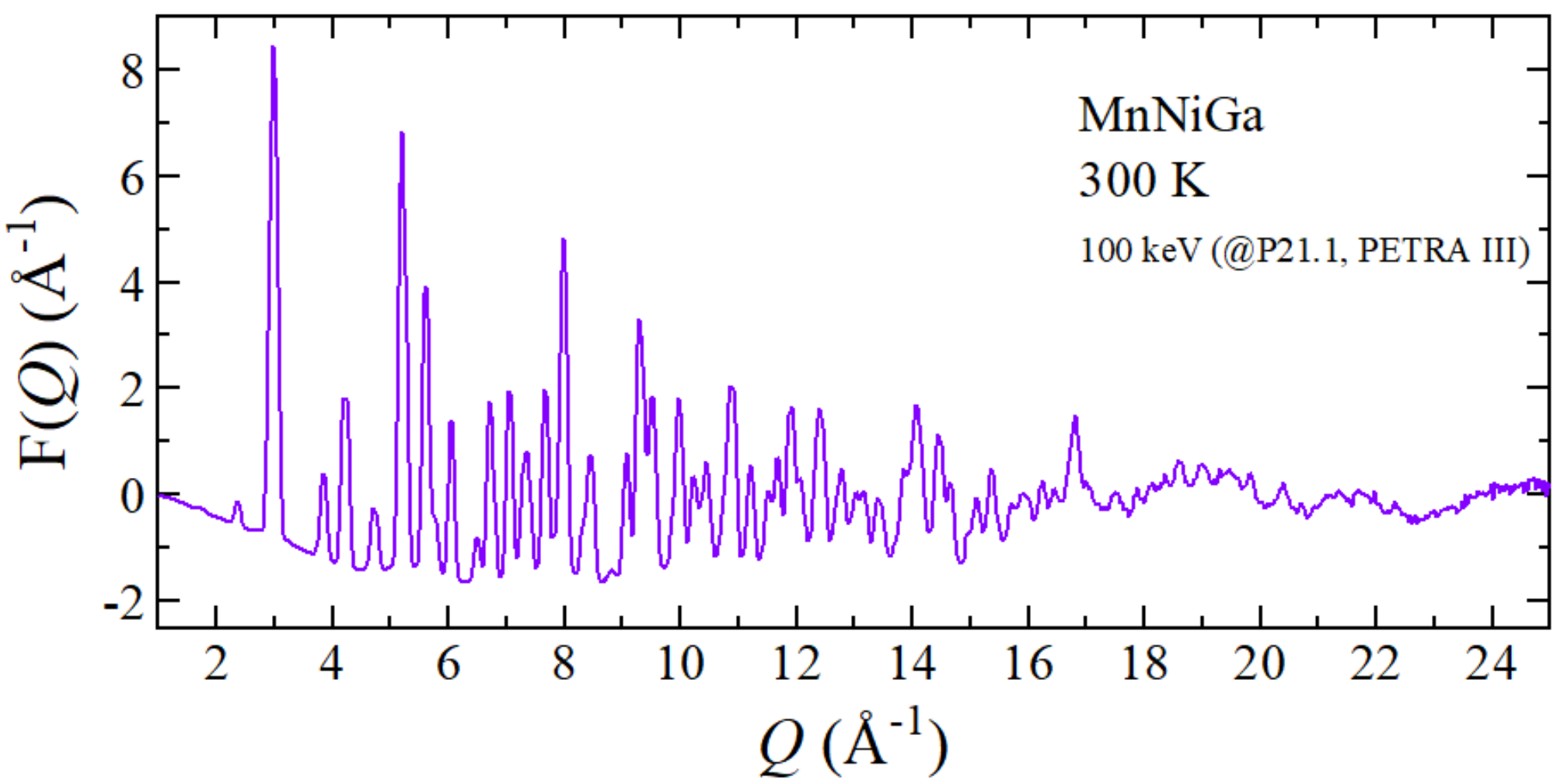

**Figure S15:** The reduced structure function F($Q$) versus $Q$ of MnNiGa at 300 K. This is obtained from the high-$Q$ SXRPD data collected at P21.1 beamline of PETRA-III.

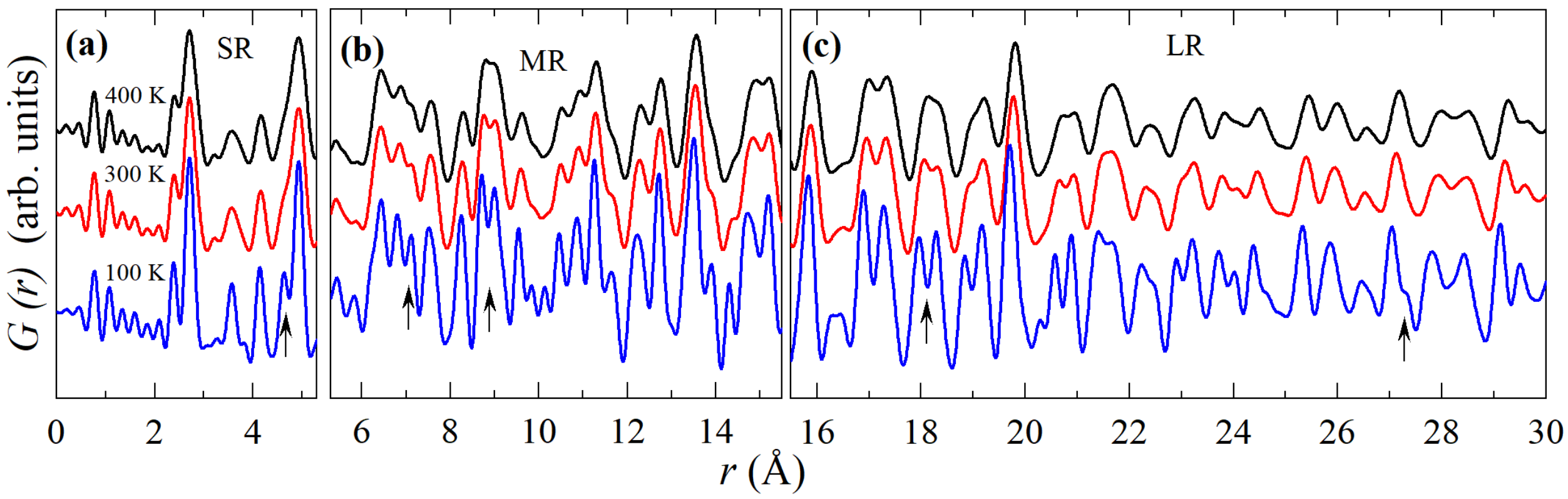

**Figure S16:** Experimental atomic PDF $G(r)$ of MnNiGa on the vertically-spaced scale at the three selected temperatures (400, 300 and 100 K) in the **(a)** short-range (SR), **(b)** medium-range (MR) and **(c)** long-range (LR) regimes. This is obtained from the high-Q SXRPD data collected at P21.1 beamline of PETRA-III. The arrow marks in **(a)-(c)** highlight some of the peaks, which gets sharpen/splitted at low temperature. Fourier ripples are clearly visible below the first physical atomic distance $r$~2.38 Å.

Subsequently, the reduced structure function F($Q$) is obtained after the standard corrections and normalization to the high-$Q$ SXRPD data (see section IV for more detail) and depicted in Figure S15. The F($Q$) intensity diminishes significantly towards higher $Q$ values due to the dominance of the diffuse scattering [19]. Following the similar procedure as described above in section IV, the atomic PDF is obtained by taking the Fourier transform of F($Q$) up to $Q_{max}$ = 25 Å$^{-1}$. Such PDFs in the SR, MR and LR regimes at the three selected temperatures are shown in Figure S16a-S16c which looks in nice agreement with the PDFs obtained from the P02.1 beamline (see Figure S5). Due to the higher $Q_{max}$-value accessible at the P21.1, the real-space PDF peak resolution is better compared to the PDFs obtained from the P02 as sharper PDF peaks can be seen in Figure S16 compared to Figure S5.

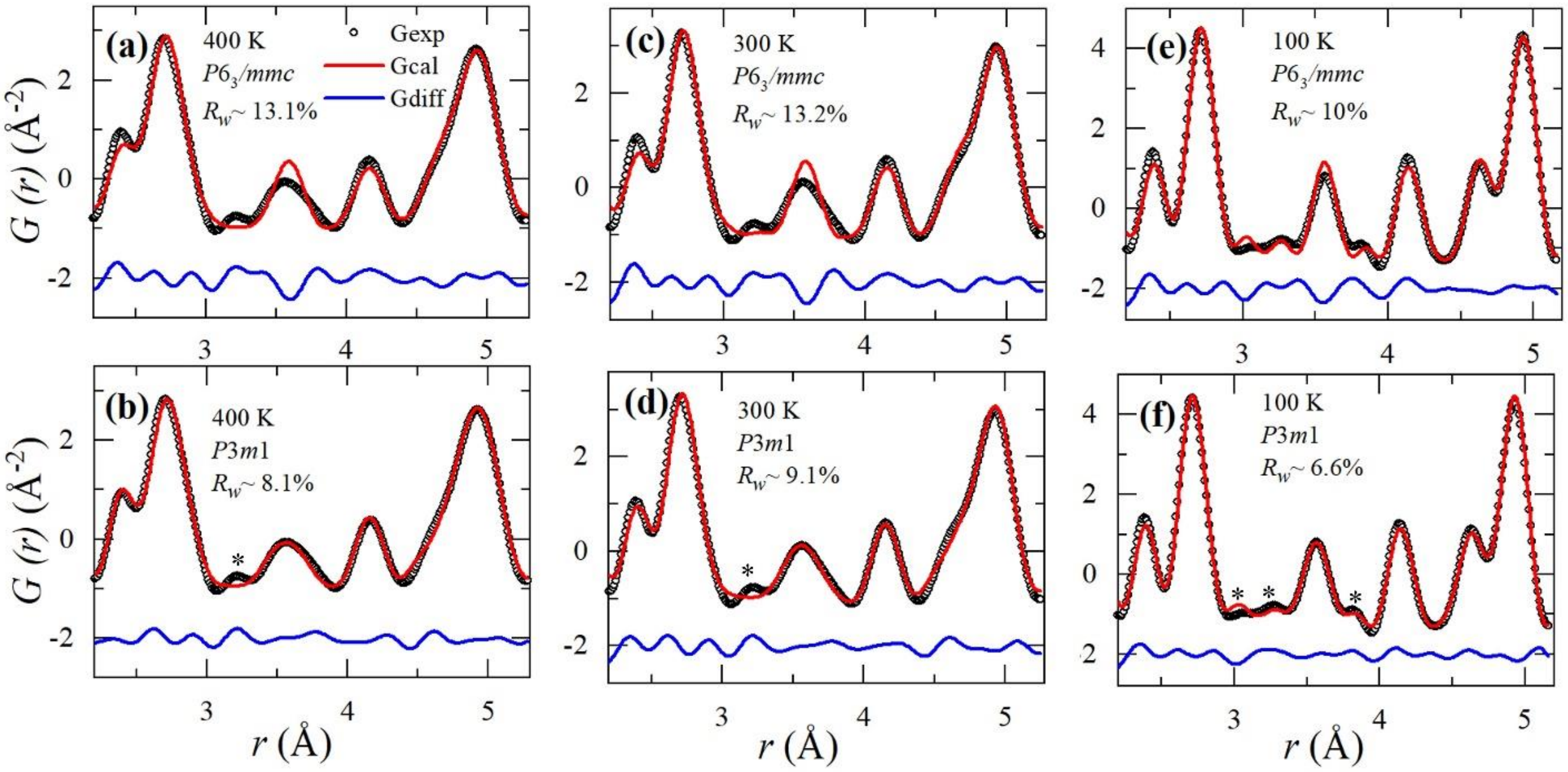

**Figure S17:** The experimental PDF (black circles), calculated PDF (red line), their difference (blue line at the bottom) and weighted agreement factor ($R_w$) obtained by real-space structure refinements in the SR regime using the $P6_3/mmc$ and $P3m1$ space groups for MnNiGa at the three indicated temperatures **(a-b)** 400 K, **(c-d)** 300 K and **(e-f)** 100 K. The "*" marks in **(b)**, **(d)** and **(f)** highlight some of the Fourier ripples are extended above the first interatomic distance.

In the next step, following the similar procedure described above in section VI, the real-space structure refinement is carried out using the $P6_3/mmc$ and $P3m1$ space groups. The results of the such refinements in the SR regime at the three selected temperatures (400, 300 and 100 K) are shown in Figure S17a-S17f which reveals significantly better fit for the $P3m1$ with considerably lower $R_w$-value in comparison to the $P6_3/mmc$. All these results confirm local structure of MnNiGa is trigonal with the $P3m1$ space group. These atomic PDFs and fits are in nice agreement with the results obtained from the data collected at P02.1 (see Figure 3 of the main text and Figure S10). We note that there are some tiny Fourier termination ripples appearing even above the first physical interatomic distance $r$~2.38 Å in the SR regime as marked by "*" in Figure S17. This is due to fact that the higher $Q_{max}$-value available at P21.1 allows to achieve a better spatial PDF peak resolution, which reveals some tiny Fourier ripples hidden in the SR-range regime. It is worth to emphasize here that the intensity of such Fourier ripples with a wavelength ~$2\pi/Q_{max}$ is known to die off exponentially with distance $r$ [19], however it may travel up to a few interatomic distances in the SR regime depending on the $Q_{max}$-value [19]. However, since the real-space structure refinement program PDFgui, employed in this study, takes into account the Fourier ripples [26], thereby the misfits appearing between the observed and calculated PDFs are genuinely part of the atomic crystal structure and provide a direct information about the atomic crystal structure.

## X. Theoretical Calculations

We performed first-principles based theoretical calculations for both MnNiGa and MnPtGa systems to understand their ground state structure. For both cases, we studied a scenario in which the material has a globally high symmetry structure with space group $P6_3/mmc$, but due to local distortion, there is also a locally deformed low symmetric structure with space group $P3m1$ present within it. We refer it as the mixed structure. It should be noted that such local distortions have been seen experimentally in the case of metal chalcogenides where the local-off centering of the metal atom creates local dipoles [45] or softening of the phonon modes [46,47] leading to the lowering of the thermal conductivity.

In the present study, to simulate such mixed structure, we consider a supercell that is formed by $3 \times 3 \times 3$ and $5 \times 5 \times 5$ $P6_3/mmc$ supercells and the unit cell at the center was replaced by one-unit cell of $P3m1$ structure. This is indicated by the light green region in the mixed structure shown in Figure 6 of the main text for the $3 \times 3 \times 3$ supercell and in Figure S18 for the $5 \times 5 \times 5$ supercell. For the calculation with the pure structures ($P6_3/mmc$ and $P3m1$), experimentally obtained lattice parameters as well as atomic positions have been used. However, for the mixed structure (one with local distortion), the $P3m1$ structure along with one outer layer of atoms (light gray region in the mixed structure), are allowed to relax while keeping the outer layers fixed in their experimentally obtained lattice parameters. The result of the calculation with the $5 \times 5 \times 5$ supercell is shown in Figure S18 (see Figure 6 of the main text for the $3 \times 3 \times 3$ supercell). Similar results for both MnPtGa and MnNiGa structures are found. In both cases (MnPtGa and MnNiGa), the mixed structure has a lower ground state energy in comparison with the pure $P3m1$ and $P6_3/mmc$ structures.

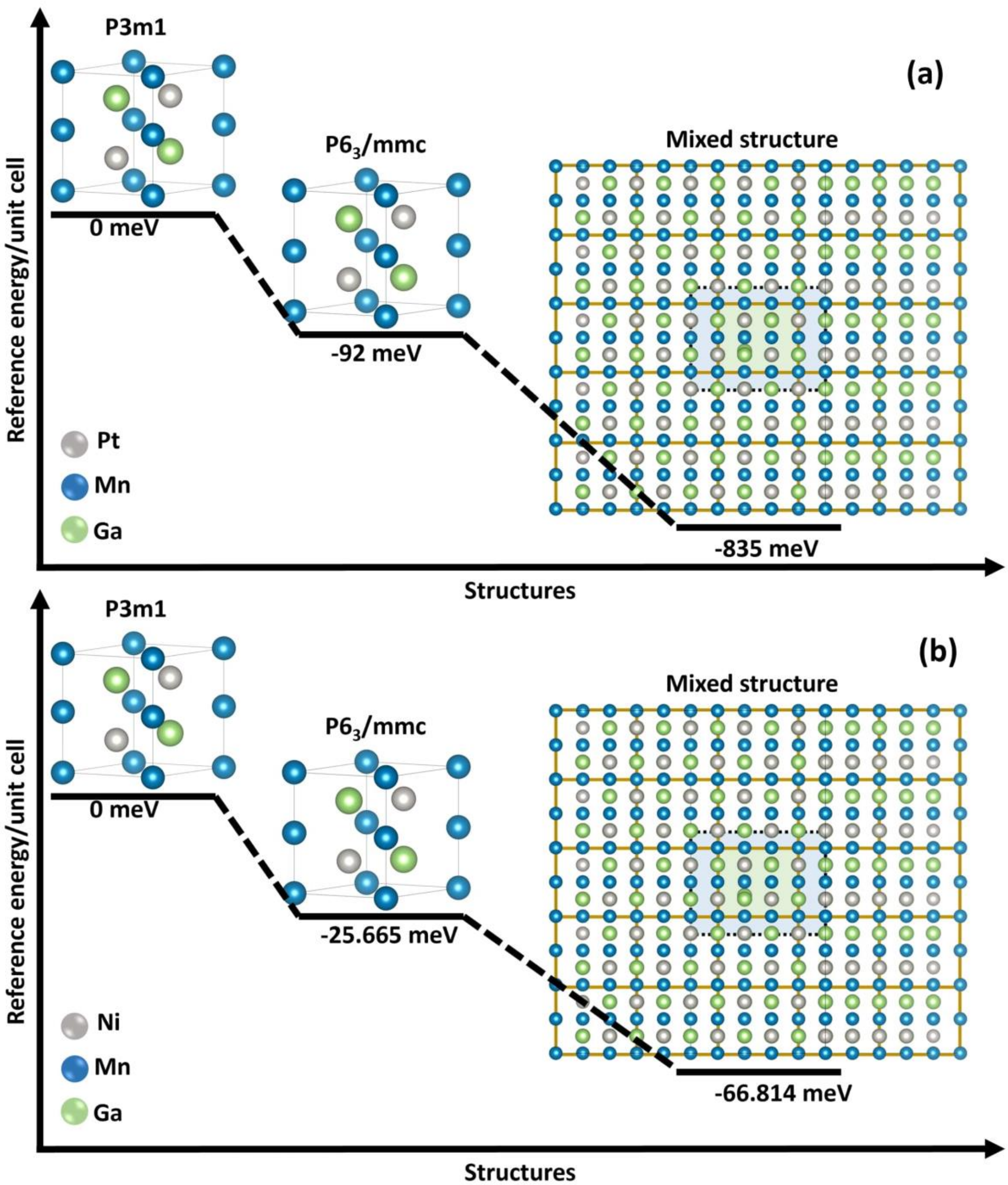

**Figure S18:** Schematic diagram for the comparison of ground state energy per unit cell for different structures for **(a)** MnPtGa and **(b)** MnNiGa. The $P3m1$ and $P6_3/mmc$ structures are considered in a single unit cell, and the mixed structure is considered as $5 \times 5 \times 5$ supercell, wherein the light green region shows one-unit cell of trigonal $P3m1$ structure replaced in place of the unit cell at the center of the $P6_3/mmc$ supercell. The light gray region in the mixed structure shows one outer layer of atoms which is allowed to relax with P3$m$1 (see text for more detail). Blue, green, and white balls correspond to Mn, Ga, and Pt/Ni, respectively.

For the case of MnPtGa, the ground state energy of the mixed structure is found to be lower than those for the $P6_3/mmc$ and $P3m1$ structures by 743 and 835 meV/f.u., respectively (see Figure S18a). In case of MnNiGa, the ground state energy of the mixed structure is lower than those for the $P6_3/mmc$ and $P3m1$ structures by 41.149 and 66.814 meV/f.u., respectively (see Figure S18b). This manifest that the mixed structure is more stable than the global $P3m1$ and $P6_3/mmc$ symmetry structures in both MnNiGa and MnPtGa compounds. Thus, these calculations further support our experimental findings of local trigonal ($P3m1$) and global hexagonal ($P6_3/mmc$) structure in these compounds.

In addition, we found that spin-canting (non-collinearity) consideration in the $P3m1$ cell of the mixed structure for the MnPtGa $5 \times 5 \times 5$ supercell led to further lowering of the energy to -837 meV/f.u., i.e., providing 2 meV/f.u. energy gain relative to the mixed structure without canting. This manifest stabilization of non-collinear vortex-type spin textures upon consideration of the spin-canting in the mixed structure of MnPtGa. For the MnNiGa $5 \times 5 \times 5$ supercell, the consideration of spin-canting did not provide any considerable drop in total energy relative to the mixed structure. This is probably due to the fact of small spin-orbit coupling for MnNiGa compared to MnPtGa as discussed in the main text. However, in both compounds the mixed structure along with the non-collinearity consideration contains significant energy lowering compared to the $P6_3/mmc$ and $P3m1$ structures, manifesting the presence of non-collinear mixed structure.

We have performed all the calculations using the density functional theory (DFT) as implemented in the Vienna Ab initio Software Package (VASP) code [48,49]. For the core and valence electrons, PAW (projected augmented wave) potentials and plane-wave basis sets were used with the Perdew–Burke–Ernzerhof (PBE) functional to get the electronic energy [50,51]. All energies converged within a cutoff of 450 eV. The conjugate gradient algorithm is used for structural optimization [52]. The convergence criteria for energy and force are $10^{-4}$ eV and -0.05 eVÅ$^{-1}$, respectively. In all calculations, spin polarization was enabled.

## References

[1] A. K. Singh, P. Devi, A. K. Jena, U. Modanwal, S.-C. Lee, S. Bhattacharjee, B. Joseph, and S. Singh, Physica Status Solidi (RRL)–Rapid Research Letters **16**, 2200057 (2022).
[2] B. Holt, J. Diaz, J. Huber, and C. A. Luengo, Revista Brasileira de Fisica **8** (1978).
[3] L. A. Turnbull *et al.*, ACS nano (2020).

[4] A.-C. Dippel, H.-P. Liermann, J. T. Delitz, P. Walter, H. Schulte-Schrepping, O. H. Seeck, and H. Franz, Journal of Synchrotron Radiation **22**, 675 (2015).
[5] M. Brunelli, J. P. Wright, G. B. Vaughan, A. J. Mora, and A. N. Fitch, Angewandte Chemie International Edition **42**, 2029 (2003).
[6] O. Ivashko *et al.*, Synchrotron Radiation **32** (2025).
[7] M. Hoelzel, A. Senyshyn, N. Juenke, H. Boysen, W. Schmahl, and H. Fuess, Nuclear Instruments and Methods in Physics Research Section A: Accelerators, Spectrometers, Detectors and Associated Equipment **667**, 32 (2012).
[8] R. A. Young, *The Rietveld Method* (Oxford university press Oxford, 1993), Vol. 6.
[9] J. Rodrigues-Carvajal, FULLPROF, a Rietveld refinement and pattern matching analysis program version 2016, Laboratoire Leon Brillouin, CEA-CNRS, France http://www.ill.eu/sites/fullprof/.
[10] W. Wang *et al.*, Adv. Mater. **28**, 6887 (2016).
[11] G. Xu *et al.*, Physical Review B **100**, 054416 (2019).
[12] G. E. Bacon, *X-Ray and Neutron Diffraction: The Commonwealth and International Library: Selected Readings in Physics* (Elsevier, 2013).
[13] K. Buschow and P. van Engen, Phys. Status Solidi (a) **76**, 615 (1983).
[14] K. Buschow and D. De Mooij, Journal of the Less Common Metals **99**, 125 (1984).
[15] J. A. Cooley, J. D. Bocarsly, E. C. Schueller, E. E. Levin, E. E. Rodriguez, A. Huq, S. H. Lapidus, S. D. Wilson, and R. Seshadri, Physical Review Materials **4**, 044405 (2020).
[16] R. Ibarra, E. Lesne, B. Sabir, J. Gayles, C. Felser, and A. Markou, Advanced Materials Interfaces, 2201562 (2022).
[17] R. Ibarra *et al.*, Applied Physics Letters **120**, 172403 (2022).
[18] G. Dwari *et al.*, Physical Review B **110**, 045111 (2024).
[19] T. Egami and S. J. Billinge, *Underneath the Bragg Peaks: Structural Analysis of Complex Materials* (Elsevier, 2003).
[20] V. Petkov, Characterization of Materials, 1 (2002).
[21] T. Proffen, S. Billinge, T. Egami, and D. Louca, Zeitschrift für Kristallographie-Crystalline Materials **218**, 132 (2003).
[22] P. Juhás, T. Davis, C. L. Farrow, and S. J. Billinge, Journal of Applied Crystallography **46**, 560 (2013).
[23] I.-K. Jeong, Physical Review B—Condensed Matter and Materials Physics **79**, 052101 (2009).
[24] I.-K. Jeong, J. Ahn, B. Kim, S. Yoon, S. P. Singh, and D. Pandey, Physical Review B **83**, 064108 (2011).
[25] P. F. Peterson, E. S. Božin, T. Proffen, and S. J. Billinge, Journal of applied crystallography **36**, 53 (2003).
[26] C. Farrow, P. Juhas, J. Liu, D. Bryndin, E. Božin, J. Bloch, T. Proffen, and S. Billinge, Journal of Physics: Condensed Matter **19**, 335219 (2007).
[27] I.-K. Jeong, R. Heffner, M. Graf, and S. Billinge, Physical Review B **67**, 104301 (2003).
[28] R. J. Koch *et al.*, Physical Review Letters **126**, 186402 (2021).
[29] A. S. Masadeh, M. T. Shatnawi, G. Adawi, and Y. Ren, Modern Physics Letters B **33**, 1950410 (2019).
[30] A. Masadeh, E. Božin, C. Farrow, G. Paglia, P. Juhas, S. Billinge, A. Karkamkar, and M. Kanatzidis, Physical Review B **76**, 115413 (2007).

[31] K. Knox, E. Bozin, C. Malliakas, M. Kanatzidis, and S. Billinge, Physical Review B **89**, 014102 (2014).
[32] M. Dutta, K. Pal, M. Etter, U. V. Waghmare, and K. Biswas, Journal of the American Chemical Society **143**, 16839 (2021).
[33] A. Vasdev, M. Dutta, S. Mishra, V. Kaur, H. Kaur, K. Biswas, and G. Sheet, Scientific Reports **11**, 17190 (2021).
[34] G. Oliveira *et al.*, Physical Review B **86**, 224418 (2012).
[35] K. Trueblood, H.-B. Bürgi, H. Burzlaff, J. Dunitz, C. Gramaccioli, H. Schulz, U. Shmueli, and S. Abrahams, Acta Crystallographica Section A: Foundations of Crystallography **52**, 770 (1996).
[36] D. M. H. H. T. Stokes, and B. J. Campbell, ISOTROPY Software Suite, iso.byu.edu.
[37] C. R. Haines, C. J. Howard, R. J. Harrison, and M. A. Carpenter, Acta Crystallographica Section B: Structural Science, Crystal Engineering and Materials **75**, 1208 (2019).
[38] M. I. Aroyo, J. M. Perez-Mato, C. Capillas, E. Kroumova, S. Ivantchev, G. Madariaga, A. Kirov, and H. Wondratschek, Zeitschrift für Kristallographie-Crystalline Materials **221**, 15 (2006).
[39] E. Liu *et al.*, Nat. Commun. **3**, 873 (2012).
[40] J. Greedan, D. Gout, A. Lozano-Gorrin, S. Derahkshan, T. Proffen, H.-J. Kim, E. Božin, and S. Billinge, Physical Review B **79**, 014427 (2009).
[41] A. K. Srivastava, P. Devi, A. K. Sharma, T. Ma, H. Deniz, H. L. Meyerheim, C. Felser, and S. S. Parkin, Advanced Materials **32**, 1904327 (2020).
[42] E. S. Bozin *et al.*, Nature Communications **10**, 3638 (2019).
[43] C. Li *et al.*, Physical Review B **90**, 214303 (2014).
[44] S. A. Kimber *et al.*, Nature Materials **22**, 311 (2023).
[45] E. S. Božin, C. D. Malliakas, P. Souvatzis, T. Proffen, N. A. Spaldin, M. G. Kanatzidis, and S. J. Billinge, Science **330**, 1660 (2010).
[46] Z.-Z. Luo *et al.*, Energy & Environmental Science **11**, 3220 (2018).
[47] M. Dutta, M. V. Prasad, J. Pandey, A. Soni, U. V. Waghmare, and K. Biswas, Angewandte Chemie **134**, e202200071 (2022).
[48] G. Kresse and J. Furthmüller, Computational Materials Science **6**, 15 (1996).
[49] G. Kresse and J. Furthmüller, Physical Review B **54**, 11169 (1996).
[50] G. Kresse and D. Joubert, Physical Review b **59**, 1758 (1999).
[51] P. E. Blöchl, Physical Review B **50**, 17953 (1994).
[52] J. P. Perdew, K. Burke, and M. Ernzerhof, Physical Review Letters **77**, 3865 (1996).